\documentclass[11pt]{article}

\usepackage{graphicx}
\usepackage{amsmath,amsthm}
\usepackage{amssymb,latexsym}
\usepackage{enumerate}
\usepackage{epigraph}
\usepackage[authoryear]{natbib}
\usepackage{mathrsfs}
\usepackage{verbatim}
\usepackage{mathabx}
\usepackage{relsize}
\usepackage{mathtools}
\usepackage{accents}
\usepackage{bm}
\usepackage[titletoc,title]{appendix}
\usepackage{fancyhdr}
\usepackage[shortlabels]{enumitem}
\usepackage{tikz}
\usepackage{lipsum}
\usepackage[toc]{glossaries}
\usepackage{makeidx}
\usepackage{url}
\usepackage{imakeidx}
\usepackage[letterpaper, margin=1in]{geometry}

\makeglossaries
\makeindex
\indexsetup{othercode=\small}

\theoremstyle{plain}

\theoremstyle{definition}

\theoremstyle{remark}

\newtheorem{exmp}{Example}[section]

\newglossaryentry{1/N.rule}
{
  name={$1/N$ rule},
  description={A probability rule that assigns to each of $N$ possibilities a probability of $1/N$ when very little is known about them. Same as \gls{law.of.equal.ignorance}}
}

\newglossaryentry{911}
{
  name={$9$/$11$},
  description={An abbreviation for the date of the World Trade Center attacks of September 11, 2001. Apart from its political and historical significance, 9/11 had a significant impact on the U.S. economy}
}

\newglossaryentry{accuracy}
{
  name={accuracy},
  description={In the theory of statistical estimation, an estimator is accurate if the expected value of its difference from the parameter is small, otherwise, inaccurate}
}
\newglossaryentry{active.portfolio}
{
  name={active portfolio},
  description={The active portfolio $A$ of portfolio $W$ against \gls{benchmark.portfolio} $B$ is the portfolio $A = W - B$}
}
\newglossaryentry{adjusted.nominal.gain}
{
  name={adjusted nominal gain},
  description={In terminology developed for this book, the \emph{adjusted nominal gain} is the total amount won (or if lost, a negative value) by a trading system when a method for determining trade size is used. See also \gls{nominal.gain}}
}
\newglossaryentry{ADR}
{
  name={ADR},
  description={An acronym for \emph{American Depository Receipts}. These are stocks of foreign companies that have listings on American stock exchanges and trade in dollars. Cash flows of these stocks are converted into dollars at the prevailing foreign exchange rate}
}
\newglossaryentry{affect}
{
  name={affect},
  description={Affect refers a specific quality of ``goodness'' or ``badness'' produced by \gls{System1} and in the words of Paul Slovic and colleagues, ``(i) experienced as a feeling state (with or without consciousness) and (ii) demarcating a positive or negative quality of a stimulus.'' Ordinary decision makers make heavy use of affect to make ``snap judgments'' based on feelings of like or dislike. Such feelings are conditioned by experience, and even a single episode is enough to condition an affect response}
}
\newglossaryentry{affect.heuristic}
{
  name={affect heuristic},
  description={A decision rule in which outcomes are parsed into the simple categories ``good' and ``bad'' using \gls{System1} heuristics. especially emotions (e.g., liking, hating), or moods (e.g. happy, sad, mad). In the common vernacular, ``gut feelings'' is the closest correspondent}
}
\newglossaryentry{aggregation}
{
  name={aggregation},
  description={As used in the \glslink{efficient.market.hypothesis}{EMH}, refers to the process of information dissemination when some players are more informed than others, i.e. to the manner in which prices adjust to reflect the information of a minority, the \glspl{.insider}. Also called \gls{information.aggregation}}
}
\newglossaryentry{agreement}
{
  name={agreement},
  description={In the \glspl{efficient.market.hypothesis}, the requirement that investors must agree on the correct prices of assets}
}
\newglossaryentry{algorithmic.strategy}
{
  name={algorithmic strategy},
  description={A \gls{strategy} that a part of a \gls{mechanical.trading.system}}
}
\newglossaryentry{algorithmic.trade}
{
  name={algorithmic trade},
  description={(1) In a \gls{PPGS.renewal.process}, all the \glspl{transaction} occuring from the time of the \gls{founding.transaction} until that transaction is closed, and (2) the part of a \gls{mechanical.trading.system}, usually computer-assisted, that performs computations to buy or sell}
}
\newglossaryentry{algorithmize}
{
  name={algorithmize},
  description={The process of taking a verbal or symbolic ``idea'' and converting it to an algorithm. For example: ``sort a set of numbers'' is algorithmized by a bubble sort}
}
\newglossaryentry{ambiguity.aversion}
{
  name={ambiguity aversion},
  description={\emph{Ambiguity aversion} is a preference for known risks over unknown risks, e.g. the Ellsberg paradox. Also known as \emph{uncertainty aversion}}
}
\newglossaryentry{anchor}
{
  name={anchor},
  description={A decision heuristic that uses some piece of information, even irrelevant information, as a reference point for a decision}
}
\newglossaryentry{anchoring.and.adjustment}
{
  name={anchoring \& adjustment},
  description={An effect in which the \gls{associative.machine} uses ambient information, including irrelevant information, to make a decision}
}
\newglossaryentry{anomaly}
{
  name={anomaly},
  description={In the context of the \gls{efficient.market.hypothesis}, a market phenomenon that has no apparent \acrshort{aEMH} explanation, i.e., an apparent contradition to the \acrshort{aEMH}},
  plural={anomalies}
}
\newglossaryentry{arb}
{
  name={arb},
  description={Nickname for an \gls{arbitrageur}}
}
\newglossaryentry{arbbed.out}
{
  name={arbbed out},
  description={Referring to a mispricing that has been corrected through \gls{arbitrage}}
}
\newglossaryentry{arbbing}
{
  name={arbbing},
  description={The process of engaging in \gls{arbitrage}}
}
\newglossaryentry{arbitrage}
{
  name={arbitrage},
  description={A strategy that buys and sells the same or statistically similar securities at advantageously different prices, e.g. index \gls{arbitrage}, \glslink{pairs.trading.strategy}{pairs trading}, options delta-neutral hedging. See also \gls{pure.arbitrage} and \gls{statistical.arbitrage}}
}
\newglossaryentry{Arbitrage.Pricing.Theory}
{
  name={Arbitrage Pricing Theory},
  description={A theory that extends the \glslink{Capital.Asset.Pricing.Model}{CAPM} by assuming that several factors besides the market are needed to model asset returns}
}
\newglossaryentry{arbitrageur}
{
  name={arbitrageur},
  description={A trader whose dominant strategies are \glspl{arbitrage}}
}
\newglossaryentry{ariely}
{
  name={Dan Ariely},
  description={Well-known researcher in behavioral finance, author of ``Predictably Irrational''}
}
\newglossaryentry{artificial.mathematical.game}
{
  name={artificial mathematical game},
  description={A stuctured contest between several mathematical entities (players) who engage strategically to determine their payoffs. See also \gls{game.theory}}
}
\newglossaryentry{asexual}
{
  name={asexual},
  description={An asexual organism does not require two parents to reproduce, but clones itself instead}
}
\newglossaryentry{aspiration.level}
{
  name={aspiration level},
  description={In the \gls{spa} theory, a means of assessing the attractiveness of a lottery using the probability that it is not below an aspiration level $\alpha$: $A = P[ V \ge \alpha]$, where $V$ equals the security-potential value}
}
\newglossaryentry{associative.machine}
{
  name={associative machine},
  description={A term used in this book in reference to the \gls{System1} method of storing memories in a network where they are connected with other memories that occurred at about the same time, at about the same place, or with respect to other salient characteristics}
}
\newglossaryentry{asymptotic.growth.rate}
{
  name={asymptotic growth rate},
  description={A \gls{geometric.growth.rate} which a time series or stochastic process approaches almost surely}
}
\newglossaryentry{attribute.substitution}
{
  name={attribute substitution},
  description={In the making of intuitive judgments, the operation of substituting one attribute for another in order to facilitate a decision. For example, instead of answering ``Which stock has the greatest prospects?", the question could morph into ``Which stock to I like the most?" See also \gls{availability} and the \gls{affect.heuristic}}
}
\newglossaryentry{Axiom.Of.Archimedes}
{
  name={Axiom of Archimedes},
  description={An axiom of utility theory (also called the \emph{\gls{Continuity.Axiom}}) which asserts that if $B$ is an event that is ``sandwiched'' between $A$ and $B$ ($A \succ B$ and $B \succ C$), then there exists a $p \in (0,1)$ such that $B$ and the prospect $(p, A; (1-p), C)$ are equivalent}
}
\newglossaryentry{availability}
{
  name={availability},
  description={In the psychology of decision making, referring to the (possibly biased) information used as input to decision making}
}
\newglossaryentry{availability.heuristic}
{
  name={availability heuristic},
  description={A decision heuristic that chooses based on readily available information}
}

\newglossaryentry{back.end}
{
  name={back end},
  description={A term used for the longer maturity instruments of the yield curve, typically ones with a maturity of $5$ years or more}
}
\newglossaryentry{backtesting}
{
  name={backtesting},
  description={The process of applying quantative methods to a trading system's historical or simulated performance to determine if it is worth trading}
}
\newglossaryentry{BnL.purely.idiosyncratic.model}
{
  name={B\&L purely idiosyncratic model},
  description={A \acrlong{abnl} evolutionary model in which each parent in each generation has a random number of offspring independently of all others}
}
\newglossaryentry{BnL.systematic.model}
{
  name={B\&L systematic model},
  description={A \acrlong{abnl} evolutionary model in which each parent has the same number of offspring as all others of that generation}
}
\newglossaryentry{bankroll}
{
  name={bankroll},
  description={Literally, in the vulgar vernacular of gamblers and grifters, a bulging roll of paper money used to finance wagers. Polite company drops the reference to paper money, preferring the euphemism \emph{investment capital}}
}
\newglossaryentry{bargaining.problem}
{
  name={bargaining problem},
  description={A game theory problem in which two players negotiate a fair price when there exists an interval of prices favorable to each. A solution was given by John Nash in $1950$}
}
\newglossaryentry{basis.point}
{
  name={basis point},
  description={In fixed income parlance with respect to interest rates, one one-hundredth of a percent, $0.01\% = 0.0001$}
}
\newglossaryentry{basket.of.stocks}
{
  name={basket of stocks},
  description={A portfolio of stocks that typically tracks a benchmark stock index, and which can be sold as part of a single trade}
}
\newglossaryentry{basket.trade}
{
  name={basket trade},
  description={A \gls{trade} of a \gls{basket.of.stocks}. Market makers specializing in trading entire baskets are available to instutional traders. Exghange traded funds \acrshort{aETF}'s are similar to \emph{basket trades} except that the ownership of the underlying securities is held in trust, not by the owner of \emph{ETF} shares}
}
\newglossaryentry{Bayes'rule}
{
  name={Bayes' rule},
  description={A property of probabilities that for events $A$ and $B$, relates $P[B|A]$ to $P[A]$, $P[B]$ and $P[A|B]$ by the formula, 
\[
  P[B|A] = P[A|B] P[B] / P[A]
\]},
}
\newglossaryentry{behavioral.economics}
{
  name={behavioral economics},
  description={The study of economic phenomena that arise from the joint behavior of human economic actors as opposed to rational automatons. BF has roots in anthropology, marketing, psychology, social psychology and sociology}
}
\newglossaryentry{behavioral.finance}
{
  name={behavioral finance},
  description={The study of finance which examines how market phenomena arise from the interactions of its human participants. BF has untrustworthy roots in anthropology, psychology, social psychology and sociology}
}
\newglossaryentry{belief.revision.bias}
{
  name={belief revision bias},
  description={The tendency of people to revise prior beliefs slower than would be optimal by \gls{Bayes'rule}. Also known as the \gls{conservatism.bias}}
}
\newglossaryentry{benchmark}
{
  name={benchmark},
  description={In finance, a numerical standard against which the performance an investment vehicle is measured, e.g., the S\&P 500 Industrial Index, the Russell 2000 Index or LIBOR. In applications, benchmarks are usually diversified indices created by financial exchanges, investment houses or financial information providers}
}
\newglossaryentry{benchmark.portfolio}
{
  name={benchmark portfolio},
  description={In finance, a portfolio of tradable securities that serves a benchmark} 
}
\newglossaryentry{bias}
{
  name={bias},
  description={In the theory of statistical estimation, an estimator is \emph{unbiased} if its expected value equals the parameter being estimated, \emph{biased} otherwise}
}
\newglossaryentry{binary.channel}
{
  name={binary channel},
  description={In \glslink{shannon}{Shannon's} theory of communication, a system that has an input alphabet with two symbols, and output alphabet with two symbols, a ``channel'' that sends input to output symbols, and conditional probabilities that each output symbol will be received given that an input symbol is sent}
}
\newglossaryentry{binary.relation}
{
  name={binary relation},
  description={In mathematics, a binary relation on a set $\Omega$ is a set of ordered pairs $(x,y)$, where $x,y \in \Omega$}
}
\newglossaryentry{bouchaud}
{
  name={Jean-Phillipe Bouchaud},
  description={A French physicist born in 1962. He is founder and Chairman of Capital Fund Management (CFM) and professor of physics at \'{E}cole polytechnique}
}
\newglossaryentry{boundary.bias}
{
  name={boundary bias},
  description={A general problem with methods that perform smoothing of data in a bounded region. Toward the boundaries (edges) of the region, the dearth of data in a symmetric region around the prediction point reduces accuracy. In time series, for example, the ``present'' is a boundary, and the bias occurs because the future values are not available to regularize the smooth},
  plural={boundary biases}
}
\newglossaryentry{bounded.rationality}
{
  name={bounded rationality},
  description={A viewpoint that humans mostly behave rationally, but are constrained by limited capacities for mental computing and by difficulties in quantifying uncertainty}
}
\newglossaryentry{Buffett.Warren}
{
  name={Warren Buffett},
  description={The must famous investor of the latter part of the $20^{th}$ century and early $21^{st}$ century}
}

\newglossaryentry{calendar.time}
{
  name={calendar time},
  description={Referring to time measured in calendar units, e.g., July 2, 2014 at 11:06:42}
}
\newglossaryentry{cap.and.trade}
{
  name={cap-and-trade},
  description={A system of allocating pollution rights by issuing tradable licenses to pollute}
}
\newglossaryentry{capacity}
{
  name={capacity},
  description={Let $X$ be a set and $\mathscr{E}(X)$ be the set of subsets of $X$ (its power set). A capacity $W$ on $X$ is a function that maps $\mathscr{E}(X)$ to $[0,1]$ and has properties  $W(\phi) = 0$, $W(X) = 1$ and $W(A) \le W(B)$ if $A \subset B$}
}
\newglossaryentry{Capital.Asset.Pricing.Model}
{
  name={Capital Asset Pricing Model},
  description={An efficient markets model for pricing securities}
}
\newglossaryentry{capital.gains.overhang}
{
  name={capital gains overhang},
  description={For a tradable asset, the aggregate gains or losses, i.e., the sum over all (unrealized) differences between the current price and investors' purchase prices}
}
\newglossaryentry{CAR}
{
  name={CAR},
  description={Cumulative Abnormal Return}
}
\newglossaryentry{carry}
{
  name={carry},
  description={The \emph{carry} of an asset is the return obtained from holding it. Carry is positive if owning the asset earns money and negative if it costs money. For example, owning a note that pays $3\%$ on principal has a carry of $3\%$ minus the usually negligible costs (i.e. service charges on a brokerage account) of maintaining the position}
}
\newglossaryentry{carry.trade}
{
  name={carry trade},
  description={A strategy that finances an pspeculative by trades borrowing in a currency that has relatively low interest rates}
}
\newglossaryentry{cardinal.utility}
{
  name={cardinal utility},
  description={See \gls{utility.theory}}
}
\newglossaryentry{certainty.equivalent}
{
  name={certainty equivalent},
  description={In utility theory involving monetary prizes, the minimum cash amount that is acceptable to forego the lottery}
}
\newglossaryentry{cheater}
{
  name={cheater},
  description={A person or trading entity that engages in illegal activity. ``Unethical'' activity is excluded since in financial theory, such activity does not exist}
}
\newglossaryentry{choice.architecture}
{
  name={choice architecture},
  description={A concept in applied \gls{behavioral.finance} that highlights the impact of the framing of choices on the outcomes. Choice architecture is an application of \glslink{prospect.theory}{prospect theory's} demonstration that framing of a \gls{decision.problem} can decisively affect the outcome}
}
\newglossaryentry{closed-end.discount.puzzle}
{
  name={Closed End Discount Puzzle},
  description={An efficient market anomaly that has no good rational explanation. Closed end funds trade at steep discounts to book value most of the time, have increased issuance during bull markets and converge to book upon close-ending}
}
\newglossaryentry{closing.transaction}
{
  name={closing transaction},
  description={In a \gls{PPGS.renewal.process}, a \gls{transaction} that exits an \gls{algorithmic.trade} resulting in no residual position}
}
\newglossaryentry{cluster.analysis}
{
  name={cluster analysis},
  description={An analysis that uses a data-analytic methods for identifying groups within a multidimensional dataset}
}
\newglossaryentry{cognitive.dissonance}
{
  name={cognitive dissonance},
  description={The mental stress caused by either holding, or being confronted by, a belief that is incoherent with those of the \gls{associative.machine}}
}
\newglossaryentry{coefficient.of.absolute.risk.aversion}
{
  name={coefficient of absolute risk aversion},
  description={For \gls{utility.function} $U$, the measure $r(x) = -U''(x) / U'x)$}
}
\newglossaryentry{coefficient.of.relative.risk.aversion}
{
  name={coefficient of relative risk aversion},
  description={For \gls{utility.function} $U$, the measure $r(x) = -x U''(x) / U'x)$}
}
\newglossaryentry{cognition}
{
  name={cognition},
  description={In psychology, referring to individuals' mental processes that process information, which includes, inter alia those of attention, memory, comprehending, learning and using language, calculating, reasoning, problem solving, and decision making}
}
\newglossaryentry{coherent}
{
  name={coherent},
  description={In \gls{System1} perception, pattern recognition intended to match most closely the existing mental constructs of the \gls{associative.machine}. Coherence is fundamental to \glslink{System1}{System 1's} decision processes}
}
\newglossaryentry{competition.for.the.second.move}
{
  name={competition for the second move},
  description={Since there are often benefits to moving second in two-player games, due to the information revealed in the first mover's choice, there can be a ``competition'' to move second}
}
\newglossaryentry{Complete.Ordering}
{
  name={Complete Ordering},
  description={An axiom of utility theory that requires that for any two events $A$ and $B$, at least one of $A \succeq B$ and $B \succeq A$ must be true}
}
\newglossaryentry{complexity}
{
  name={complexity},
  description={The quality or state of a system that is not simple; in this work, of a dynamic system whose laws of motion are too complicated to be quantified using known laws}
}
\newglossaryentry{component.decomposition}
{
  name={component decomposition},
  description={A functional decomposition of multidimensional data into simpler, usually $1$-dimensional factors. Principal components analysis is an example}
}
\newglossaryentry{Compound.Equivalence}
{
  name={Compound Equivalence},
  description={An axiom of utility theory that requires that lotteries of lotteries must follow the multiplication law of probabilities}
}
\newglossaryentry{concave.strategy}
{
  name={concave strategy},
  description={A \gls{strategy} that sells when price is going up and buys when price is going down},
  plural={concave strategies}
}
\newglossaryentry{confirmation.bias}
{
  name={confirmation bias},
  description={The tendency to process new information in a way that confirms one's point of view. The bias arises from the \glslink{associative.machine}{associative machine's} preference for \gls{coherent} explanations}
}
\newglossaryentry{conservatism.bias}
{
  name={conservatism bias},
  description={The tendency of people to revise prior beliefs slower than would be optimal by \gls{Bayes'rule}. Also known as the \gls{belief.revision.bias}}
}
\newglossaryentry{Continuity.Axiom}
{
  name={Continuity Axiom},
  description={An axiom of utility theory (also called the \emph{Archimedian axiom}) which asserts that if $B$ is an event that is ``sandwiched'' between $A$ and $B$ ($A \succ B$ and $B \succ C$), then there exists a $p \in (0,1)$ such that $B$ and the prospect $(p, A; (1-p), B)$ are equivalent}
}
\newglossaryentry{continuous.auction.market}
{
  name={continuous auction market},
  description={A market in which participants may place orders to buy, sell or cancel at any time during the venue's hours of operation. Transactions in a continuous auction market occur only when a buyer meets a seller's price or seller meets a buyer's price}
}
\newglossaryentry{contrarian}
{
  name={contrarian},
  description={An investor who follows contrarian strategies. Contrarian strategies posit that financial instruments go through periods of being \gls{overbought} and \gls{oversold} and that these can be detected, yielding trading \glspl{edge}}
}
\newglossaryentry{convex.strategy}
{
  name={convex strategy},
  description={A \gls{strategy} that buys when price is going up and sells when price is going down},
  plural={convex strategies}
}
\newglossaryentry{coordination.game}
{
  name={coordination game},
  description={A game in which individuals must cooperate to achieve a jointly favorable outcome, but in any agreement to cooperate, most achieve less than their maximal outcome}
}
\newglossaryentry{cost.of.carry}
{
  name={cost of carry},
  description={The negative of \gls{carry}. For example, selling a note that pays $3\%$ on principal has a cost of carry of $3\%$ plus the usually negligible costs (i.e. service charges on a brokerage account) of maintaining the position}
}
\newglossaryentry{counter.cumulative.distribution.function}
{
  name={counter-cumulative distribution function},
  description={The function $S(x) = F_{\!_{>}}(x) = 1 - F_{\!_{\le}}(x)$, where $F_{\!_{\le}}(x)$ is a \gls{cumulative.distribution.function}. Also called a \gls{survival.function}}
}
\newglossaryentry{crowding}
{
  name={crowding},
  description={Crowding is a phenomenon in which ``too many'' \glspl{arb} attempt to exploit a mispricing, resulting in a ``tragedy of the commons'' effect. In some cases, the result is a market reversal in which the \gls{arbitrage} moves violently against the  \glspl{arb}, inflicting upon them significant losses. Some cases in which this was thought to have happened: (1) the collapse of the hedge fund \gls{LTCM} and (2) the Hedge Fund Crisis of August, 2007}
}
\newglossaryentry{cumulative.distribution.function}
{
  name={cumulative distribution function},
  description={The function $F_{\!_{\le}}(x) = P[X \le x]$. Often called simply the \gls{distribution.function}}
}
\newglossaryentry{cumulative.prospect.theory}
{
  name={cumulative prospect theory},
  description={Cumulative prospect theory was introduced by Kahneman and Tversky in 1992 to address certain problems with the original \gls{prospect.theory}, most notably its failure to preserve stochastic dominance}
}
\newglossaryentry{curse.of.dimensionality}
{
  name={curse of dimensionality},
  description={A term that refers to various problems that arise when analyzing data from high-dimensional spaces. For example, when variables representing dimensions are stochastically independent, then almost all the probability within a ball of radius $R$ will lie in a thin shell between $R$ and $R - \epsilon$}
}

\newglossaryentry{data.mining}
{
  name={data mining},
  description={In data science, techniques that attempt to find patterns in (generally large) datasets. Such techniques are succeptible to the \gls{problem.of.multiplicity}}
}
\newglossaryentry{decision.science}
{
  name={decision science},
  description={The study of human decision making in which decisions of other parties can be ignored, often called ``games against nature.'' Three branches are often distinguished: (1) normative, the ``correct'' way to make decisions, (2) descriptive, the way humans decide in reality, and (3) prescriptive, the study of improving decision making}
}
\newglossaryentry{decision.problem}
{
  name={decision problem},
  description={A mathematical problem that abstracts decision making for a player when no strategic behavior by other players is possible}
}
\newglossaryentry{decision.theory}
{
  name={decision theory},
  description={A term for normative \gls{decision.science}}
}
\newglossaryentry{decoding}
{
  name={decoding},
  description={The process of recovering an encoded message}
}
\newglossaryentry{decrementing.transaction}
{
  name={decrementing transaction},
  description={In a \gls{PPGS.renewal.process}, a \gls{transaction} for which the number of shares, units or contracts of the prior \gls{position} is decreased}
}
\newglossaryentry{decumulative.distribution.function}
{
  name={decumulative distribution function},
  description={The function $D(x) = P[ X \ge x]$. It is related to the \gls{counter.cumulative.distribution.function} $S(x)$ by $D(x) = S(x) + P[X = x]$ The decumulative distribution function differs from the \gls{counter.cumulative.distribution.function} only if there are probability mass points, e.g. discrete distributions}
}
\newglossaryentry{defined.benefit.plan}
{
  name={defined-benefit plan},
  description={An annuity that promises payments calculated by a specific formula, usually based on years of service and income earned during that time. Persons in a defined-benefit plan have little choice about how to invest; their main decision is when to begin withdrawing benefits}
}
\newglossaryentry{defined.contribution.plan}
{
  name={defined-contribution plan},
  description={An completely portable savings annuity that requires an employee to elect how much to contribute, what investments to select and when to begin withdrawing}
}
\newglossaryentry{degenerate.random.variable}
{
  name={degenerate random variable},
  description={A \gls{random.variable} that assumes a single value with probability $1$}
}
\newglossaryentry{degree.of.diffusion}
{
  name={degree of diffusion},
  description={An informal term that describes the extent to which a strategy is known by the community of traders. A strategy with low degree of diffusion has few followers, but the \gls{strategy.capacity} must be sufficient to present \gls{arbitrage} opportunities}
}
\newglossaryentry{descriptive.decision.science}
{
  name={descriptive decision science},
  description={See \gls{decision.science}}
}
\newglossaryentry{dimensional.reduction}
{
  name={dimensional reduction},
  description={The reduction of the complexity of a dataset or problem that has a large number of variables by ideally identifying a small number of functions of those variables that explain salient features}
}
\newglossaryentry{direction}
{
  name={direction},
  description={In mathematics, a vector having unit length}
}
\newglossaryentry{discretionary.trade}
{
  name={discretionary trade},
  description={A \gls{trade} initiated by a human trader using judgment, not a purely mechanical algorithm}
}
\newglossaryentry{disposition.effect}
{
  name={disposition effect},
  description={The tendency of investors to sell winners to early and hold losers too long}
}
\newglossaryentry{distortion.factor}
{
  name={distortion factor},
  description={Any exogenous news, constraints on trading, widely held beliefs, or widely used strategies that effect \glspl{price.impact}. For example, the fact that mutual funds cannot short stocks suggests that prices might be higher than they would be if shorting were allowed. Same as \gls{price.distorter}}
}
\newglossaryentry{distribution.function}
{
  name={distribution function},
  description={The function $F_{\!_{\le}}(x) = P[X \le x]$. Also called the \gls{cumulative.distribution.function}}
}
\newglossaryentry{Dot-com.Bubble}
{
  name={Dot-com Bubble},
  description={A remarkable modern bubble ($1996$-$2000$) powered by the fantasy of a ``new economy.'' The story line was that ``.com'' (read: ``dot com'') companies would in time eradicate conventional retail storefronts by offering products online. In a scant four years, this craze caused the Nasdaq composite index to quadruple, going from about $1,200$ to over $5,000$. But remarkably, an abrupt crash did not occur; instead, there occurred a slow, steady and punishing decline back to about $1,500$}
}
\newglossaryentry{dominant.paradigm}
{
  name={dominant paradigm},
  description={A \gls{market.paradigm} which is the best expression of market sentiment},
}
\newglossaryentry{dominated}
{
  name={dominate},
  description={In game theory, a strategy for a player is said to be \emph{dominated} is there is another strategy that always produces better payoffs. One solution method in games consists of successive elimination of dominated strategies, and chosing one of the remaining ones. Thie  method has limited usefulness for solving games because it seldom produces a unique solution}
}
\newglossaryentry{downsizing.transaction}
{
  name={downsizing transaction},
  description={In a \gls{PPGS.renewal.process}, a \gls{transaction} that decreases the absolute value of a current, nonzero \gls{position} without reducing it to zero}
}
\newglossaryentry{drawdown}
{
  name={drawdown},
  description={If not explicitly stated otherwise, a peak-to-trough drawdown of a financial time series is its greatest loss from a previous high usually expressed as a percentage. For example, a series that had a previous high of $1000$ and a lowest low later of $800$ would have at least a $20\%$ peak-to-trough drawdown}
}
\newglossaryentry{drawdown.loss.function}
{
  name={drawdown loss function},
  description={For a real-valued time series $\{X_s\}_{s=1}^{s=t}$ of gains, the largest cumulative loss $\mathcal{L}(X,t)$ from a previous high, i.e. for $V_s = \sum_{u=1}^{u=s} X_u$, $ s \ge 1$ and $V_0 = 0$,
\[
  \mathcal{L}(X,t) = \underset{0 \le s \le t}{max} \left\{ (\underset{0 \le u \le s}{max} V_u) \, - \, V_s ) \right\}.
\]
Note that $\mathcal{L}(X,t) \ge 0$}
}
\newglossaryentry{driftless.random.walk}
{
  name={driftless random walk},
  description={A random walk $X_t = \mu + X_{t-1} + U_t$ for which $\mu = 0$}
}
\newglossaryentry{DSSW.Model}
{
  name={DSSW Model},
  description={A model of bubbles in which (1) \glspl{fundamentalist} correct a mispricing, creating the illusion of a trend, (2) \glspl{rational.speculator} push price away from fundamental value after fundamental value is restored, and continue until they can offload it to \glspl{noise.trader}. In this work, called \gls{legal.pump.and.dump}}
}
\newglossaryentry{durability.bias}
{
  name={durability bias},
  description={Refers to the tendency for people to overestimate the persistence of a positive or negative event, episode or circumstance}
}

\newglossaryentry{ecology}
{
  name={ecology},
  description={In finance, an ecology at a moment in time includes both the \gls{strategic.ecology} and market structure, in the sense of a game that has rules, venues and technology within which strategies can be enacted. Also called \gls{market.ecology}}
}
\newglossaryentry{econ}
{
  name={Econ},
  description={A \acrshort{avnm} financial decision maker. Also called \gls{homo.economicus} (\acrshort{aHE})}
}
\newglossaryentry{econophysics}
{
  name={econophysics},
  description={The study of economic phenomena by persons trained in the sciences, whether physicists or not, using techniques borrowed from those sciences}
}
\newglossaryentry{edge}
{
  name={edge},
  description={A gambling term for the expected gain or loss from a bet. A bet with positve expectation has ``edge'', and one that is not positive, has no ``edge''}
}
\newglossaryentry{efficient.market.hypothesis}
{
  name={Efficient Market Hypothesis},
  description={The economic theory that markets incorporate all available information making it impossible for investors to consistently earn excess profits}
}
\newglossaryentry{EMH.1}
{
  name={EMH.1},
  description={Rationality: Investors are mostly rational}
}
\newglossaryentry{EMH.2}
{
  name={EMH.2},
  description={Cancellation: To the extent that investors are not rational, they make random trading decisions that in aggregate produce random deviations from fair prices}
}
\newglossaryentry{EMH.3}
{
  name={EMH.3},
  description={Arbitrage: To the extent that the random deviations of \gls{EMH.2} are sufficiently large, rational traders enter the market and correct them}
}
\newglossaryentry{EMH-S}
{
  name={EMH-S},
  description={Strong form of the \gls{efficient.market.hypothesis}: It is impossible to earn superior \glspl{risk-adjusted.return} given public (i\gls{EMH-W} and \gls{EMH-SS}) and private (\emph{\gls{.insider}}) information}
}
\newglossaryentry{EMH-SS}
{
  name={EMH-SS},
  description={Semi-strong form of the \gls{efficient.market.hypothesis}: It is impossible to earn superior \glspl{risk-adjusted.return} given publicly available information, which includes \gls{EMH-W} information plus data such as \gls{.insider} trader SEC reports of the purchases and sales of their stocks by corporate officers, earnings and dividends surprises, and accounting and auditing reports}
}
\newglossaryentry{EMH-W}
{
  name={EMH-W},
  description={Weak form of the \gls{efficient.market.hypothesis}: It is impossible to earn superior \glspl{risk-adjusted.return} given knowledge of past prices, returns, volume and open interest alone}
}
\newglossaryentry{empirical.distribution.function}
{
  name={empirical distribution function},
  description={For real numbers $x_1, x_2, \ldots, x_n$, the empirical distribution function $\hat{F}(x)$ is defined by
  \[
    \hat{F}(x) = \frac{1}{n} \sum_{i=1}^{i=n} I( x \ge x_i ) = \frac{1}{n} \{ \# i \, | \, x_i \le x \},
  \]
  where $I(A)$ is the indicator function of $A$. When $X_1, X_2, \ldots, X_n$ is a random sample from \gls{cumulative.distribution.function} $F(x)$, $\hat{F}(x)$ converves uniformly to $F(x)$ (Glivenko-Cantelli Theorem)}
}
\newglossaryentry{encoding}
{
  name={encoding},
  description={The transformation of a message into a code}
}
\newglossaryentry{endogenous.event}
{
  name={endogenous event},
  description={In systems theory, an event that can be explained by the system; it occurs ``inside'' it, or rather, as a consequence of it. Opposed to an \gls{exogenous.event}}
}
\newglossaryentry{endogenous.market.distortion}
{
  name={endogenous market distortion},
  description={A predictable behavior of markets due to the interaction of strategies. Example: bubbles, crashes and \gls{crowding}}
}
\newglossaryentry{endowment.effect}
{
  name={endowment effect},
  description={A bias in which a person values the same object more if it is owned than if it is not}
}
\newglossaryentry{ensemble}
{
  name={ensemble},
  description={The set of all paths (histories) that a stochastic process can follow}
}
\newglossaryentry{equivalence}
{
  name={equivalence},
  description={In \gls{utility.theory}, equivalence of events $A$ and $B$ is expressed as the requirement that $A \succeq B$ and $B \succeq A$. Such events are also said to be \emph{equivalent}. See also \gls{indifference}}
}
\newglossaryentry{ergodic}
{
  name={ergodic},
  description={A stochastic process whose ensemble averages are almost surely the same as its asymptotic time averages}
}
\newglossaryentry{event-precipitated.trend}
{
  name={event-precipitated trend},
  description={A trend that occurs due to an event}
}
\newglossaryentry{event.time}
{
  name={event time},
  description={Referring to time measured by the occurrence of events}
}
\newglossaryentry{exchange.traded.fund}
{
  name={exchange traded fund},
  description={A type of mutual fund in which shares are claims on a published portfolio of stocks. These stocks are held in trust, and the value of shares determined in the usual way from the values of those stocks}
}
\newglossaryentry{exchange}
{
  name={exchange},
  description={A marketplace in which financial instruments such as equities, bonds, commodities, options and futures are traded; securities that trade on exchanges, which by law are required to report trading prices, are called \emph{listed}. In the U.S., the New York Stock Exchange (\acrshort{aNYSE}), American Stock Exchange (\acrshort{aAMEX}), and National Association of Securities Dealers Automatic Quotation System (\acrshort{anasdaq}) are the most established stock exchanges, but many new electronic exchanges have arisen since the $1990$'s. There is also an informal market that deals in over-the-counter securities}
}
\newglossaryentry{exogenous.event}
{
  name={exogenous event},
  description={In systems theory, an event that unexplained by the system; it occurs ``outside'' it. An earthquake that has adverse economic consequences, is an example of an exogenous event for financial markets. Opposed to an \gls{endogenous.event}}
}
\newglossaryentry{exogenous.market.distortion}
{
  name={exogenous market distortion},
  description={A \emph{exogenous market distortion} is a potentially predictable market behavior that results from constraint(s) due to  beliefs, strategies or institutional arrangements}
}
\newglossaryentry{exogenous.shock}
{
  name={exogenous shock},
  description={In financial theory, an event that occurs ``outside'' the market, and which therefore, is at least partly unanticipatible. Bad weather, eqrthquakes and tsunamis that have significant economic consequences are examples}
}
\newglossaryentry{exponential.distribution}
{
  name={exponential distribution},
  description={The exponential distribution having density $\lambda e^{- \lambda x}$ for $x > 0$}
}
\newglossaryentry{expected.utility}
{
  name={expected utility},
  description={In this work, the same as \gls{vNM.expected.utility.theory}}
}
\newglossaryentry{experimental.economics}
{
  name={experimental economics},
  description={A field developed by 2002 Nobel Laureate in Economics, Vernon Smith. Experimental econometricians conduct laboratory experiments to test the predictions of economic theories}
}
\newglossaryentry{experimental.data}
{
  name={experimental data},
  description={Data acquired in a controlled scientific experiment}
}
\newglossaryentry{expiration.pin.risk}
{
  name={expiration pin risk},
  description={The risk associated with an option conversion or reversal position when the underlying expiration price is its strike}
}
\newglossaryentry{exponential.moving.average}
{
  name={exponential.moving average},
  description={An exponential moving average with parameter $\beta$, $0 < \beta \le 1$ is a \gls{moving.average} having $\alpha_s = \beta (1 - \beta)^^s$, where $EMA_t = \sum_{s=0}^{s=\infty} \alpha_s X_{t-s}$}
}
\newglossaryentry{extreme.percentile.price.trending}
{
  name={extreme percentile price trending},
  description={A pricing anomaly in which stocks in extreme percentiles of past returns either continue or reverse those trends. An example of the former is stock \gls{momentum}, of the latter, the De Bondt-Thaler 3-5 year trend reversal in stocks}
}

\newglossaryentry{factor.model}
{
  name={factor model},
  description={A multiple regression or factor analysis model that identifies components, usually linear, that reduce the dataset dimensionality}
}
\newglossaryentry{falsifiability}
{
  name={falsifiability},
  description={As used with respect to a scientific hypothesis, the requirement that there exist experimental methods to show that with high probability, it is untrue}
}
\newglossaryentry{farmer}
{
  name={Doyne Farmer},
  description={Prominent physicist, econophysicist and entrepreneur formerly of the Santa Fe Institute, currently Co-Director, Complexity Economics, The Institute for New Economic Thinking at the Oxford Martin School}
}
\newglossaryentry{Fast.Frugal.Heuristics}
{
  name={Fast \& Frugal Heuristics},
  description={A term originated by \gls{gigerenzer} to describe domain-specific ``mental shortcuts'' for making decisions, e.g. the \emph{\gls{recognition.heuristic}} and the \emph{\gls{pick.the.best.heuristic}}}
}
\newglossaryentry{fertility}
{
  name={fertility},
  description={In population biology and demographics, the study of patterns of reproduction by members of a species}
}
\newglossaryentry{FIFO}
{
  name={FIFO},
  description={Abbreviation for \emph{First in, first out}. Used in queuing theory}
}
\newglossaryentry{founding.transaction}
{
  name={founding transaction},
  description={In a \gls{PPGS.renewal.process}, a \gls{transaction} that initiates a position, i.e. when the prior position is zero}
}
\newglossaryentry{float}
{
  name={float},
  description={For a stock, its float is the number of shares outstanding available without restrictions to the public}
}
\newglossaryentry{fractional.return}
{
  name={fractional return},
  description={In finance, given price $p$ at time $t'$ and $q$ at time $t'' > t'$, the fractional return is $(q-p)/p$}
}
\newglossaryentry{fragile}
{
  name={fragile},
  description={A condition in which a financial instrument or market has a non-trivial chance of a very large move. Not \glslink{robustness}{robust}}
}
\newglossaryentry{framing}
{
  name={framing},
  description={In \gls{prospect.theory}, a statement of the \gls{decision.problem} in terms of gains, losses, or neither. \acrshort{aKnT} showed framing to be critical in the decision process, and could even lead to preference reversals. For example, stating a problem in terms of gains could result in the opposite decisions from one stated in terms of losses}
}
\newglossaryentry{free.rider}
{
  name={free rider},
  description={In a public goods game, a player who contributes little or nothing but receives the benefits nonetheless}
}
\newglossaryentry{front.end}
{
  name={front end},
  description={A term used for the shorter maturity instruments of the yield curve, typically ones with a maturity of less than $5$ years}
}
\newglossaryentry{frontrunning}
{
  name={front running},
  description={The practice of brokers who place their orders ahead of large customer orders. The practice is illegal in many markets, a notable exception being foreign currency markets}
}
\newglossaryentry{full.disclosure}
{
  name={full disclosure},
  description={A non-standard term in \gls{game.theory} in which a player discloses that he or she will pursue a certain strategy with a probability of either $0$ or $1$}
}
\newglossaryentry{fundamental.analysis}
{
  name={fundamental analysis},
  description={The valuation of stocks based on financial statements, business analysis and industry comparisons},
  plural={fundamental analyses}
}
\newglossaryentry{fundamentalist}
{
  name={fundamentalist},
  description={In the \gls{DSSW.Model} of bubbles, a trader who buys and sells according to \gls{fundamental.analysis}. Many fundamentalists are precluded from shorting, e.g. mutual funds, while others are unwilling, since it is quite risky in real markets}
}
\newglossaryentry{Fundamental.Laws.of.Gambling}
{
  name={Fundamental Laws of Gambling},
  description={There are two Fundamental Laws of Gambling: (1) The Fundamental Law - Never bet unless you think you have an \gls{edge}, and (2) The Wagering Principle - Bet an amount appropriate for the edge}
}

\newglossaryentry{gambler's.fallacy}
{
  name={gambler's fallacy},
  description={The mistaken belief that if an event happens frequently in an \gls{i.i.d.} series, it is less likely to happen in the future} 
}
\newglossaryentry{gambler's.ruin}
{
  name={gambler's ruin},
  description={A gambler is said to be \emph{ruined} if his wagering capital is exhausted. In probability theory, \emph{gambler's ruin} problems involve one or more players with finite funds wagering repeatedly until one of them goes broke. In the simplest cases, gambler's ruin problems have closed form solutions (Feller, 1957)}
}
\newglossaryentry{gambling}
{
  name={gambling},
  description={The low enterprise of wagering on the outcomes of matters both small and large, and on events from the most trivial to the most consequential, with little regard for the outcomes excepting the gain or loss incurred. Not to be confused with \emph{investing}, which hides the enterprise behind a fa\c{c}ade of respectability}
}
\newglossaryentry{gambling.irrational}
{
  name={gambling irrational},
  description={Same as \gls{girrational}}
}
\newglossaryentry{gambling.rational}
{
  name={gambling rational},
  description={Same as \gls{grational}}
}
\newglossaryentry{gambling.system}
{
  name={gambling system},
  description={A gambling system for $\{X_t\}$ with respect to $\{Z_t\}$ is a family of real-valued functions $f_t(Z_t, Z_{t-1} , . . .)$ that specify how much to bet at time $t$ on the outcome at time $t + 1$ given the history $Z_t , Z_{t-1}, \ldots$}
}
\newglossaryentry{game.analysis}
{
  name={game analysis},
  description={The study of the best way to play rules-based, possibly stochastic, contests in which two or more participants (players) make decisions according to the rules, and receive payoffs as a function of a terminal outcome. Unlike \gls{game.theory}, game analysis is concerned with winning at games, not with theoretically perfect play}
}
\newglossaryentry{game.theory}
{
  name={game theory},
  description={The study of rules-based, possibly stochastic, contests in which two or more participants (players) make decisions according to the rules, and receive payoffs as a function of a terminal outcome. A \emph{solution} in game theory consists of strategies for the players that are in some sense optimal}
}
\newglossaryentry{gaussian.distribution}
{
  name={Gaussian distribution},
  description={The Gaussian (normal) distribution $N(\mu,\sigma^2)$ is a two-parameter distribution with $\mu$ the mean and $\sigma > 0$ the standard deviation. Its density is $\phi(x; \mu, \sigma) = (2 \pi \sigma^2)^{-1/2} exp(- ( (x - \mu)/ \sigma )^2)$, $x \in \mathbb{R}$},
  plural={Gaussian density}
}
\newglossaryentry{Gaussian.random.walk}
{
  name={Gaussian random walk},
  description={A random walk $X_t = \mu + X_{t-1} + U_t$ for which $U_t \sim N(\mu,\sigma^2)$}
}
\newglossaryentry{Gaussian.white.noise}
{
  name={Gaussian white noise},
  description={A \gls{white.noise} $\{X_t\}$ in which all $X_t$ have Gaussian distributions}
}
\newglossaryentry{geometric.growth.rate}
{
  name={geometric growth rate},
  description={For a positive time series $\{X_t\}$, $t = 0, 1, \ldots$, the almost sure limit $G$ defined by 
\[
  ln(X_n/X_0)/n \rightarrow G, \text{ as } n \rightarrow \infty, 
\]
provided it exists}
}
\newglossaryentry{Soros.George}
{
  name={George Soros},
  description={Arguably the most successful, but unquestionably the most famous hedge fund manager of the latter $20^{th}$ and early $21^{st}$ centuries}
}
\newglossaryentry{generalized.prisoner's.dilemma} 
{
  name={generalized prisoner's dilemma},
  description={A generalization of a 2 person \gls{prisoners.dilemma} to an $N$ player game in which a coalition $M$ faces a prisoner's dilemma against $N \ M$}
}
\newglossaryentry{gigerenzer}
{
  name={Gerd Gigerenzer},
  description={Psychologist who developed Fast \& Frugal Heuristics}
}
\newglossaryentry{girrational}
{
  name={girrational},
  description={Said of an investor who is not \gls{grational}}
}
\newglossaryentry{Global.Financial.Crisis}
{
  name={Global Financial Crisis},
  description={The financial conflagration that was clearly discernible by $2007$ when the subprime mortgage market melted down, by the fall of $2008$ becoming a worldwide crisis of all financial markets, finally ending in the spring of $2009$. The worst crisis since $1929$}
}
\newglossaryentry{granularity}
{
  name={granularity},
  description={Referring to the size of the time interval in regularly sampled data; \emph{coarse granularity} indicates sampling at long intervals, such as days, and \emph{fine granularity} indicates sampling at short intervals such as seconds}
}
\newglossaryentry{grational}
{
  name={grational},
  description={Said of an investor whose criterion for trades is that they have expected return of at least $r$ and probability of \glspl{drawdown} exceeding $d$ no more that $p$. See also \gls{girrational}}
}
\newglossaryentry{Great.Depression}
{
  name={Great Depression},
  description={The worldwide financial collapse that followed the stock market crash of $1929$, and lasted for nearly a decade},
}
\newglossaryentry{Grossman-Stiglitz.paradox}
{
  name={Grossman-Stiglitz paradox},
  description={An EMH paradox attributed to S. Grossman and J. Stiglitz which argues that costly research is (theoretically) unnecessary for an efficient market, but a market can't be made efficient without such research}
}

\newglossaryentry{Half.Cubic.Law}
{
  name={Half-Cubic Law},
  description={Laws discovered by the Stanley econophysicist group that relate aggregate stock impact, quantity traded and number of shares traded as power laws with exponents of either $3/2$ or $3$}
}
\newglossaryentry{halo.effect}
{
  name={halo effect},
  description={The tendency to view more favorably things associated with something or someone you like}
}
\newglossaryentry{heavy.tail}
{
  name={heavy tail},
  description={Referring to the large values of a positive probability distribution (the ``tail'') that has a slower than exponential die-off, i.e. $\underset{x \rightarrow \infty} {lim} \; e^{\lambda x} P[X >x] = \infty$. For distributions with both positive and negative values, a \emph{heavy upper tail} and \emph{heavy lower tail} are defined analogously}
}
\newglossaryentry{hedge.fund}
{
  name={hedge fund},
  description={A pooled investment vehicle whose legal structure allows the use of high  \gls{leverage} and shorting and which generally invests in liquid assets. Hedge funds differ from \glspl{mutual.fund}, which have limits on  \gls{leverage} and cannot short, and from private equity funds, which often invest in relatively illiquid assets}
}
\newglossaryentry{herding}
{
  name={herding},
  description={Imitative behavior by groups of animals or humans, where all or most individuals make the same decision at about the same time, move in the same direction at the same time, etc. During market crashes, for example, most people try to sell at the same time}
}
\newglossaryentry{high-frequency}
{
  name={high-frequency},
  description={Data having short, usually irregular, time spans between successive observations. In this work, it refers to data occurring, or sampled at, time periods less than or equal to $15$ minutes}
}
\newglossaryentry{high.kurtosis}
{
  name={high kurtosis},
  description={Said of a distribution that has a third central moment greater than $3$ (the kurtotis of a normal distribution)}
}
\newglossaryentry{hindsight.bias}
{
  name={hindsight bias},
  description={The tendency to view an event retrospectively as having greater probability than it was thought to have at the time. Also known as \gls{knew.it.all.along.effect}}
}
\newglossaryentry{holistic.perception}
{
  name={holistic perception},
  description={Perception that emphasizes a whole over its parts. As opposed to \gls{rule.based.perception}}
}
\newglossaryentry{homo.economicus}
{
  name={Homo Economicus},
  description={A \acrshort{avnm} financial decision maker. Also called \gls{econ}, abbreviated \acrshort{aHE}}
}
\newglossaryentry{homogeneity}
{
  name={homogeneity},
  description={In a time series, the property of having the same statistical behavior at all scales; self-similarity}
}
\newglossaryentry{homogeneous.random.walk}
{
  name={homogeneous random walk},
  description={A random walk $X_t = \mu + X_{t-1} + U_t$ with unrestricted $\mu$}
}
\newglossaryentry{hot.hand.fallacy}
{
  name={hot hand fallacy},
  description={The belief that someone who makes frequent correct choices will continue to do so in the future. This belief is obviously mistaken if the choices arise from an \gls{i.i.d.} series}
}
\newglossaryentry{hypothesis-forming.stage}
{
  name={hypothesis-forming stage},
  description={The stage of scientific inquiry in which hypotheses are proposed to explain experimental data}
}

\newglossaryentry{incrementing.transaction}
{
  name={incrementing transaction},
  description={In a \gls{PPGS.renewal.process}, a \gls{transaction} for which the number of shares, units or contracts of the prior \gls{position} is increased}
}
\newglossaryentry{identically.distributed}
{
  name={identically distributed},
  description={A set of random variables $\{X_i\}$ are 
               identically distributed if $P[X_i \le x] = 
               P[X_j \le x]$ for all $i \ne j$}
}
\newglossaryentry{illusory.correlation.causation}
{
  name={illusory correlation and causation},
  description={The commonly held (false) view that correlated events entail 
               causative relationships. It often results from a misunderstanding of regression analysis}
}
\newglossaryentry{Immediate.Adjustment}
{
  name={Immediate Adjustment},
  description={In the \glspl{efficient.market.hypothesis}, the requirement that prices adjust immediately and correctly to unanticipated news}
}
\newglossaryentry{independence}
{
  name={independence},
  description={A reference to a set of random variables that are \gls{independent}. Same as \gls{statistical.independence}}
}
\newglossaryentry{independent}
{
  name={independent},
  description={In its technical meaning from probability 
               theory, random variables $\{X_i\}$ are 
               (jointly) independent if the distribution 
               function of any finite subset $(X_{i_1}, 
               X_{i_2}, \ldots, X_{i_n})$ can be expressed
               as a product of marginal distributions:
               \[ 
                  P[X_1 \le x_1, X_2 \le x_2, \ldots, X_n \le x_n] = 
                  \prod_{i=1}^{i=n} P[X_i \le x_i]
               \]
               }
}
\newglossaryentry{independent.components.analysis}
{
  name={independent components analysis},
  description={A method of decomposing a multivariate 
               dataset into components that are 
               maximally independent. There are several 
               different methods of calculating such 
               components, depending on the measure of 
               independence used}
}
\newglossaryentry{independent.white.noise}
{
  name={independent white noise},
  description={A \gls{white.noise} $\{X_t\}$ in which
               for any choice of $t_1 < t_2 < \cdots < t_n$, 
               $X_{t_1}, X_{t_2}, \ldots, X_{t_n}$ are 
               jointly independent}
}
\newglossaryentry{information.aggregation}
{
  name={information aggregation},
  description={As used in the \glslink{efficient.market.hypothesis}{EMH}, refers to the process of information dissemination when some players are more informed than others, i.e. to the manner in which prices adjust to reflect the information of a minority, the \glspl{.insider}. Also called simply \gls{aggregation}}
}
\newglossaryentry{i.i.d.white.noise}
{
  name={i.i.d. white noise},
  description={A \gls{white.noise} $\{X_t\}$ for which 
               all $X_t$ have the same distribution and 
               for any choice of $t_1 < t_2 < \cdots < t_n$, 
               $X_{t_1}, X_{t_2}, \ldots, X_{t_n}$ are 
               jointly independent}
}
\newglossaryentry{i.i.d.}
{
  name={i.i.d.},
  description={An abbreviation meaning ``independent and 
               identically distributed,'' referring to a 
               random sample or time series  $\{X_t\}$ 
               for which each $X_t$ has the same probability 
               distribution and any subset $X_{t_1}, X_{t_2}, 
               \ldots, X_{t_k}$ of variables are 
               independent}
}
\newglossaryentry{indifference}
{
  name={indifference},
  description={In \gls{utility.theory}, indifference between events $A$ and $B$ is expressed as the requirement that $A \succeq B$ and $B \succeq A$. An individual is said to be \emph{indifferent} to such events. See also \gls{equivalence}}
}
\newglossaryentry{information.ratio}
{
  name={information ratio},
  description={The \emph{information ratio} for return series $R_t$,
               $t = 1, 2, \ldots, n$ is its mean divided by its 
               standard deviation. The information ratio is similar to the 
               \gls{Sharpe.ratio} but is more useful in the Grinold \& 
               Kahn evaluation of active management}
}
\newglossaryentry{information.set}
{
  name={information set},
  description={For a time series $\{X_t\}$ at time $t'$, 
               a set $I_{t'}$ that gives data known at $t'$.  
               Generally used to form a conditional 
               distribution $X_{t'+1} \, | \, I_{t'}$ for the 
               next-period value, $X_{t'+1}$}
}
\newglossaryentry{.insider}
{
  name={insider},
  description={An individual who has access to non-public information, generally through channels of position, power or wealth}
}
\newglossaryentry{insider.trading}
{
  name={insider trading},
  description={The deplorable practice of trading on non-public information in advance of its market moving impact. At this writing, insider trading is illegal in the United States, but not in Monaco}
}
\newglossaryentry{intermittency}
{
  name={intermittency},
  description={In the context of this book, intermittency refers to the irregular onset of price volatility, in which there are alternating periods of high and low volatility.}
}
\newglossaryentry{intrinsic.value}
{
  name={intrinsic value},
  description={The value of a company or assed based on a balanced appraisal of all its characteristics and aspects, both tangible and intangible. \Gls{fundamental.analysis} strategies base their trading on estimated intrinsic values compared to market values}
}
\newglossaryentry{IPO}
{
  name={IPO},
  description={An acronym for \emph{Initial Public Offering}. An IPO is a stock offering of a previously privately-held company for public investment. See also \gls{SEO}}
}
\newglossaryentry{irrational}
{
  name={irrational},
  description={In the theory of financial decision making, a decision maker who is not \gls{vNM.rational}}
}

\newglossaryentry{January.effect}
{
  name={January effect},
  description={In U.S. equity markets, a calendar phenomenon in which abnormal positive returns occur for many stocks at the turn of a year. Prior to its discovery, its effect on small stocks was larger that mid-cap or large cap stocks. After its discovery, it failed to perform as well in succeeding years}
}

\newglossaryentry{kahneman}
{
  name={Daniel Kahneman},
  description={Nobel Laureate in Economics, 2002. The ``K'' in the acronym ``\acrshort{aKnT}''}
}
\newglossaryentry{Karl.Popper}
{
  name={Karl Popper},
  description={Famous philosopher of science, mentor of \gls{George.Soros}},
  sort={Popper}
}
\newglossaryentry{Kelly.Criterion}
{
  name={Kelly Criterion},
  description={A method of determining an investment fraction that maximizes the growth rate of wealth. Also called \gls{Optimal.f}}
}
\newglossaryentry{kernel.smoother}
{
  name={kernel smoother},
  description={A smoothing method that produces an approximating function $\hat{f}(x)$ for a dataset $(x_i,y_i)$ by using a kernel function for weights and a Nadaraya-Watson estimator for $\hat{f}(x)$}
}
\newglossaryentry{k-nearest.neighbors}
{
  name={k-nearest neighbors},
  description={A smoothing method for a scatterplot $x_i,y_i)$ , $i = 1, 2, \ldots, N$ which for a given $x_0$ finds the $k$ nearest $x_i$, and averages the $y_i$ associated with them. See also \gls{loess}}
}
\newglossaryentry{knew.it.all.along.effect}
{
  name={knew-it-all-along effect},
  description={The tendency to recollect the probability of an event after its occurrence as more likely than before its ocurrence. Also known as the \gls{hindsight.bias}}
}
\newglossaryentry{KS-test}
{
  name={KS-test},
  description={The ``KS'' stands for Kolmogorov-Smirnov, the creators of two eponymous tests, (1) the KS one sample test, and (2) the KS two sample test. In the one sample test of sample data coming from a hypothesized distribution $F(x)$, the statistic of interest is 
\[
     D = \underset{x}{max} | \hat{F}(x) - F(x) |, 
\]
  where $\hat{F}(x)$ is the \gls{empirical.distribution.function}. The 2-sample test compares two \glspl{empirical.distribution.function}. Small sample distributions are available for the 1-sample test, whereas only asympotic distributions are available for the 2-sample test}
}

\newglossaryentry{law.of.equal.ignorance}
{
  name={law of equal ignorance},
  description={A probability rule that assigns to each of $N$ possibilities a probability of $1/N$ when very little is known about them. Same as \gls{1/N.rule}}
}
\newglossaryentry{legal.pump.and.dump}
{
  name={legal pump-and-dump},
  description={Legal versions of pump-and-dump, e.g. \gls{strategic.pump.and.dump} and \gls{speculative.pump.and.dump}}
}
\newglossaryentry{leverage}
{
  name={leverage},
  description={Refers to the common financial practice of purchasing or shorting financial instruments by pledging only a fraction of their nominal value. The difference of the cost of the purchase or short is generally supplied at interest from a bank or institution that specializes in lending to trading entities}
}
\newglossaryentry{leveraged.buyout}
{
  name={leveraged buyout},
  description={The takeover of a corporation by \glspl{arbitrageur} who \gls{leverage} the purchase. Part of the \gls{mergersandacquisitions} game}
}
\newglossaryentry{likelihood.function}
{
  name={likelihood function},
  description={Given a sample $x = \{ x_1, x_2, \ldots, x_n \}$ and a family of distributions $F(x ; \theta)$, $\theta \in \Theta$, the function $\mathscr{L}(\theta ; x) = F(x;\theta)$ considered as a function of $\theta$ with $x$ fixed}
}
\newglossaryentry{limit.order}
{
  name={limit order},
  description={An order to buy or sell at a given price (the limit price) or better, optionally with a time to leave the order open. See also \gls{market.order} and \gls{stop.order}}
}
\newglossaryentry{loess}
{
  name={loess},
  description={A smoothing method for a scatterplot $x_i,y_i)$ , $i = 1, 2, \ldots, N$ which for a given $x_0$ finds the $k$ nearest $x_i$ (where $ k = \alpha N$), and averages the $y_i$ associated with them. See also \gls{k-nearest.neighbors}}
}
\newglossaryentry{LOR}
{
  name={Leland O'Brien Rubinstein Associates},
  description={The firm whose principals developed, advocated and marketed \gls{portfolio.insurance} prior to the \acrlong{aWSMC87}. They subsequently admitted the likelihood that portfolio insurance was a major cause of that Crash}
}
\newglossaryentry{log-relative}
{
  name={log-relative},
  description={In finance, given price $p$ at time $t'$ and $q$ at time $t'' > t'$, the log-relative is $log(q/p)$}
}
\newglossaryentry{logarithmic.return}
{
  name={logarithmic return},
  description={log-relative}
}
\newglossaryentry{lognormal.distribution}
{
  name={lognormal distribution},
  description={$Y$ has a lognormal distribution with parameters $\mu$ and $\sigma^2$ if $y \sim exp(X)$, where $X \sim N(\mu,\sigma^2)$ has a Gaussian distribution with mean $\mu$ and variance $\sigma^2$}
}
\newglossaryentry{log.return}
{
  name={log return},
  description={A logarithmic return}
}
\newglossaryentry{long.memory}
{
  name={long memory},
  description={A technical term for a stochastic process $X_t$ that requires its ``correlations to die off slowly.'' One definition requires that for each time $t$, the sum of correlations $\sum_{s=1}^{\infty} \rho_{t+s}$ diverges}
}
\newglossaryentry{long-tailed}
{
  name={long-tailed},
  description={Referring to a distribution with the property
  \[
     \underset{ \rightarrow \infty}{lim} \frac{P \left[ X > x + y \right]}{P \left[ X > x \right]} = 1
  \]
  }
}
\newglossaryentry{lookahead.bias}
{
  name={lookahead bias},
  description={In trading system development, the (often inadvertent) development and testing of a trading system using information assumed to be known at trade time, but in actuality known only at some future time. Some real life examples: (1) using a sentiment index dated on Friday of each week, but not made available until the following Tuesday, and (2) simulating a trade in the middle of a period based on the mean price of the entire period, but then in practice using the mean available from the beginning of the period to the trade time}
}
\newglossaryentry{lookback.horizon}
{
  name={lookback horizon},
  description={Trading system jargon for the period of history needed to make a trading decision. For example, if market history for the past week is all that is needed to decide to buy, sell or do nothing, then the lookback horizon is one week}
}
\newglossaryentry{loss.aversion}
{
  name={loss aversion},
  description={A bias in which people attempt to avoid losses, even if such action gives away considerable \gls{edge}}
}
\newglossaryentry{lottery}
{
  name={lottery},
  description={In \gls{utility.theory}, a gamble that returns one and only one prize $A_1, A_2, \ldots, A_n$ with known probabilities $p_1, p_2, \ldots, p_n$. Notation: 
\begin{equation*}
  ( p_1,A_1; \; p_2, A_2, \; \ldots, \; p_n, A_n )
\end{equation*}
}}
\newglossaryentry{LTCM}
{
  name={Long Term Capital Management},
  description={A \gls{hedge.fund} formed in $1994$ by ex-Solomon Brothers ``whiz kids,'' featuring in addition two Nobel Laureates in Economics, who had developed and advocated the ``new finance \gls{arbitrage} strategies'' that it allegedly used. After its notorous collapse in $1998$, the Federal Reserve, fearful of a run on shadow banking, engineered a rescue by a consortium of $14$ banks. A clear demonstration of the Federal Reserve's real mandate, not the sanitized official version}
}

\newglossaryentry{Manager.of.Managers}
{
  name={Manager of Managers},
  description={In the nonstandard usage of this work, a investment fund that does not itself trade its funds, but hires other managers for that purpose. See also \gls{Trader/Manager}}
}
\newglossaryentry{market.capacity}
{
  name={market capacity},
  description={An informal concept which expresses the idea that fundamental values in a security cannot be maintained when the demand for the security too large. When the demand greatly exceeds the supply, the market can become severly \emph{\gls{overbought}} leading to the likelihood of a correction later}
}
\newglossaryentry{market.ecology}
{
  name={market ecology},
  description={A market ecology at a moment in time includes both the \gls{strategic.ecology} and market structure, in the sense of a game that has rules, venues and technology within which strategies can be enacted. Also called \gls{ecology}}
}
\newglossaryentry{market.fundamentalism}
{
  name={market fundamentalism},
  description={A theory decocting matters of great financial complexity into a simplistic recipe --- that matters of greatest importance to societal welfare are always best served in a laissez-faire stew}
}
\newglossaryentry{market.maker}
{
  name={market maker},
  description={A trader who specializes in providing liquidity to markets by buying and selling to counterparties at either posted or negotiated prices. Market makers are usually exchange-sanctioned and pay low fees to trade. In \glspl{continuous.auction.market}, market makers post bids and offers that the public can trade against. Market makers perform well in non-volatile markets, but tend to withdraw bids and offers in volatile ones and thereby, may exacerbate large price moves}
}
\newglossaryentry{market.order}
{
  name={market order},
  description={An order to buy or sell at the best available price. See also \gls{limit.order} and \gls{stop.order}}
}
\newglossaryentry{Market.Neutral.Strategy}
{
  name={Market Neutral Strategy},
  description={A \emph{Market Neutral Strategy (\acrshort{aMNS})} is any strategy that selects from a universe of assets a portfolio \(P\) such that the \(\beta\) of \(P\) with respect to that universe is nearly zero},
  plural={Market Neutral Strategies}
}
\newglossaryentry{market.paradigm}
{
  name={market paradigm},
  description={A collection of beliefs about the current state of the market that leads to one of three judgments: the market is (1) undervalued, (2) overvalued or (3) neither (1) nor (2)},
}
\newglossaryentry{market.sentiment}
{
  name={market sentiment},
  description={A term that refers to the range of opinions and attitudes of the players in a market. In financial markets, diverse attitudes are often simplified to: bullish, bearish and neutral; in real estate, to prices ``too high'', ``too low'' or ``afforable.'' Sentiment has been believed to be important in selecting market tops and bottoms --- low bullish sentiment predicting an advancing market, and high bullish sentiment predicting a declining market. Same as \gls{sentiment}}
}
\newglossaryentry{market.tone}
{
  name={market tone},
  description={A term that refers to the way that a market reacts to positive and negative news. A market has good tone if it has a positive reaction to good news and negative reaction to negative news. It has poor tone if it has no or a negative reaction to good news and a negative reaction to negative news. \emph{Market tone} is a soft, rather subjective \gls{sentiment.indicator}}
}
\newglossaryentry{martingale}
{
  name={martingale},
  description={In this work, a time series $X_t$ for which $E[X_t] < \infty$ and $E[X_{t+1} \, | \, X_t ] = X_t$ for all meaningful $t$}
}
\newglossaryentry{martingale.difference.series}
{
  name={martingale difference series},
  description={In this work, a time series $X_t$ is a martingale difference series with respect to \glspl{information.set} $\{\mathcal{I}_t\}$ if for all meaningful $t$, $E[X_t \, | \, \mathcal{I}_{t-1}]  < \infty$ and $E[X_{t} \, | \, \mathcal{I}_{t-1} ] = 0$. When information sets are not specified, it is understood that each $\mathcal{I}_t = \{X_t, X_{t-1}, X_{t-2}, \ldots\}$ is the history of the series up to time $t$}
}
\newglossaryentry{maximum.likelihood}
{
  name={maximum likelihood},
  description={A method of producing estimators from a sample $x = \{ x_1, x_2, \ldots, x_n \}$ assmuing that the data arise from a family of distributions $F(\theta)$ indexed by $\theta \in \Theta$. The \gls{likelihood.function} $\mathscr{L}(\theta ; x) = F(x;\theta)$ is considered to be a function of $\theta$ with the sample $x$ fixed. The \gls{maximum.likelihood.estimator} is the distribution(s) $F(\theta_{max}$ at which the \gls{likelihood.function} is maximized as a function of $\theta$}
}
\newglossaryentry{maximum.likelihood.estimator}
{
  name={maximum likelihood estimator},
  description={Given a sample $x = \{ x_1, x_2, \ldots, x_n \}$ and a family of distributions $F(x ; \theta)$, $\theta \in \Theta$, the distribution $F(\theta_{max})$ for which the \gls{likelihood.function} $\mathscr{L}(\theta ; x) = F(x;\theta)$ is maximized as a function of $\theta$}
}
\newglossaryentry{mccauley}
{
  name={Joseph L. McCauley},
  description={Well-known specialist in chaos theory and econophysics}
}
\newglossaryentry{Mean.Variance.Portfolio.Theory}
{
  name={Mean-Variance Portfolio Theory},
  description={A theory that develops a notion of efficient portfolios as ones belonging to an efficient frontier. The development requires only the means and covariances of instruments in a portfolio}
}
\newglossaryentry{mean-variance.world}
{
  name={Mean-Variance World},
  description={A fictitious world featuring the curious analysis of complex stochastic data assuming that only the first two moments of the distributions are necessary. Inhabited only by economists and their lackeys}
}
\newglossaryentry{mechanical.trading.system}
{
  name={mechanical trading system},
  description={In trading and investment, a collection of well-defined rules that specify when to enter the market, the size and side (buy or sell) for entry, and when to exit all or part of an existing position. In this work, the same as a \gls{trading.system}}
}
\newglossaryentry{mental.accounting}
{
  name={mental accounting},
  description={A compartmentalization of financial activity into \emph{mental acounts} in which gains or losses in any one account are not aggregated with gains or losses in the others}
}
\newglossaryentry{mergersandacquisitions}
{
  name={mergers \& acquisitions},
  description={The investment banking and large funds business of corporate restructuring by mergers and \glspl{leveraged.buyout}}
}
\newglossaryentry{Miller's.paradox}
{
  name={Miller's paradox},
  description={An \acrshort{aEMH} paradox due to Edward Miller, 1977. He applied \gls{winner's.curse} arguments to market auctions and concluded, inter alia, that (1) prices will not be fair and investment capital will be inefficiently allocated, and (2) the riskiest stocks will be the most overpriced. These conclusions are so embarrassing to EMH theory that they are seldom written about by academic economists}
}
\newglossaryentry{minskymoment}
{
  name={minsky moment},
  description={A time in the business cycle at which financial values suddenly collapse}
}
\newglossaryentry{mixed.strategy}
{
  name={mixed strategy},
  description={In \gls{game.theory}, a strategy in which each player independently chooses a strategy using a private probability distribution. Also known as a \gls{randomized.strategy}},
  plural={mixed strategies}
}
\newglossaryentry{momentum}
{
  name={momentum},
  description={In academic finance, a \gls{statistical.regularity} in which a portfolio that has appreciated has the tendency to continue appreciating, and a portfolio that has depreciated has a tendency to continue to depreciate. Note that this usage does not correspond directly to ``momentum'' in technical analysis, but rather to ``\glslink{trend}{trending}.'' The term came into widespread use following the discoveries by Jegadeesh \& Titman \cite{citeulike:940976} that the top decile of stocks ranked on 1-year returns contined to appreciate for about another year, and the bottom decile continued to drop for another year. In \gls{technical.analysis}, called \glslink{trend}{trending}}
}
\newglossaryentry{Momentum.Puzzle}
{
  name={Momentum Puzzle},
  description={The findings of two \acrlong{aJnT} famous studies, \cite{citeulike:940976} (1993) and \cite{citeulike:4057155} (2001) showed convincingly that the \gls{efficient.market.hypothesis} \acrshort{aEMH} is untenable. It destroyed all three pillars of the \acrshort{aEMH} by exhibiting that a simple \gls{momentum} strategy produces excess returns, and that \gls{arbitrage} between $1993$ and $2001$ failed to eliminate it. The puzzle is
\begin{enumerate}
\item Why do U.S. stock returns that have recent \gls{momentum} tend to persist over periods averaging a year?
\item Why do U.S. stock returns over $3$ years tend to revert over the next $3$ years?
\item Why did \gls{arbitrage} fail to eliminate the \gls{momentum} anomaly after the publication of the first \acrshort{aJnT} study, i.e. between $1993$ and $2001$?
\end{enumerate} }
}
\newglossaryentry{moving.average}
{
  name={moving average},
  description={A moving average is a time series method of averaging past observations, usually for purposes of prediction. For a univariate or multivariabe time series $\{X_t\}$, a generic moving average has the form $MA_t = \sum_{s=0}^{s=\infty} \alpha_s X_{t-s}$ where $\alpha_s$ are fixed weights for which $\sum \alpha_s = 1$. There are various types of moving average: \gls{simple.moving.average}, \gls{weighted.moving.average} and \gls{exponential.moving.average}. See glossary entries for definitions of these types}
}
\newglossaryentry{Muller-Lyer.Illusion}
{
  name={M\"uller-Lyer Illusion},
  description={A famous visual illusion from Gestalt psychology in which two lines of equal length, one with inward-pointing arrows at the ends, the other with outward-pointing arrows, are perceived as having different lengths}
}
\newglossaryentry{musical.chairs}
{
  name={musical chairs},
  description={A popular child's game in which $n$ children walk in a circle with $n-1$ chairs close by while music plays. When the music stops, each rushes to sit in a chair, leaving one hapless child without a chair. That child is eliminated, and the game continues, always with one chair less than the number of children, until some prescribed number is left. Also reported to be played by some (presumably feeble-minded) adults. A variant occurs in the \gls{DSSW.Model}: (1) the music stops permanently at some random point, and (2) the goal is not to win, but to be eliminated as close to the end as possible}
}
\newglossaryentry{mutation}
{
  name={mutation},
  description={A process whereby some parent genes are copied inexactly}
}
\newglossaryentry{mutual.fund}
{
  name={mutual fund},
  description={A mutual fund is an investment company that sells shares in a portfolio it manages using the pooled monies from investors. In the United States, mutual funds are registered with the Securities and Exchange Commission and are regulated by the Investment Company Act of $1940$. There are three types of mutual funds in the United States: open-end funds, closed-end funds and unit investment trusts. Open-end funds are publicly traded and are required to offer new shares and redeem existing ones at certain times only, usually at the end of each business day. Closed end funds at creation offer shares which are not redeemable until the fund either liquidates or converts to an open end fund. Unit investment trusts form fixed portfolios at creation, with a specific termination date, and in a public offering sell shares to the public. While investors in closed end funds and unit investment trusts cannot redeem their shares prior to ending, secondary markets often allow investors to sell shares prior to that time. Mutual funds may be either actively or passively managed}
}
\newglossaryentry{myopically.rational}
{
  name={myopically rational},
  description={Referring to na\"{i}ve, putatively self-interested strategies pursued by many investors. Such strategies can cause major market inefficiencies such as bubbles and crashes},
}

\newglossaryentry{Nash.equilibrium}
{
  name={Nash equilibrium},
  description={A concept introduced by John Nash in $1950$ in which players' strategies cannot be unilaterally improved upon. Nash showed that under general conditions, such strategies always exist. See Myerson, \cite{myerson1997game} for a presentation},
  plural={Nash equilibria}
}
\newglossaryentry{noisy.channel.model}
{
  name={Noisy Channel Model},
  description={A model (the \acrshort{ancm}) due to Martin Hilbert that explains eight behavioral biases using a parsimonious model based on Claude \glslink{shannon}{Shannon's} information theory}
}
\newglossaryentry{negative.sum}
{
  name={negative-sum},
  description={A real or artificial mathematical game in which the sum of payoffs to the players is negative. Negative-sum games typically occur because a referee or game organizer charges a fee to play}
}
\newglossaryentry{neuroeconomics}
{
  name={neuroeconomics},
  description={The study of brain activity during economic decision making, such as buying a car or investing in a stock. When only financial instruments are involved, called \emph{neurofinance}}
}
\newglossaryentry{New.Deal}
{
  name={New Deal},
  description={The slogan adopted by U.S. President Franklin Roosevelt for his progressive program during the \gls{Great.Depression}. In financial markets, the New Deal involved regulation of the financial system to punish undue speculation and afford a measure of protection to average investors}
}
\newglossaryentry{no.arbitrage.principle}
{
  name={no-arbitrage principle},
  description={For pure \gls{arbitrage}, the principle that arbitrage cannot yield a sure profit. For risk arbitrage, the principle that arbitrage cannot lead to excess expected risk-adjusted returns. The principle is ordinarily invoked to show that arbitrage cannot lead to consistent profits, which recalls the joke about the economist and the statistician \ldots in Volume 2,}
}
\newglossaryentry{noise.trader}
{
  name={noise trader},
  description={(1) In the \emph{noise trader model of efficient markets}, an investor whose decisions may be treated as random. In this model, all investors are either rational or noise traders. See also \gls{efficient.market.hypothesis} (2) In the \gls{DSSW.Model} of bubbles and crashes, a na\"{i}ve \gls{trend.follower} who is left unseated in that game of \gls{musical.chairs}}
}
\newglossaryentry{nominal.gain}
{
  name={nominal gain},
  description={In terminology developed for this book, the \emph{nominal gain} is the total amount won (or if lost, a negative value) by a trading system when the same amounts are bet at every trade. See also \gls{adjusted.nominal.gain}}
}
\newglossaryentry{nonlinear.regression}
{
  name={nonlinear regression},
  description={A model of the form $y = f(x,e)$, where $y$ is dependent variable, $x$ is an independent variable, $e$ is an ``error'' or ``fluctuation,'' and $f$ is a nonlinear function of $(x,e)$. $x$ and $y$ may be multidimensional vectors}
}
\newglossaryentry{nonparametric.model}
{
  name={nonparametric model},
  description={In statistics, a \emph{nonparametric model} is one that makes no explicit assumption about a finite-dimensional family of distributions. Some examples are nonparametric regression, quantile regression, generalized additive models such as \acrshort{aACE}, \acrshort{aAVAS} and \acrshort{aGAM}}
}
\newglossaryentry{normal.form}
{
  name={normal form},
  description={An \gls{artificial.mathematical.game} in which the \glspl{strategy} available to the players are indicated by elements of a set $C$. This set in simple cases is finite, but in most games is infinite, and may represent a complete specification of every choice a player might make in some game contingency. Also called the \gls{strategic.form}}
}
\newglossaryentry{normative.decision.science}
{
  name={normative decision science},
  description={See \gls{decision.science}}
}
\newglossaryentry{normative.to.descriptive.step}
{
  name={normative-to-descriptive step},
  description={The requirement that studies be performed to verify that a normative economic model adequately describes real economic behavior (Michael Bacharach \cite{bacharach1977economics})}
}

\newglossaryentry{observational.data}
{
  name={observational data},
  description={Data acquired through observation without the use of experimental methods}
}
\newglossaryentry{observational.experimental.stage}
{
  name={observational/experimental stage},
  description={The stage of scientific inquiry that collects experimental data}
}
\newglossaryentry{Optimal.f}
{
  name={Optimal $f$},
  description={A method of determining an investment fraction that maximizes the growth rate of wealth. Also called the \gls{Kelly.Criterion}}
}
\newglossaryentry{order}
{
  name={order},
  description={In trading, an \emph{order} is a request, possibly contingent, to buy or sell a financial instrument. \emph{Orders} are time-stamped requests that specify quantity and contingently, price limits, immediate execution, cancellation if not filled, and so on}
}
\newglossaryentry{Ordinal.Utility}
{
  name={Ordinal Utility},
  description={See \gls{utility.theory}}
}
\newglossaryentry{outlier}
{
  name={outlier},
  description={A unusually large or small obsevation in a data sample}
}
\newglossaryentry{outlier-proneness}
{
  name={outlier-proneness},
  description={In dynamic systems, the tendency to produce outliers}
}
\newglossaryentry{outstanding.shares}
{
  name={outstanding shares},
  description={The number of shares issued by a company, which includes restricted shares and the \gls{float}}
}
\newglossaryentry{overbought}
{
  name={overbought},
  description={Said of a financial instrument that is priced too high due excessive ``buying pressure.'' See also \gls{oversold}}
}
\newglossaryentry{overreaction}
{
  name={overreaction},
  description={In financial markets, overreaction refers to an excessively large price move in response to an event. See also \gls{underreaction}}
}
\newglossaryentry{oversold}
{
  name={oversold},
  description={Said of a financial instrument that is priced too low due to excessive ``selling pressure.'' See also \gls{overbought}}
}

\newglossaryentry{pairs.trading.strategy}
{
  name={pairs trading strategy},
  description={A \emph{pairs trading strategy (PTS)} is any strategy that selects from some universe of stocks those price-pairs that have a putatively overvalued and undervalued member, and nearly simultaneously buy the undervalued and  sell the overvalued. The pair so-formed is then liquidated when prices are judged to return to ``equilibrium.''},
  plural={pairs trading strategies}
}
\newglossaryentry{parametric.model}
{
  name={parametric model},
  description={In statistics, a model in which the data are generated by a distribution drawn from a family of distributions, $\{ F_{\theta} \, | \, \theta \in \Theta \}$, where $\Theta$ is a finite dimensional parameter space. Estimation in parametric models requires finding the parameter $\theta_0 \in \Theta$ that in some sense provides a best fit for observed data. Some standard fitting approaches are \gls{maximum.likelihood}, Bayesian methods and least squares estimation with normally distributed errors}
}
\newglossaryentry{pareto}
{
  name={Villfredo Pareto},
  description={Italian engineer, sociologist, economist, political scientist and philosopher who was one of the first to apply scientific analysis to markets. (1848~--~1923)}
}
\newglossaryentry{Pareto.efficient}
{
  name={Pareto efficient},
  description={In decision or game theory, a \gls{strategy.profile} $s$ is Pareto efficient iff given players $N$, payoff functions $u_i$, $i \in N$, there is no $i \in N$ and \gls{strategy.profile} $s'$ such that $u_i(s') > u_i(s)$ unless there is a $j \in N$ such that $u_j(s) < u_j(s')$},
  plural={Pareto efficiency}
}
\newglossaryentry{partial.disclosure}
{
  name={partial disclosure},
  description={A non-standard term in \gls{game.theory} in which a player discloses that he or she will pursue a certain strategy with some probability $p$, $0 < p < 1$ and $p \ne 0.5$}
}
\newglossaryentry{peak.to.trough.drawdown}
{
  name={peak-to-trough drawdown},
  description={In a financial time series, the largest loss from a previous high}
}
\newglossaryentry{percentage.return}
{
  name={percentage return},
  description={In finance, given price $p$ at time $t'$ and $q$ at time $t'' > t'$, the percentage return is $100*(q-p)/p$}
}
\newglossaryentry{perfect.memory}
{
  name={perfect memory},
  description={A time series $\{X_t\}$ with information sets $\{I_t\}$ has perfect memory if $I_t \supseteq \{ X_{t}, X_{t-1}, \ldots \}$}
}
\newglossaryentry{personal.probability}
{
  name={personal probability},
  description={A theory of probability in which each person makes his or her own subjective, consistent probability estimates. Sometimes called \emph{subjective probability}}
}
\newglossaryentry{pick.the.best.heuristic}
{
  name={pick-the-best heuristic},
  description={A \gls{Fast.Frugal.Heuristics} that makes decisions by examining cues for two choices until the first one that is better is found}
}
\newglossaryentry{POPP.trend}
{
  name={POPP trend},
  description={A trend that occurs as part of a \gls{Pursuit.of.Profits.Paradigm} (POPP)}
}
\newglossaryentry{portfolio.insurance}
{
  name={portfolio insurance},
  description={A strategy for insuring against portfolio losses by dynamical replication of a put on a benchmark index. Popular before the worldwide stock market crash of $1987$, but not subsequently due to its poor performance during that crash}
}
\newglossaryentry{position}
{
  name={position},
  description={In a trading context, either a \emph{long position}, which consists of ownership of an instrument or a fungible contract used in trading, or a \emph{short position} that consists of the opposite of a long position (except for financing costs), created synthetically through stock borrowing or directly through ownership of the opposite side of a long contract position. Same as a \gls{trading.position}}
}
\newglossaryentry{positional.format}
{
  name={positional format},
  description={The representation of a trading history as time-stamped cumulative positions}
}
\newglossaryentry{positive.affine.transformation}
{
  name={positive affine transformation},
  description={A positive affine transformation of $u \in \mathbb{R}$ is a function $g(u) = a u \, + \, b$ where $a,b \in \mathbb{R}$ are constants and $a > 0$}
}
\newglossaryentry{positive.feedback}
{
  name={positive feedback},
  description={In a financial context, a process in which perturbations in the process are amplified as it progresses, which leads to system instability. Negative feedback processes damp perturbations, leading to stability}
}
\newglossaryentry{post.event.price.drift}
{
  name={post-event price drift},
  description={A market anomaly in which an event's positive or negative surprises lead to price drift up or down thereater. The best known example is the \emph{post earnings announcement drift} in which the averge drift of prices after the announcement is strictly increasing in the \gls{SUE} (Standardized Unexpected Earnings) of the announcement}
}
\newglossaryentry{potentially.profitable.gambling.system}
{
  name={potentially profitable gambling system},
  description={A potentially profitable gambling system (PPGS) is a betting system for a financial time series $\{X_t\}$ that has at least one positive expected return},
} 
\newglossaryentry{power.law}
{
  name={power law},
  description={A distribution concentrated on $(0,\infty)$ having a cumulative distribution of the form $F(x) = 1 - x^{-\alpha}$, where $\alpha > 0$. $\alpha$ is called the \emph{exponent} of the power law}
}
\newglossaryentry{power.law.tail}
{
  name={power law tail},
  description={A distribution whose upper and/or lower tails are asymptotically equivalent to $(x/x_0)^{-\alpha}$, where $x_0 > 0$ is a constant and $\alpha > 0$. As on \glspl{power.law}, $\alpha$ is called the \emph{exponent} of the power law tail}
}
\newglossaryentry{PPGS.renewal.process}
{
  name={PPGS renewal process},
  description={A potentially profitable gambling system renewal process is an algorithm that (1) produces transactions depending on the state of the market and its own trading history, (2) as a function of those inputs, produces trades each consisting of a founding transaction that initiates a position from a state having no position, a closing transaction that liquidates remaining positions leaving a zero position, and has no other transactions between founding and closing that change the founding position from long to short or short to long, and (3) produces a potentially unending stream of trades (renewal)},
  plural={PPGS renewal processes}
}
\newglossaryentry{precision}
{
  name={precision},
  description={In the theory of statistical estimation, an estimator is called \emph{precise} if its standard deviation is small and otherwise is called \emph{imprecise}}
}
\newglossaryentry{predictability}
{
  name={predictability},
  description={In a dynamic systems, the propensity to forecast its future evolution, if only stochastially. A system can be stongly predictable or weakly predictable}
}
\newglossaryentry{preference.order}
{
  name={preference order},
  description={A ranking among a set $\Omega$ of events that satisfies three axioms: \gls{Complete.Ordering}, \gls{Reflexivity.Of.Preference} and \gls{Transitivity}}
}
\newglossaryentry{preference.reversal}
{
  name={preference reversal},
  description={A \gls{decision.problem} in which the independence of irrelevant alternatives is violated so that $A \succ B$ when $C$ is present, and $B \succ A$ if $C$ is absent}
}
\newglossaryentry{present.value}
{
  name={present value},
  description={The value of a stream of future cash payments using the method of discounting with a rate schedule}
}
\newglossaryentry{price.distorter}
{
  name={price distorter},
  description={Any exogenous news, constraints on trading, widely held beliefs, or widely used strategies that effect \glspl{price.impact}. For example, the fact that mutual funds cannot short stocks suggests that prices might be higher than they would be if shorting were allowed. Same as \gls{distortion.factor}}
}
\newglossaryentry{price.impact}
{
  name={price impact},
  description={The change in price due to a trade or sequence of trades, as a function of price, quantity and market depth. Note that concept is ill-defined, in the sense that it assumes a counterfactual price that would have obtained in the absence of that trade or trade sequence. Since markets are replete with ``nuisance variables'' that are demonstrably important in determining price impact, in practice one usually settles for estimates of average impact using statistical models}
}
\newglossaryentry{Price.Impact.Law}
{
  name={Price Impact Law},
  description={A part of the \glspl{Half.Cubic.Law} which states that the average normalized price impact is approximately proportional to the square root of the average normalized quantity traded}
}
\newglossaryentry{price.resistance}
{
  name={price resistance},
  description={In \gls{technical.analysis}. a price that acts as a barrier to price increases. A stock has resistance at $\$50$, for example, if the price approaches that level, usually twice or more, but each time fails to exceed it. Resistance is strong if it bounces hard off that level, usually with significant size offered at that level, or weak otherwise. See also \gls{resistance}, \gls{price.support} and \gls{support}}
}
\newglossaryentry{price.support}
{
  name={price support},
  description={In \gls{technical.analysis}. a price that acts as a barrier to price decreases. A stock has support at $\$50$, for example, if the price approaches that level, usually twice or more, but each time fails to drop below it. Support is strong if it bounces hard off that level, usually with significant size offered, or weak otherwise. See also \gls{support}, \gls{price.resistance} and \gls{resistance}}
}
\newglossaryentry{priming}
{
  name={priming},
  description={An effect in which the \gls{associative.machine} uses ambient information, including irrelevant information, to make a decision}
}
\newglossaryentry{principal.components.analysis}
{
  name={principal components analysis},
  description={A method of decomposing a $n$-multivariate dataset into $m < n$ mutually orthogonal \glspl{direction} such that the sum of the variances asociated with the directions is maximal over all other sets of $m$ directions. \acrshort{aPCA} results from performing an eigenvalue decomposition of the covariance matrix and selecting eigenvectors associated with the $m$ largest eigenvalues},
  plural={principal components analyses}
}
\newglossaryentry{prisoners.dilemma}
{
  name={prisoner's dilemma},
  description={A famous two player artificial mathematical game in which the unique Nash equilibrium is not Pareto efficient. Often called a paradox, but that is incorrect. It is a consequence of the requirements of non-cooperative play that precludes the players from cooperating}
}
\newglossaryentry{prescriptive.decision.science}
{
  name={prescriptive decision science},
  description={See \gls{decision.science}}
}
\newglossaryentry{problem.of.multiplicity}
{
  name={problem of multiplicity},
  description={If tests are independent, then testing each of $n$ hypotheses at a significance level $p$ results in a probability $1 - (1 - p)^n$ of rejecting at least one, a quantity that approaches 1 as $n \rightarrow \infty$. For example, if 40 independent hypotheses are each tested at a $5\%$ level of significance level, then the probability of at least one rejection is $1 - 0.95^40 = 0.87$}
}
\newglossaryentry{probability.matching}
{
  name={probability matching},
  description={A behavior that occurs in experiments when a reward has probabilities known to experimenters but not to experimental subjects. In such experiments, both animals and humans eventually randomize their choices in alignment with those probabilities. Such behavior is irreconcilable with \acrlong{arct}}
}
\newglossaryentry{probability.space}
{
  name={probability space},
  description={A probability space $\mathcal{U}$ consists of (1) a set of \emph{outcomes} $\Omega$,
(2) a sigma-field $\mathcal{E}$ of \emph{events} on $\Omega$, and (3) a \emph{probability} function $P: \mathcal{E} \rightarrow [0,1]$ for which (i) $P[\Omega] = 1$, and (ii) for any disjoint family $\{E_i\}_{i=1}^{i=\infty} \subset \mathcal{E}$, $P[\bigcup_{i=1}^{i=\infty}] = \sum_{i=1}^{i=\infty} P[E_i]$}
}
\newglossaryentry{professional.gamblers}
{
  name={professional gamblers},
  description={A person whose livelihood consists of enlightened risk taking. In the economic theory of markets, the set of all professional gamblers in financial markets is empty}
}
\newglossaryentry{profile}
{
  name={profile},
  description={In the technical trading terminology ot this book, a \emph{profile} or \gls{signal.profile} is a collection of vectors $v_t \in \mathbb{R}^n$, that at each time $t$ gives a trading signal for trading variables $p_t \in  \mathbb{R}^n$ as a scalar product $v_t \cdot p_t$. If $v_t = v$ is constant for all $t$, then the \emph{profile} is called \emph{stationary}. For the most part, methods used in this book produce stationary profiles}
}
\newglossaryentry{profit.and.loss}
{
  name={profit \& loss},
  description={and abbreviation for \emph{profit and loss}},
  plural={profits \& losses}
}
\newglossaryentry{prospect}
{
  name={prospect},
  description={A lottery}
}
\newglossaryentry{prospect.theory}
{
  name={prospect theory},
  description={A descriptive model of human decision making in risky and uncertain situations}
}
\newglossaryentry{pure.arbitrage}
{
  name={pure arbitrage},
  description={An \gls{arbitrage} having such small risk that a profit is almost certain, e.g. the near simultaneous purchase and sale of the same security on two different exchanges at advantageously different prices}
}
\newglossaryentry{pure.strategy}
{
  name={pure strategy},
  description={In \gls{game.theory}, a strategy in a player's strategy set},
  plural={pure strategies}
}
\newglossaryentry{pure.strategy.profile}
{
  name={pure strategy profile},
  description={In \gls{game.theory}, a strategy in the Cartesian product of players' strategy sets}
}
\newglossaryentry{Pursuit.of.Profits.Paradigm}
{
  name={Pursuit of Profits Paradigm},
  description={A type of \gls{strategic.evolution} in which a ``chase after riches'' leads to boom-bust cycles in markets}
}

\newglossaryentry{random.variable}
{
  name={random variable},
  description={A mapping from a \gls{probability.space} to a space of values. Intuitively, a varying quantity which may be thought of as arising from an underlying and unseen probabilistic process. A real-valued random variable is one whose values are real numbers}
}
\newglossaryentry{random.walk}
{
  name={random walk},
  description={A time series $X_t$, $t \in \mathbb{N}$ such that $X_t = \mu + X_{t-1} + U_t$, where $\{U_t\}$ is an \gls{i.i.d.white.noise}}
}
\newglossaryentry{randomized.game}
{
  name={randomized game},
  description={In \gls{game.theory}, a game which is derived from a basic game by extending strategies to randomized strategies and payoffs to expected utilities of \glspl{strategy.profile}}
}
\newglossaryentry{randomized.strategy}
{
  name={randomized strategy},
  description={In \gls{game.theory}, a strategy in which each player independently chooses a strategy with a known private probability distribution. Also known as a \gls{mixed.strategy}},
  plural={randomized strategies}
}
\newglossaryentry{rational}
{
  name={rational},
  description={Referring to a being in an enlightened state of \gls{rationality}. It is an open question, however, as to whether a being of such intellect would choose to operate as if guided by a \gls{utility.function}}
}
\newglossaryentry{rational.actor}
{
  name={rational actor},
  description={A decision maker who applies a von Neumann/Morgenstern (vNM) utility function to \glspl{decision.problem}}
}
\newglossaryentry{rational.choice.theory}
{
  name={rational choice theory},
  description={The astounding proposal that everybody acts at each and every waking moment as though they seek the most cost-effective way to achieve their goals, irrespective of the worthiness of those goals --- definitely good for rapacious psychopaths, bad for everybody else}
}
\newglossaryentry{rational.expectations}
{
  name={rational expectations},
  description={The curious proposal that the average human is a vNM automaton, or failing that, that an average human can be treated as a vNM automaton, or if that be unacceptable, that average groups of humans act as though they are vNM automatons, and in the event that this is found wanting, that these assumptions are close enough for economic work \ldots justifying the momentous conclusion that the best guess about the future is people's average opinion today}
}
\newglossaryentry{rational.speculator}
{
  name={rational speculator},
  description={In the \gls{DSSW.Model} of bubbles, a trader who acts with other like-minded traders to continue \glspl{trend} beyond an assets' fundamental values. These traders need to be nimble in order to exit before price corrects --- essentially a game of \gls{musical.chairs}. This strategy can be called \gls{legal.pump.and.dump}}
}
\newglossaryentry{rationality}
{
  name={rationality},
  description={Referring to an elightened state of being characterized by making all, or at least all financial, decisions as if guided by a \gls{utility.function}}
}
\newglossaryentry{R.language}
{
  name={R Statistical Language},
  description={A statistical and data analysis software package available at URL: \url{www.r-project.org}}
}
\newglossaryentry{reading.people}
{
  name={reading people},
  description={In gambling, the art of predicting from a person's gestures, mannerisms and expressions the mistakes they're about to make}
}
\newglossaryentry{real.game}
{
  name={real game},
  description={A real stuctured contest between several persons who engage strategically to determine their payoffs. See also \gls{game.theory} and \gls{artificial.mathematical.game}}
}
\newglossaryentry{rebalancing}
{
  name={rebalancing},
  description={In finance, the act of modifying portfolio allocations by trading only the differences between the current and target allocaions. All public funds perform periodic rebalancing to provide for addition or withdrawl of invested funds or to modify the current portfolio}
}
\newglossaryentry{recency}
{
  name={recency},
  description={In psychology, the strong tendency to give more recent events greater weight in decisions. Is important in the \Gls{availability} Heuristic}
}
\newglossaryentry{recognition.heuristic}
{
  name={recognition heuristic},
  description={A \gls{Fast.Frugal.Heuristics} that selects from a menu of alternatives the choice that is most ``recognizable''}
}
\newglossaryentry{Reflexivity.Of.Preference}
{
  name={Reflexivity of Preference},
  description={An axiom of utility theory that requires for each event $A$, $A \succeq A$}
}
\newglossaryentry{reflexivity}
{
  name={reflexivity},
  description={A philosophical framework for understanding social phenomena, but especially financial markets, developed by hedge fund manager \gls{George.Soros}}
}
\newglossaryentry{regret}
{
  name={regret},
  description={\emph{Regret} is used in two senses in this work: (1) as the feeling or emotion of sorrow for a past action or choice, and (2) as a rule in a \gls{decision.problem} that selects the choice which minimizes the probability of worst loss}
}
\newglossaryentry{relative.value.strategy}
{
  name={relative value strategy},
  description={Any strategy that shorts an ``overvalued'' portfolio and buys and ``undervalued'' one. The entire portfolio can be \emph{market neutral}, meaning that it has little expected gain or loss when the market moves. It can also be \emph{dollar neutral}, meaning that it requires a net zero investment ignoring fees},
  plural={relative value strategies}
}
\newglossaryentry{representativeness}
{
  name={representativeness},
  description={A decision heuristic that makes the most representative choice}
}
\newglossaryentry{resistance}
{
  name={resistance},
  description={Same as \gls{price.resistance}. See also \gls{price.support} and \gls{support}}
}
\newglossaryentry{return}
{
  name={return},
  description={In finance, a measure of the gain or loss of an investment that does not depend on price. Given an investment purchases at price $p$ and later valued at $q$, the fractional return over that holding period $(q-p)/p$, the percentage return is $100*(q-p)/p$, and the logarithmic return is $log(q/p)$}
}
\newglossaryentry{riding.the.yield.curve}
{
  name={riding-the-yield-curve},
  description={A type of \gls{carry.trade} in which short maturity notes are sold and long maturity ones purchased, and as short maturity ones mature are replaced by new ones. When the yield curve is upward sloping, the trade has positive returns; when it inverts, though, the trade loses}
}
\newglossaryentry{risk}
{
  name={uncertainty},
  description={In financial decision theory, referring to situations whose non-deterministic events are known with known probabilities. See also \gls{uncertainty}}
}
\newglossaryentry{risk-adjusted.return}
{
  name={risk-adjusted return},
  description={An asset return that is ``adjusted'' for risk. The most common method consists of dividing an asset's excess return (the difference of the return and the risk-free rate) by the asset's standard deviation (Sharpe ratio), but other methods of risk-adjusting also exist: the Treynor ratio, the Sortino Ratio, the Sterling Ratio and Jensen's Alpha}
}
\newglossaryentry{risk-free.rate}
{
  name={risk-free rate},
  description={A default-free rate of interest}
}
\newglossaryentry{risky.decision.problem}
{
  name={risky decision problem},
  description={A \gls{decision.problem} that involves only events with known probabilities. See also \gls{uncertain.decision.problem}},
  plural={Risky Decision Problem}
}
\newglossaryentry{risk.arbitrage}
{
  name={risk arbitrage},
  description={Statistical \gls{arbitrage}. See also \emph{pure arbitrage}}
}
\newglossaryentry{risk.averse} 
{
  name={risk averse},
  description={An individual is risk averse at $x$ if their utility function's absolute or relative coefficient of risk aversion at $x$ is positive, $r(x) > 0$. He or she is risk averse if $r(x) > 0$ for all $x$}
}
\newglossaryentry{risk.neutral}
{
  name={risk neutral},
  description={An individual is risk neutral at $x$ if their utility function's absolute or relative coefficient of risk aversion at $x$ is zero, $r(x) = 0$. He or she is risk neutral if $r(x) = 0$ for all $x$}
}
\newglossaryentry{risk.seeking}
{
  name={risk seeking},
  description={An individual is risk seeking at $x$ if their utility function's absolute or relative coefficient of risk aversion at $x$ is negative, $r(x) < 0$. He or she is risk seeking if $r(x) < 0$ for all $x$}
}
\newglossaryentry{robust-yet-fragile}
{
  name={robust-yet-fragile},
  description={A property financial networks in which small shocks have low probabilities of turning into cascades (\gls{robustness}) but there is a small chance of very large cascade (\glslink{fragile}{fragility})}
}
\newglossaryentry{robustness}
{
  name={robustness},
  description={The property of a dynamic system in which the probability of large and rapid changes have very low probability. Not \gls{fragile}}
}
\newglossaryentry{roehner}
{
  name={Bertrand M. Roehner},
  description={Prominent early contributor (physicist) to econophysics}
}
\newglossaryentry{Royal.Dutch.Company}
{
  name={Royal Dutch Company},
  description={A petroleum company headquarted in London, but which shares all profits with the Dutch company, \gls{Shell}. Deviation of the prices of these two companies is the source of a serious EMH \gls{anomaly}}
}
\newglossaryentry{rule.based.perception}
{
  name={rule-based perception},
  description={Perception using rules which generally involves decomposing scenarios into components. As opposed to \gls{holistic.perception}}
}
\newglossaryentry{rule.of.seemingly.unrelated.accidents}
{
  name={rule of seemingly unrelated accidents},
  description={A rule by which the occurrence of frequent, sporadic, seemingly unrelated ``accidents'' in markets augur a crisis, the mecahism being some hidden contagion-provoking connections among apparently unrelated segments of the markets}
}

\newglossaryentry{salience}
{
  name={salience},
  description={In psychology, the strong tendency to give greater weight to observations that for some reason ``stand out'' from others. For example, a red dot in a field of black ones draws attention to itself and is remembered better than the black ones, a traumatic event will be remembered easily and a moving object will stand out against a stationary background. Is important in the \Gls{availability} Heuristic}
}
\newglossaryentry{SAFM}
{
  name={SAFM},
  description={An acronym for the \emph{Strategic Analysis of Financial Markets} framework}
}
\newglossaryentry{satisficing}
{
  name={satisficing},
  description={A term originated by \gls{simon} for \gls{bounded.rationality}. A boundedly rational decision maker, unlike a Superhuman, has limited time and resources, so Simon suggested that a search would consider choices only until finding one that is ``good enough''; this process of making a choice is called \emph{satisficing}}
}
\newglossaryentry{save.more.tomorrow}
{
  name={Save More Tomorrow},
  description={A savings plan designed by Shlomo Benarzi and Richard Thaler based on behavioral principles, that offers employees a program that commits only raises to contributions and changes the default choice to the Save More Tomorrow plan}
}
\newglossaryentry{security.level}
{
  name={security level},
  description={A minimum acceptable wealth level in the \gls{spa} theory}
}
\newglossaryentry{security.potential.criterion}
{
  name={security-potential criterion}, 
  description={In the \gls{spa} theory a decumulatively weighted value rule of the form $SP = \sum_i h(D_i)(v_i - v_{i-1})$, where the decumulative probabilities are $D_i = \sum_{j=i}^{i=n} p_j$, and the function $h$ has the form $h(D) = w D^{q_s + 1} + (1-w)(1 - (1 - D)^{q_p + 1}$}
}
\newglossaryentry{self-control}
{
  name={self-control},
  description={In general usage, restraint exercised over one's own impulses, emotions, or desires. In finance, referring to the ability to commit to future courses of action. Weak self-control implies that an individual is impulsive, incapable of honoring self-imposed rules, e.g. following a diet. Strong self-control implies that an individual is able to make rules and to follow them}
}
\newglossaryentry{semelparous}
{
  name={semelparous},
  description={A mode of reproduction in which parents die when offspring are born}
}
\newglossaryentry{semiparametric.model}
{
  name={semiparametric model},
  description={In statistics, a model that has parametric and nonparametric (infinite dimensional parameter space) parts}
}
\newglossaryentry{sentiment}
{
  name={sentiment},
  description={A term that refers to the range of opinions and attitudes of the players in a market. In financial markets, diverse attitudes are often simplified to: bullish, bearish and neutral; in real estate, to prices ``too high'', ``too low'' or ``afforable.'' Sentiment has been believed to be important in selecting market tops and bottoms --- low bullish sentiment predicting an advancing market, and high bullish sentiment predicting a declining market. Same as \gls{market.sentiment}}
}
\newglossaryentry{sentiment.indicator}
{
  name={sentiment indicator},
  description={A statistic that purports to measure bullish and bearish \gls{sentiment} in a market}
}
\newglossaryentry{SEO}
{
  name={SEO},
  description={An acronym for \emph{Seasoned Equity Offering}, which offers shares of stock denominated in dollars on an existing foreign publicly-held company. See also \gls{IPO}}
}
\newglossaryentry{shannon}
{
  name={Claude Shannon},
  description={A scientist who developed the theory of communication and information theory}
}
\newglossaryentry{shares.outstanding}
{
  name={shares outstanding},
  description={Same as \gls{outstanding.shares}}
}
\newglossaryentry{Sharpe.ratio}
{
  name={Sharpe ratio},
  description={The \emph{Sharpe ratio} for return series $R_t$, $t = 1, 2, \ldots, n$ equals the difference of its mean and the risk free rate divided by its standard deviation. The \gls{information.ratio} is similar but does not deduct the \gls{risk-free.rate} from the mean return}
}
\newglossaryentry{Shefrin.Hersh}
{
  name={Hersh Shefrin},
  description={Author of seminal research in behavioral finance explaining, inter alia, the disposition effect and investors' preferences for dividends}
}
\newglossaryentry{Shell}
{
  name={Shell},
  description={A petroleum company headquarted in Amsterdam, but which shares all profits with the English company, \gls{Royal.Dutch.Company}. Deviation of the prices of these two companies is the source of a serious EMH \gls{anomaly}}
}
\newglossaryentry{signal}
{
  name={signal},
  description={In the terminology used in this work, a \emph{signal} at time $t$ is a statistic calculated from data available on, or prior to $t$, for use in making trading decisions}
}
\newglossaryentry{signal.profile}
{
  name={signal profile},
  description={In the technical trading terminology ot this book, a \emph{signal.profile} or just \gls{profile} is a collection of vectors $v_t \in \mathbb{R}^n$, that at each time $t$ gives a trading signal for trading variables $p_t \in  \mathbb{R}^n$ as a scalar product $v_t \cdot p_t$. If $v_t = v$ is constant for all $t$, then the \emph{profile} is called \emph{stationary}. For the most part, methods used in this book produce stationary profiles}
}
\newglossaryentry{simon}
{
  name={Herbert Simon},
  description={Nobel Laureate in Economics, 1978}
}
\newglossaryentry{simple.moving.average}
{
  name={simple moving average},
  description={A simple moving average with memory $k$ is a \gls{moving.average} having $\alpha_s = 1/k$ for $0 \le s \le k-1$ and $0$ for $s ge k$, where $SMA_t = \sum_{s=0}^{s=\infty} \alpha_s X_{t-s}$}
}
\newglossaryentry{slippage}
{
  name={slippage},
  description={In the context of trade execution, the difference between the target price of a trade and the actual price of execution. For example, a purchase of $1,000$ shares at target price price $\$10.50$ might be executed instead at $\$10.60$, in which case the slippage is $\$0.10$. In general, the slippage will be greater the larger the trade, and this must be accounted for in the backtesting of any trading system}
}
\newglossaryentry{vernon.smith}
{
  name={Vernon Smith},
  description={Nobel Laureate in Economics, 2002 for his contributions to experimental econometrics. Smith conducted laboratory experiments to test the predictions of economic theories},
  sort={Smith}
}
\newglossaryentry{smoothing}
{
  name={smoothing},
  description={A method of producing from a training dataset $(x_i,y_i)$ sampled from a density satisfying a functional relationship $f(x) = E[ Y | X=x]$, an estimator $\hat{f(x)}$ that had controlled variation, e.g. that is, ``smoothness''}
}
\newglossaryentry{smoothing.spline}
{
  name={smoothing spline},
  description={A smoothing method for a spline that uses a penalty for non-smoothness (smoothness regularization) to produce an approximating function. It is useful when the data have noise or there are multiple y-values for each x-value}
}
\newglossaryentry{social.dilemma}
{
  name={social dilemma},
  description={A game in which all individuals gain if all cooperate, but any one of them gains by not cooperating when all the others do}
}
\newglossaryentry{sornette}
{
  name={Didier Sornette},
  description={Prominent geophysicist and econophysicist, author or coauthor on crash models (LPPL), power laws in economics and complex systems. Currently Chair of Entrepeneurial Risks at ETH, Switzerland}
}
\newglossaryentry{spa}
{
  name={SP/A},
  description={An acronym for \emph{Security-Potential/Aspiration}, a theory of choice under uncertainty proposed by Lola Lopes. Uncertainty leads to fear which is manifested as need for security. Lower perceived undertainty encourages hope which manifests as ``perceived potential.'' All decisions are constrained by aspirations}
}
\newglossaryentry{speculative.pump.and.dump}
{
  name={speculative pump-and-dump},
  description={A rational trading strategy that consists of buying (selling) in concert with other like-minded traders in order to create a self-sustaining trend that can later be liquidated profitably to uninformed traders. This strategy relies on some traders (the more, the better) who follow trends without recourse to fundamental information}
}
\newglossaryentry{standardization}
{
  name={standardization},
  description={A term that applies to a sample $X_1, X_2, \ldots, X_n$. The $X_i$ are said to be standardized (to $X_i'$) by $X_i' = \frac{X_i - \, \bar{X}}{S}$, where $\bar{X} = n^{-1} \sum X_i$ and $S = (n-1)^{-1} \sum (X_i - \bar{X})^2$ }
}
\newglossaryentry{stanley}
{
  name={H. Eugene Stanley},
  description={Prominent physicist (statistical mechanics) and indisciplinary scientist at Boston University}
}
\newglossaryentry{stationary}
{
  name={stationary},
  description={A property of stochastic processes that requires the finite dimensional distributions to be invariant under time shifts}
}
\newglossaryentry{statistical.arbitrage}
{
  name={statistical arbitrage},
  description={An non-pure \gls{arbitrage} having a positive expected return, e.g. index \gls{arbitrage}, \glslink{pairs.trading.strategy}{pairs trading}, options delta-neutral hedging}
}
\newglossaryentry{statistical.independence}
{
  name={statistical independence},
  description={A reference to a set of random variables that are \gls{independent}. Same as \gls{independence}}
}
\newglossaryentry{statistical.regularity}
{
  name={statistical regularity},
  description={A recurring pattern in the market data for tradable assets},
  plural={statistical regularities}
}
\newglossaryentry{Statman.Meir}
{
  name={Meir Statman},
  description={Author of seminal research in behavioral finance explaining, inter alia, the disposition effect and investors' preferences for dividends}
}
\newglossaryentry{status.quo.bias}
{
  name={status quo bias},
  description={A bias in which a person values the status quo (no change) more highly than superior alternatives},
  plural={status quo biases}
}
\newglossaryentry{stochastic.process}
{
  name={stochastic process},
  description={A collection $\{X_t\}$ indexed by times $t \in T$. The time domain $T$ can be discrete or continuous},
  plural={stochastic processes}
}
\newglossaryentry{stop.order}
{
  name={stop order},
  description={A market order to buy (sell) at a \emph{stop price} or higher (lower). If the order is not executed at the stop price, it becomes a \gls{market.order}. See also \gls{market.order} and {limit.order}}
}
\newglossaryentry{Strategic.Analysis.of.Markets.Method}
{
  name={Strategic Analysis of Markets Method},
  description={The \emph{\acrshort{aSAMM}} is a framework for developing trading systems by using game theoretic, strategic and statistical analysis. As such, the SAMM is one degree removed from flesh-and-blood humans, but in its favor, is amenable to game theoretic analysis}
}
\newglossaryentry{strategic.ecology}
{
  name={strategic ecology},
  description={A strategic ecology at a moment in time is a collection of trading strategies existant at that time},
  plural={strategic ecologies}
}
\newglossaryentry{strategic.evolution}
{
  name={strategic evolution},
  description={A process in which strategies change, often in a patterned way, over time. The Minsky-Kindleberger Model is one example of strategic evolution}
}
\newglossaryentry{strategic.form}
{
  name={strategic form},
  description={An \gls{artificial.mathematical.game} in which the \glspl{strategy} available to the players are elements of strategy sets $C_i$. These sets in simple cases are finite, but in many games are infinite, and may represent a complete specification of every choice a player might make in some game contingency. Also called the \gls{normal.form}}
}
\newglossaryentry{strategic.plan}
{
  name={strategic plan},
  description={An incomplete trading scheme or idea that has parameters which if specified convert it to a \gls{trading.algorithm}. A plan such as ``Buy in a bull market, sell in a bear market by ,'' is not a strategic plan as it stands, but could be one if parameters type=\{bull, bear, neither\}, asset=\{set of tradable assets\}, execution=\{open,close\}, and any other variables that are required to decide what and when to buy and sell, are added to its statement}
}
\newglossaryentry{strategic.plan.capacity}
{
  name={strategic plan capacity},
  description={A return schedule for the aggregate return of strategies that implement a \gls{strategic.plan}. The idea is that while a \gls{strategic.plan} admits polymorphic implementations, the commonality of parameters and the commmonality of estimation methods will cause all to be exposed to common factors}
}
\newglossaryentry{strategic.pump.and.dump}
{
  name={strategic pump-and-dump},
  description={A rational trading strategy that consists of buying (selling) from a fundamentalist-initiated trend in concert with other like-minded traders in order to create a self-sustaining trend that can later be liquidated profitably to uninformed traders. This strategy relies on some traders (the more, the better) who follow trends without recourse to fundamental information}
}
\newglossaryentry{strategy}
{
  name={strategy},
  description={(a) In \gls{game.theory}, a complete specification of the choices that a player would make under every possible game contingency, (b) In trading, a set of guidelines, a set of rules or an algorithm that a trader, investor or computer uses to buy and sell. Some strategies are purely mechanical and can be executed by a computer. Some are purely discretionary, and require one to treat human decisions to buy and sell as an algorithm, albeit one that cannot be programmed},
 plural={strategies}
}
\newglossaryentry{strategy.capacity}
{
  name={strategy capacity},
  description={A schedule that indicates a strategy's return degredation as a function of the aggregate position size}
}
\newglossaryentry{strategy.diversity}
{
  name={strategy diversity},
  description={In the \gls{Pursuit.of.Profits.Paradigm}, the degree of heterogeneity among strategies that implement its \gls{strategic.plan}}
}
\newglossaryentry{strategy.linkage}
{
  name={strategy linkage},
  description={In the \gls{Pursuit.of.Profits.Paradigm}, the degree of potentially contagious connection to other strategies}
}
\newglossaryentry{strategy.profile}
{
  name={strategy profile},
  description={In \gls{game.theory}, given strategy sets $C_i$ for $i =1, 2, \ldots, n$, a vector of strategy choices $c = (c_1, c_2, \ldots, c_n)$}
}
\newglossaryentry{strict.ordering}
{
  name={strict ordering},
  description={In utility theory, a strict ordering $\succ$ is derived from a weak ordering $\succeq$ by  
\begin{equation*}
  A \succ B \mbox{ } := \mbox{ } A \succeq B \mbox{ and } A \not\sim B
\end{equation*}
}
}
\newglossaryentry{t.distribution}
{
  name={Student's t distribution},
  description={Random variable $X$ has a Student's t distribution with $\nu$ degrees of freedom if its density is 
\[
  \frac{\Gamma(\frac{\nu+1}{2})}{\sqrt{\nu \pi} \, \Gamma(\frac{\nu}{2})} \, \left[ 1 + \frac{x^2}{\nu} \right]^{-\frac{\nu + 1}{2}} \text{ for } x \in (-\infty,\infty)
\]}
}
\newglossaryentry{stylized.fact}
{
  name={stylized fact},
  description={A \gls{statistical.regularity}}
}
\newglossaryentry{subadditivity.bias}
{
  name={subadditivity bias},
  description={The tendency to value a whole less than the sum of its parts}
}
\newglossaryentry{subdominant.paradigm}
{
  name={subdominant paradigm},
  description={A \gls{market.paradigm} which is a competitor to the \gls{dominant.paradigm}},
}
\newglossaryentry{subexponential}
{
  name={subexponential},
  description={A distribution that satisfies the asymptotic tail condition: 
  \[
    P \left[ (X_1 + X_2 + \cdots + X_n) > x \right] \sim n P\left[ X_1 > x \right]
  \]
  }
}
\newglossaryentry{submartingale}
{
  name={submartingale},
  description={In this work, a time series $X_t$ for which $E[X_t] < \infty$ and $E[X_{t+1} \, | \, X_t ] \ge X_t$ for all meaningful $t$. A series for which the inequality is strict is called a \emph{strict submartingale}}
}
\newglossaryentry{Substitution}
{
  name={Substitution},
  description={An axiom of \gls{cardinal.utility} theory which requires that for $A \sim B$ and any event $C$, the ithe lotteries $(p,A; (1-p)C)$ and $(p,B; (1-p)C)$ are equivalent for any $p \in [0,1]$, $(p,A; (1-p)C) \sim (p,B; (1-p)C)$}
}
\newglossaryentry{SUE}
{
  name={SUE},
  description={An acronym for \emph{Standardized Unexpected Earnings}, defined as the 
             current earnings minus those of a year ago (YoY earnings) divided by 
             their standard deviation}
}
\newglossaryentry{superior.active.portfolio}
{
  name={superior active portfolio},
  description={An active portfolio $A$ for which $E[A] > 0$}
}
\newglossaryentry{supermartingale}
{
  name={supermartingale},
  description={In this work, a time series $X_t$ for which $E[X_t] < \infty$ and $E[X_{t+1} \, | \, X_t ] \le X_t$ for all meaningful $t$. A series for which the inequality is strict is called a \emph{strict supermartingale}}
}
\newglossaryentry{support}
{
  name={support},
  description={Same as \gls{price.support}. See also \gls{price.resistance} and \gls{resistance}}
}
\newglossaryentry{suppressed.ambiguity}
{
  name={suppressed ambiguity},
  description={A mental process in which facts viewed more favorably, suppress facts that are in conflict with, or contradictory to them. The \gls{halo.effect} can lead to suppressed ambiguity}
}
\newglossaryentry{surprise}
{
  name={surprise},
  description={In the classification of market events, an unscheduled, unanticipated event that has market-moving impact. See also \gls{suspense}}
}
\newglossaryentry{survival.function}
{
  name={survival function},
  description={The function $S(x) = F_{\!_{>}}(x) = 1 - F_{\!_{\le}}(x)$, where $F_{\!_{\le}}(x)$ is a \gls{cumulative.distribution.function}. Also called the \gls{counter.cumulative.distribution.function}}
}
\newglossaryentry{suspense}
{
  name={suspense},
  description={In the classification of market events, one for which the time and place of occurrence is known, even inexactly, but not the exact outcome. In general, {suspense} events can cause short-term moves in a market due to the fact that most traders withdraw their orders before the event. Resting orders on the ``wrong'' side of the market can be ``picked off'' after the event. See also \gls{surprise}}
}
\newglossaryentry{System1}
{
  name={System 1},
  description={The mind's ``reflexive'' cognitive subsystem; massively parallel, quick, hard to reprogram, effortless}
}
\newglossaryentry{System2}
{
  name={System 2},
  description={The mind's ``reflective'' cognitive subsystem; single focussed, slow, relatively easy to reprogram, effortful}
}

\newglossaryentry{technical.analysis}
{
  name={technical analysis},
  description={The study of predictive patterns in financial markets involving only time, open, high, low, close, volume and open interest},
  plural={technical analyses}
}
\newglossaryentry{tell}
{
  name={tell},
  description={A giveaway mannerism, gesture or expression that often portends a rather unfortunate futurei, emphatically not related to \emph{fortune telling}}
}
\newglossaryentry{temporal.construal}
{
  name={temporal construal},
  description={The tendency to represent events near in time more concretely and in greater detail than events further removed}
}
\newglossaryentry{Thaler.Richard}
{
  name={Richard Thaler},
  description={Colleague of Daniel Kahneman and descoverer of behavioral finance terms --- loss aversion, the status quo bias and mental accounting}
}
\newglossaryentry{tick}
{
  name={tick},
  description={A unit of change in a futures contract. In Eurodollars priced to two decimal places, a tick is a change of $0.01$, e.g. $94.32$ to $94.33$. Generally, a tick is a minimum change, but that rule has many exceptions. Some Eurodollars, for example, are prices in \emph{half-ticks} ($0.005$) or \emph{quarter-ticks} ($0.0025$)}
}
\newglossaryentry{timestamped}
{
  name={timestamped},
  description={Referring to an event that is recorded as having occurred at a particular time, the ``timestamped'' time, of course}
}
\newglossaryentry{time.series}
{
  name={time series},
  description={A probabilistic process that produces observations at deterministic or random times}
}
\newglossaryentry{tower.property}
{
  name={tower property},
  description={Information sets $\{I_t\}$ have the tower property if $I_{t+1} \supset I_t$ for all meaningful $t$}
}
\newglossaryentry{tracking.error}
{
  name={tracking error},
  description={The deviation of the value of a \gls{basket.of.stocks} from a benchmark index}
}
\newglossaryentry{trade}
{
  name={trade},
  description={In this work, a \gls{founding.transaction} followed (eventually) by a \gls{closing.transaction} that exits the entire position created by the founding and any subsequent transactions that maintain a position on the same side of the market. This concept deviates slightly from trader's parlance, in that there may be many reductions in the size of the opening transaction before its complete closing. When refering to historical closed trades, is called a \emph{closed trade}, while when ongoing (not yet closed), is called an \emph{open trade}}
}
\newglossaryentry{Trader/Manager}
{
  name={Trader/Manager},
  description={A nonstandard term used in this work for a firm that invests client funds directly into tradable assets such as bonds, stocks and commodities. See also \gls{Manager.of.Managers}}
}
\newglossaryentry{trading.algorithm}
{
  name={trading algorithm},
  description={In trading and investment, a collection of rules for trading that are specific enough to be implemented as a computer program, requiring no human intervention beyond decisions of when to use them}
}
\newglossaryentry{trading.position}
{
  name={trading position},
  description={In a trading context, either a \emph{long position} in an instrument consists of ownership of the instrument or fungible contract that can be traded, or a \emph{short position} that consists of the opposite of a long position (except for financing costs). In stocks, a short position is  created synthetically through stock borrowing; in instruments that are two-sided tontracts, the opposite side of a long contract position. When the context is clear, abbreviated to \gls{position}. For a trading entity, its position is the collection of all positions in instruments}
}
\newglossaryentry{trading.system}
{
  name={trading system},
  description={In mechanical trading and investment, a collection of well-defined rules that specify when to enter the market, the size and side (buy or sell) for entry, and when to exit all or part of an existing position. In this work, the same as a \gls{mechanical.trading.system}}
}
\newglossaryentry{Trading.System.Summary}
{
  name={Trading System Summary},
  description={A statistical report that summarizes the historical activity of a trading system in order to evaluate its performance},
  plural={Trading System Summaries}
}
\newglossaryentry{trading.time}
{
  name={trading time},
  description={Market parlance for time elapsed only during periods in which trading is possible. The daytime trading session at the \acrshort{aNYSE}, for example, is from $9$:$30$ to $4$:$30$ EST, Mondays through Fridays, excepting designated holidays and market closings due to extraordinary events}
}
\newglossaryentry{tragedy.of.the.commons}
{
  name={tragedy of the commons},
  description={A game paradigm in which many players of a game compete for a free or semi-free good in scarce supply, resulting in its uncoordinated depletion as each player attempts to maximize profit independently of other players}
}
\newglossaryentry{transaction}
{
  name={transaction},
  description={A \gls{transaction} is an executed or partially executed order, that is, one in which a quantity of a financial instrument bought or sold at a particular time and place}
}
\newglossaryentry{transactional.data}
{
  name={transactional data},
  description={Observational data of a \gls{transaction} event, in this work a buy or sell, that occurs at a particular time and is timestamped. Other market events, such as the posting of a buy or sell order or cancellation of same is not transactional data, although it becomes so if and when an actual purchase or sale is realized}
}
\newglossaryentry{transactional.format}
{
  name={transactional format},
  description={The representation of a trading history as a series of time-stamped transactions}
}
\newglossaryentry{Transitivity}
{
  name={Transitivity},
  description={An axiom of utility theory that requires of events $A \succ B$ and $B \succ C$, that $A \succ C$}
}
\newglossaryentry{trend}
{
  name={trend},
  description={In \gls{technical.analysis}, a venerated \gls{statistical.regularity} in which many portfolios that have appreciated have the tendency to continue appreciating, and many that have depreciated have a tendency to continue depreciating}
}
\newglossaryentry{trend.follower}
{
  name={trend follower},
  description={A trader that follows trends, that is, who buys when price is going up and/or sells when price is going down}
}
\newglossaryentry{trend.following}
{
  name={trend following},
  description={A strategy premised on the belief that profits are possible by detecting trends in progress and following them}
}
\newglossaryentry{TRIN}
{
  name={TRIN},
  description={The TRader's INdex (or ARMS index), which is calculated as the number of advancing issues divided by the number of declining issues}
}
\newglossaryentry{turnover.ratio}
{
  name={turnover ratio},
  description={For stocks, the ratio of the quantity traded over a specified period to the shares outstanding. For example, the daily turnover ratio of stock XYZ is the daily volume of XYZ divided by its shares outstanding}
}
\newglossaryentry{tversky}
{
  name={Amos Tversky},
  description={Amos Tversky, co-investigator with 2002 Nobel Laureate Daniel Kahneman. The ``T'' in the acronym ``K\&T''}
}

\newglossaryentry{unanticipated.news}
{
  name={unanticipated news},
  description={In financial markets, a piece of news, some portion of which was not anticipated. The term is often used to explain why some news stories have little market impact, while other seemingly innocuous stories have great impact. In this view, the market discounts all the ``anticipatible'' parts of news, and is affected only by the ``unanticipatible'' part}
}
\newglossaryentry{uncertain.decision.problem}
{
  name={uncertain decision problem},
  description={A \gls{decision.problem} that has some events with unknown probabilities. See also \gls{risky.decision.problem}}
}
\newglossaryentry{uncertainty}
{
  name={uncertainty},
  description={In financial decision theory, referring to situations affected by events whose probabilities are unknown. See also \gls{risk}}
}
\newglossaryentry{uncorrelated.white.noise}
{
  name={uncorrelated white noise},
  description={Same as \gls{white.noise}}
}
\newglossaryentry{underreaction}
{
  name={underreaction},
  description={In financial markets, underreaction refers to an insufficient price move in response to an event. See also \gls{overreaction}}
}
\newglossaryentry{upsizing.transaction}
{
  name={upsizing transaction},
  description={In a \gls{PPGS.renewal.process}, a \gls{transaction} that increases the absolute value of a current, nonzero \gls{position}}
}
\newglossaryentry{utiles}
{
  name={utiles},
  description={The units of a \gls{utility.function}}
}
\newglossaryentry{utility.function}
{
  name={utility function},
  description={A function $U(A)$ that expresses the six axioms of \Gls{cardinal.utility}. Such a function is unique up to a positive affine transformation. The expected values of such a function are the basis of \gls{vNM.expected.utility.theory}. See \gls{utility.theory}}
}
\newglossaryentry{utility.theory}
{
  name={utility theory},
  description={A theory of consistent choice making which has two types: (1) ordinal, and (2) cardinal. Ordinal utility theory has axioms of \gls{Complete.Ordering}, \gls{Reflexivity.Of.Preference} and \gls{Transitivity}. Cardinal utility theory requires the ordinal axioms and those of \gls{Compound.Equivalence}, \gls{Substitution} and the \gls{Continuity.Axiom}}
}

\newglossaryentry{valuation}
{
  name={valuation},
  description={In general, the process of assigning a value, e.g. in finance, valuation of a financial instrument, a portfolio, an outcome, etc. In \gls{prospect.theory}, the process of valuing a \glslink{framing}{framed} prospect using a \gls{value.function} and \gls{weight.function}}
}
\newglossaryentry{value}
{
  name={value},
  description={In game theory, the average payoffs that players achieve with best play. \acrshort{avnm} proved that every \gls{zero.sum} randomized game has a value, although some non-zero sum games have no well-defined one}
}
\newglossaryentry{value.function}
{
  name={value function},
  description={In \gls{prospect.theory}, a function that describes value of a decision relative to a reference point, which \acrshort{aKnT} demonstrated empirically to be convex below the reference point and concave above it}
}
\newglossaryentry{value.investing}
{
  name={value investing},
  description={An method of security analysis that uses market fundamentals to find \emph{investment value}. Made famous by Benjamin Graham in his landmark book, ``The Intelligent Investor'' \cite{graham2003intelligent}}
}
\newglossaryentry{verification.stage}
{
  name={verification stage},
  description={The stage of scientific inquiry that confirms hypotheses by collecting new data and showing that predictions match theory}
}
\newglossaryentry{vNM.expected.utility.theory}
{
  name={von Neumann/Morgenstern Expected Utility Theory},
  description={The axiomatic expected utility theory of John von Neumann and Oskar Morgenstern, which assumes that decision making is preference-consistent, extensible to lotteries and Archimedian}
}
\newglossaryentry{vNM.rational}
{
  name={vNM rational},
  description={An actor is vNM rational if he, she or it uses \gls{vNM.expected.utility.theory} to make decisions. In our usage, a vNM rational actor is simply called ``rational''}
}
\newglossaryentry{volatility.clustering}
{
  name={volatility clustering},
  description={Referring to episodic low and high volatility in financial time series in financial time series. Same as volatility \gls{intermittency}}
}

\newglossaryentry{weak.ordering}
{
  name={weak ordering},
  description={In utility theory, a relation that satisfies the three axioms: \gls{Complete.Ordering}, \gls{Reflexivity.Of.Preference} and \gls{Transitivity}. In this work, called a \gls{preference.order}}
}
\newglossaryentry{weight.function}
{
  name={weight function},
  description={In \gls{prospect.theory}, a function that maps probabilities in a prospect to weights useful in \gls{valuation}}
}
\newglossaryentry{weighted.moving.average}
{
  name={weighted moving average},
  description={A weighted moving average with memory $k$ is a \gls{moving.average} having $\alpha_s \ge 0$ for $0 \le s \le k-1$, and $\alpha = 0$ for $s ge k$, $\alpha_{k-1} > 0$ and $\sum_{s \ge 0} \alpha_s = 1$, where $WMA_t = \sum_{s=0}^{s=\infty} \alpha_s X_{t-s}$}
}
\newglossaryentry{white.noise}
{
  name={white noise},
  description={A time series $X_t$ for which $E[X_t] = 0$,
               $Var[X_t] = \sigma^2$, and $Cov[X_s,X_t] = 0$ 
               for all meaningful $s, t$ and where $\sigma$ 
               is the standard deviation of $X_1$}
}
\newglossaryentry{winner's.curse}
{
  name={winner's curse},
  description={A game-theoretic phenomenon in competitive auction markets. In effect, it states that the winner of a multi-bidder auction will be the one that is most optimistic, thus will overpay. In game theory, analogs of the winner's curse occur in all \emph{coordination games}}
}
\newglossaryentry{wn}
{
  name={WN},
  description={Abbreviation for white noise. Sometimes 
               written as $WN(\mu,\sigma^2)$, which enlarges 
               the definition to include processes that have 
               non-zero drift $\mu$}
}
\newglossaryentry{worst.loss.function}
{
  name={worst loss function},
  description={For a real-valued time series $\{X_s\}_{s=1}^{s=t}$ of gains, the largest cumulative loss $\mathcal{L}(X,t)$ to time $t$, i.e. for $V_0 = 0$, $V_s = \sum_{u=1}^{u=s} X_u$, $s \ge 1$, \[\mathcal{L}(X,t) = -\underset{1 \le s \le t}{min} V_s.\] Note that $\mathcal{L}(X,t) \ge 0$}
}

\newglossaryentry{zero.sum}
{
  name={zero-sum},
  description={A real or artificial mathematical game in which the sum of payoffs to the players is zero}
}
\newglossaryentry{Zipfs.law}
{
  name={Zipf's law},
  description={A power law distribution having positive lower threshold and exponent $1$}
}

\newacronym{aaa}{$A\&A$}{Anchoring \& Adjustment}
\newacronym{aACE}{ACE}{Alternating Conditional Expectations Model}
\newacronym{aAMEX}{AMEX}{American Stock Exchange}
\newacronym{aarch}{$ARCH$}{Autoregressive Conditional Heteroschedasticity}
\newacronym{aarima}{$ARIMA$}{Autoregressive Integrated Moving Average}
\newacronym{aarma}{$ARMA$}{Autoregressive Moving Average}
\newacronym{aAMH}{AMH}{Adaptive Markets Hypothesis}
\newacronym{aAVAS}{AVAS}{Additivity and  Variance Stabilizing Transformations Model}
\newacronym{abf}{BF}{behavioral finance}
\newacronym{abp}{b.p.}{basis point}
\newacronym{abnl}{B\&L}{Brennan \& Lo}
\newacronym{aCAPM}{CAPM}{Capital Asset Pricing Model}
\newacronym{aCAR}{CAR}{Cumulative abnormal return}
\newacronym{aCARA}{CARA}{Constant Absolute Risk Aversion}
\newacronym{accdf}{CCDF}{counter-cumulative distribution function}
\newacronym{acdf}{CDF}{cumulative distribution function}
\newacronym{aCME}{CME}{Chicago Mercantile Exchange}
\newacronym{aCRRA}{CRRA}{Constant Relative Risk Aversion}
\newacronym{acrsp}{CRSP}{Center for Research in Security Prices}
\newacronym{aCPT}{CPT}{Cumulative Prospect Theory}
\newacronym{aDJIA}{DJIA}{Dow Jones Industrial Average}
\newacronym{adp}{DP}{Decision Problem}
\newacronym[longplural={Decision Theories}]{adt}{DT}{Decision Theory}
\newacronym{aefm}{EFM}{Evolutionary Finance Model}
\newacronym{aEMH}{EMH}{Efficient Market Hypothesis}
\newacronym{aETF}{ETF}{Exchange Traded Fund}
\newacronym{aeu}{EU}{Expected Utility}
\newacronym{afd}{FD}{Fast-decaying Distribution}
\newacronym{aFED}{FED}{Federal Reserve System}
\newacronym{aFIH}{FIH}{Financial Instability Hypothesis}
\newacronym{aFLOG}{FLOG}{Fundamental Laws of Gambling}
\newacronym{afnd}{F\&D}{Froot \& Debora}
\newacronym{afifo}{FIFO}{First In/First Out}
\newacronym{aGAM}{GAM}{Generalized Additive Model}
\newacronym{aGFC}{GFC}{Global Financial Crisis of 2007-8}
\newacronym{aGnH}{G\&H}{Grinblatt \& Han}
\newacronym{aGnT}{G\&T}{Gigerenzer \& Todd}
\newacronym{agarch}{$GARCH$}{Generalized Autoregressive Conditional Heteroschedasticity}
\newacronym{ahbm}{HBM}{Heterogeneous Beliefs Model}
\newacronym{aHE}{HE}{Homo Economicus}
\newacronym[longplural={independent components analyses}]{aICA}{ICA}{independent components analysis}
\newacronym{aIRA}{IRA}{Individual Retirement Account}
\newacronym{aJnT}{J\&T}{Jegadeesh \& Titman}
\newacronym{alor}{LOR}{Leland O’Brien Rubinstein Associates}
\newacronym{ahft}{HFT}{High-frequency Trading}
\newacronym{aKnT}{K\&T}{Kahneman and Tversky}
\newacronym{aMBS}{MBS}{Mortgage-Backed Security}
\newacronym{amkm}{MKM}{Minsky-Kindleberger Model}
\newacronym{aMNS}{MNS}{Market Neurtal Strategy}
\newacronym{aNASDAQ}{NASDAQ}{National Association of Security Dealers Exchange}
\newacronym{ancm}{NCM}{Noisy Channel Model}
\newacronym{amna}{M\&A}{Mergers \& Acquisitions}
\newacronym{aNYSE}{NYSE}{New York Stock Exchange}
\newacronym[longplural={principal components analyses}]{aPCA}{PCA}{principal components analysis}
\newacronym{apd}{PD}{Prisoner's Dilemma}
\newacronym{apdf}{p.d.f}{probability density function}
\newacronym{apgg}{PGG}{Public Goods Game}
\newacronym{apl}{PL}{Power Law}
\newacronym{aPLT}{PLT}{Power Law Tail}
\newacronym{aPPGS}{PPGS}{Potentially Profitable Gambling System}
\newacronym{aptb}{PTB}{Pick-the-Best Heuristic}
\newacronym{aRE}{RE}{Rational Expectations}
\newacronym{arecap}{RECAP}{Record, Evaluate, and Compare Alternative Prices}
\newacronym[longplural={Rational Belief Equilibria}]{arbe}{RBE}{Rational Belief Equilibrium}
\newacronym{arct}{RCT}{Rational Choice Theory}
\newacronym{armm}{RMM}{Reflexive Market Model}
\newacronym{aSAMM}{SAMM}{Strategic Analysis of Markets Method}
\newacronym{asf}{s.f.}{survival function}
\newacronym{aSnP}{S\&P 500}{Standard \& Poors 500 Index}
\newacronym{aSMK}{SMK}{Soros-Minsky-Kindleberger Model}
\newacronym{aAUM}{AUM}{Assets Under Management}
\newacronym{aVIX}{VIX}{CBOE Volatility Index}
\newacronym{avnm}{vNM}{von Neumann/Morgenstern}
\newacronym{aweird}{WEIRD}{Western, Educated, Industrial, Rich and Democratic}
\newacronym{awn}{WN}{white noise}
\newacronym{aWSMC87}{Crash of 1987}{Worldwide Stock Market Crash of 1987}
\newacronym{awysiati}{WYSIATI}{What you see is all there is}

\begin{document}

\begin{titlepage}
  \centering
  {\scshape\Large {Why Markets are Inefficient:\\A Gambling ``Theory'' of Financial Markets\\For Practitioners and Theorists}\par}
  \vspace{1.5cm}
  {\scshape\Large February 22, 2017\par}
  \vspace{2cm}
  {\Large Steven D. Moffitt\textsuperscript{\textdagger}\par}
  \vfill
  \textdagger Adjunct Professor of Finance, Stuart School of Business, Illinois Institute of Technology and
              Principal, Market Pattern Research, Inc.\par
\end{titlepage}

\abstract{
\noindent 
The purpose of this article is to propose a new ``theory,'' the \emph{Strategic Analysis of Financial Markets (SAFM)} theory, that explains the operation of financial markets using the analytical perspective of an enlightened gambler. The gambler understands that all opportunities for superior performance arise from suboptimal decisions by humans, but understands also that knowledge of human decision making alone is not enough to understand market behavior --- one must still model how those decisions lead to market prices. Thus are there three parts to the model: gambling theory, human decision making and strategic problem solving. A new theory is necessary because at this writing in 2017, there is no theory of financial markets acceptable to both practitioners and theorists. Theorists' efficient market theory, for example, cannot explain bubbles and crashes nor the exceptional returns of famous investors and speculators such as Warren Buffett and George Soros.  At the same time, a new theory must be sufficiently quantitative, explain market "anomalies" and provide predictions in order to satisfy theorists. It is hoped that the SAFM framework will meet these requirements.
}

\null\vfill
\newpage

\section{Introduction}

Though we have over two centuries of financial market history and various theories of price formation, today there exists no single theory acceptable to market practitioners, yet rigorous enough to satisfy market theorists. The only major theory that purports universality is efficient market theory, which fails to explain some of the most important market phenomena, e.g. bubbles and crashes. That theory's insistence that markets can't be beaten, despite stellar careers of such luminaries as \gls{Buffett.Warren}\index{Buffett, Warren} and \gls{Soros.George}\index{Soros, George}, makes it unacceptable to practitioners. The purpose of this article is to propose the \emph{\glslink{SAFM}{Strategic Analysis of Financial Markets (SAFM)}}\index{SAFM}\index{Strategic Analysis of Financial Markets framework} framework as a ``theory'' that will be acceptable to practioners and theorists alike.

The perspective of this framework is best understood through the eyes of an enlightened gambler. That gambler uses the \emph{\gls{Fundamental.Laws.of.Gambling}}\index{Fundamental Laws of Gambling}\index{FLOG} together with knowledge of other market participants' trading strategies to identify potentially superior investments, and then uses data analysis to extract winners. The gambler acknowledges that in the market game, prices are the result of strategic action by many players, some astute, some mediocre and some fools. The gambler understands that markets evolve in response to changing conditions, and in particular, that all static trading systems eventually get \gls{arbbed.out}\index{arbbed out} and fail. 

The SAFM's development is quite different from efficient market theory's. It is a constructive theory, in the sense that its building blocks are human behavior and strategic thinking, and that it is not developed by assuming unrealistic axioms. It requires the theory of gambling, because only good gamblers succeed in the market game. It requires a strategic life cycle model (the \emph{\glslink{Pursuit.of.Profits.Paradigm}{POPP}}\index{Pursuit of Profits Paradigm}) to explain how markets evolve over time. And it requires \emph{\gls{behavioral.finance}}\index{behavioral finance}, the study of financial decision making, to explicitly involve humans. These elements are combined logically to form the SAFM framework. A great advantage of the SAFM, and a major reason for its viability, is that a method for discovering trading edges (the \emph{\gls{Strategic.Analysis.of.Markets.Method}})\index{Strategic Analysis of Markets Method (SAMM)} flows out of it naturally.

Here are the main elements of the framework, leaving explanations for undefined terms to later sections.
\begin{itemize}
  \item{\emph{\glslink{potentially.profitable.gambling.system}{Potentially Profitable Gambling Systems} (PPGS's)}.\index{Potentially Profitable Gambling System}\index{PPGS}}
  \item{The \emph{\gls{Fundamental.Laws.of.Gambling} (FLOG)}.\index{Fundamental Laws of Gambling}\index{FLOG}}
  \item{\emph{FLOG} $\longrightarrow$ \emph{\glslink{grational}{Grationality}}.\index{grationality}}
  \item{The \emph{\gls{Pursuit.of.Profits.Paradigm} (POPP)}\index{POPP} and strategic life cycles.}
  \item{Getting an Edge: Human Behavior and \emph{Price Distorters}}
  \item{The \emph{\gls{Strategic.Analysis.of.Markets.Method} (SAMM)} $\longrightarrow$ PPGS's.\index{Strategic Analysis of Markets Method}\index{SAMM}}
\end{itemize}
Brief discussions of each item in this list are contained in the following sections. But much is omitted in the interests of brevity --- for a full development, see the books on which this summary is based, ``The Strategic Analysis of Financial Markets, Volume 1: Framework'' (\cite{moffitt2017V1}) and `The Strategic Analysis of Financial Markets, Volume 2: Trading System Analytics'' (\cite{moffitt2017V2}).

\section{Potentially Profitable Gambling\\ Systems}

In efficient market theory, prices are assumed to be unpredictable given past publicly available information. For prices represented as a time series with fixed, equally spaced times, one way to state this is
\begin{align}
  E[P_t + C_t \, | \, \mathscr{I}_{t-1}] \; = \; E[ 1 + R_t \, | \, \mathscr{I}_{t-1}] \, p_{t-1} \label{RME:E:EfficientMarketHypothesisFormula-General}
\end{align}
where
\begin{align*}
  P_t               & \quad \text{ is the price random variable for any asset at time } t, \\
  C_t               & \quad \text{ is the time $t$ present value of its cash flows in } (t-1,t], \\
  R_t               & \quad \text{ is the random, security-independent return from time }\\
                    & \quad \text{ }  t-1 \text{ to time } t,  \\
  \mathscr{I}_{t-1} & \quad \text{ is an information set available at time } t-1, \\
  p_{t-1}           & \quad \text{ is the realized (i.e. not random) price at time } t-1.
\end{align*}

Condition \eqref{RME:E:EfficientMarketHypothesisFormula-General} specifies that the expected next period value, $P_t + C_t$, of $P_{t-1}$ given current $\mathscr{I}_{t-1}$, has the same rate of growth $E[ 1 + R_t \, | \, \mathscr{I}_{t-1}]$ for all securities. It follows that no system which trades at time $t-1$ can earn an expected return better than $E[ R_t \, | \, \mathscr{I}_{t-1}]$ so that it's impossible to consistently ``beat the market'' by trading some subset of securities.

Because of the nature of scientific hypothesis testing, not to mention the near infinite dimensionality of $\mathscr{I}_{t-1}$, it is impossible to ``prove'' that \eqref{RME:E:EfficientMarketHypothesisFormula-General} is false. The best one can expect is a demonstration that some trading systems have produced excess returns with high confidence. Thus the most credible evidence against efficient markets comes from revealed research showing that excess returns were earned over long periods.

We list below a few cases in which excess returns were reported. The descriptions are brief --- readers desiring more information should consult \cite{moffitt2017V1} or \cite{citeulike:4510014}. 

\small
\begin{enumerate}[label=\bfseries {(\thesection.\arabic*)},ref={(\thesection.\arabic*)}]
\item{\bf Three Year Mean Reversion.} In 1985, DeBondt and Thaler published research using 46 years of market history, which showed that excess returns occurred in a zero cost portfolio that sold at the end of each month the top decile of stocks ranked by previous 3 year returns and simultaneously bought the bottom decile. There were various criticisms of the study, including (1) whether returns were sufficiently large to offset costs of trading, and (2) the absence of of a good explanation for why the largest returns occurred in January each year (\cite{RePEc:bla:jfinan:v:40:y:1985:i:3:p:793-805}).\label{Enum:ThreeYearMEanReversion}

\item{\bf Momentum.} \cite{citeulike:940976} Titman attempted to reproduce the DeBondt \& Thaler study and uncovered a remarkable pattern. The top decile of stocks ranked by 3-6 month returns had excess (positive) returns for an additional 6-12 months, and the bottom decile had inferior returns over the same period. Thus the pattern: 1 Year Momentum followed by 3 Year Reversal. The main objections to this study were (1) that data mining might have accounted for the effect, and (2) the effect would quickly be arbbed out after this publication. Surprisingly, though, when \cite{citeulike:4057155} conducted a follow up study eight years later, they found that if anything, the momentum ``anomaly'' had strengthened. The authors ascribed that failure of arbitrage to behavioral anomalies, but under the proposed model of this paper, they were only partly correct.\label{Enum:Momentum} 

\item{\bf \gls{closed-end.discount.puzzle}.\index{Closed-end Discount Puzzle}} The reconciliation of the closed-end pricing anomaly, referred to as the \gls{closed-end.discount.puzzle}, has as yet no satisfactory EMT explanation, but a quite good behavioral explanation. The puzzle is that closed-end funds are generally initiated during bull markets, within 120 days of start up trade at a discount to book value, and continue trading at discount until fund dissolution or open ending, when they converge to book value. For details, see \cite{citeulike:4510014} which devotes an entire chapter to the topic.\label{Enum:ClosedEndDiscountPuzzle}

\item{\bf Market Return vs. Prior Year New Issues.} A striking EMT anomaly published by \cite{RePEc:bla:jfinan:v:55:y:2000:i:5:p:2219-2257} hints at cyclical stock patterns. They discovered that heightened levels of new equity issues predict poor next-year equity returns! For further discussion, see \cite{citeulike:4510014} or \cite{moffitt2017V1}.\label{Enum:AnnualRetsVsPriorNewYearIssues}
\end{enumerate}
\normalsize
None of these ``anomalies'' have good EMT explanations, but all have straightforward \gls{SAFM}\index{SAFM}\index{Strategic Analysis of Financial Markets framework} explanations. 

For present purposes, a trading system is a family of functions $\{f_t\}$ that maps a (possibly multivariate) set of prices $\{P_t\}$ and \emph{information sets} $\{\mathscr{I}_t\}$\footnote{An information set $\mathscr{I}_t$ is a set of information that is known at time $t$. This theoretical constuct mirrors the reality that trading systems make bets on future outcomes conditional on those of realized variables.} to a stream of transactions $\{f_t( P_t \; | \; \mathscr{I}_{t-1} )\}$. A  basic requirement for inefficient markets is the existence of a \emph{\glslink{potentially.profitable.gambling.system}{Potentially Profitable Gambling System} (PPGS)}\index{Potentially Profitable Gambling System}\index{PPGS} which is, simply, a trading system that is not a \emph{\gls{martingale}\index{martingale}} with respect to some choice of information sets $\{\mathscr{I}_t\}$. For a PPGS, one also needs to know whether the expected return $E[ f_t( P_t \; | \; \mathscr{I}_{t-1} ) ]$ at each time is positive, negative or zero, so that one can potentially buy in the first case, sell in the second and not trade in the third. Of course, prices need adjustment for prevailing interest rates in cases where that is an issue, but for cases the author has encountered, such adjustment has seldom been necessary. Translating into trader-speak, a PPGS is a trading system that relies on historical data to initiate a series of trades for which the expected value is $\ge 0$ for all $t$.

A PPGS may be tradable or not, depending on the sizes of its expected returns, on its fees and slippage, on its riskiness, and so on.\footnote{The \gls{SAFM}\index{SAFM}\index{Strategic Analysis of Financial Markets framework} does not use variance, except as a proxy, to measure risk.} The importance of the EMT anomalies \ref{Enum:ThreeYearMEanReversion}-\ref{Enum:AnnualRetsVsPriorNewYearIssues} is that each supposedly yielded \emph{tradable} returns that exceeded the market's. We remark in passing that publications which tout ``winning systems'' or ``tradable anomalies'' seldom show estimates of actual returns and risks. Systems developers, take note!

\section{The Fundamental Laws of Gambling (FLOG)}

Before presenting the two fundamental laws, it's important to dispel an urban myth --- that all gamblers lead roller-coaster lives unacceptable to prudent people. That may be true of compulsive gamblers, but it is definitely false for professional gamblers. Good gamblers get rich; bad gamblers go broke.

There are only two principles in our \emph{\gls{Fundamental.Laws.of.Gambling}}\index{Fundamental Laws of Gambling}\index{FLOG}. In their imperative forms, they are
\begin{enumerate}[label=\bfseries {(\thesection.\arabic*)},ref={(\thesection.\arabic*)}]
\item{\textbf{FUNDAMENTAL LAW:}} Never bet unless you think you have an edge.\label{GG:Enum:Rule:FundamentalLawOfGambling}
      \index{Fundamental Laws of Gambling!The Fundamental Law|textbf} \index{FLOG!The Fundamental Law|textbf}
\item{\textbf{WAGERING PRINCIPLE:}} Pick a fractional amount of your capital consistent with a
      \gls{drawdown}\index{drawdown} or loss condition and a minimum return condition, and use estimates of
      your edge to chose a betting fraction. \label{GG:Enum:Rule:WageringPrinciple}
      \index{Wagering Principle|textbf}\index{Fundamental Laws of Gambling!The Wagering Principle|textbf} \index{FLOG!The Wagering Principle|textbf}
\end{enumerate}
The justification of these principles is based on an analysis of fixed fraction gambling systems. To illustrate our approach, we present a simple fractional betting system developed in \cite{Moffitt:Part1:SimpleKelly:2914620}. Assume that a biased coin with probability of heads $p > 0.5$ and tails $q = 1-p$ is flipped repeatedly, and assume stochastic independence of flips.  At each flip, a fixed fraction $f$ of current wealth is wagered at odds $d > 0$\footnote{That is, you receive $d$ when you win and pay $1$ when you lose.}, so that wealth after the flip is $1 + df$ or $1 - f$ times initial wealth. With this setup, one can show

\small
\begin{enumerate}[label={\bfseries (\thesection.\alph*)},ref={(\thesection.\alph*)}]
\item For a fixed number of bets and $p > 1/(1 + d)$, the average return is greatest for a strategy
      consisting of betting everything each time. However, the chance of winning
      becomes vanishingly small as the number of bets increases, i.e. ruin is 
      asymptotically certain. Therefore, this strategy is unacceptable. 
      \label{GG:Enum:SystemAllIn}

\item The strategy that wins money at the maximum rate consists of betting a
      fraction \(f = p - q/d\) of current wealth on heads at each turn when 
      $p - q/d > 0$ and  $f = 0$ when $p - q/d \le 0$ (see Appendix 
      \ref{A:CalculationOfTheKellyFraction} for the Kelly calculation). In its 
      practical usage, betting the Kelly fraction $f = p - q/d > 0$ can lead 
      to unacceptably large drawdowns.\label{GG:Enum:SystemKelly}

\item There is a critical fraction $f_c$ such that the betting of any greater
      fraction on heads assures ultimate ruin, and the betting of any lesser 
      fraction assures ultimate unbounded growth of wealth.
      \label{GG:Enum:CriticalFraction}

\item When faced with a betting situation in which the probability of winning
      at each turn varies with random probability $P$, then the optimal fraction
      is $f = p - q/d$ with $p = E[\, P \,]$. \label{GG:Enum:StochasticP}

\item When $p$ is unknown and $p - q/d > 0$, using $f_n = \bar{X}_n - (1 - \bar{X}_n)/d$, 
      yields the same asymptotic growth rate as $f = p - q/d$. \label{GG:Enum:UnknownP}

\item Given known probabilities $p_i > 0.5$ of heads at each turn and odds $d_i$, $i=1,2,\ldots,n$ for a 
      sequence of bets, the strategy that wins money at the maximum rate consists 
      of wagering the fraction $f_i = p_i - q_i/d_i$ at the $i^{th}$ bet when 
      $f_i > 0$ and nothing otherwise. \label{GG:Enum:MixedP}

\item Volatility and drawdowns can be reduced with some loss of optimal growth 
      through the use of \emph{Fractional Kelly} betting. Fractional Kelly is 
      straightforward: bet a fraction $\alpha f$ at each turn, where 
      $0 < \alpha < 1$ and $f$ is the Kelly fraction. \label{GG:Enum:FractionalKelly}
\end{enumerate}
\normalsize
For readers who want more detail on Kelly strategies, especially fractional Kelly, the best reference at this writing is unquestionably \cite{RePEc:wsi:wsbook:7598}. 

Analogs of points \ref{GG:Enum:SystemAllIn}-\ref{GG:Enum:FractionalKelly} exist when bets involve security prices, not coin flips. Despite their relevance, we do not pursue such generalizations here and again refer interested readers to \cite{RePEc:wsi:wsbook:7598}. 

The important takeaways from Kelly Criterion betting are that (1) using it produces optimal growth, but results in huge drawdowns, (2) betting the critical fraction or greater leads to asymptotic ruin, and (3) approximate optimal growth given drawdown constraints can be achieved by fractional Kelly betting.

\section{Grationality}\index{grationality}

\glslink{grational}{Grationality} (=``gambling rationality'')\index{grationality} for simple bets is easily understood through an example. Suppose that a trader with capital $\$1,000$ at time $0$ faces the prospect of a series of independent bets each paying $+1$ with probability $p > 0.5$ and $-1$ with probability $q = 1-p$. The best long run growth of capital occurs if a fraction $f^{*} = p - q$ is bet at each opportunity. But as stated earlier, optimal growth using $f^{*}$ comes at a considerable cost: large drawdowns.

Large drawdowns are unacceptable for investors and most traders, leading to the search for alternative betting systems. The reasons are that (1) even a modest probability of markets having evolved to make a system unprofitable creates the incentive to liquidate during drawdowns, and (2) an individual whose funds are managed has \emph{agency} issues that encourage liquidation of fund shares. Is this one reason that \gls{Buffett.Warren}\index{Buffett, Warren} invests through a closed-end fund, so that these liquidation incentives do not lead to share redemptions?

Luckily, an obvious means of controlling drawdowns is available: find a fraction $f$ that maximizes expected growth, while having a probability no larger than $u$ of producing a \gls{drawdown}\index{drawdown} (loss) exceeding $d$ ($l$)? To pose the problem analytically, let $X = \{ X_1, X_2, \ldots, X_n, \ldots \}$ be \glslink{i.i.d.}{independent and identically distributed} random variables  
\[
    X_i = \begin{cases}
             1 & \text{ with probability } p \\
             0 & \text{ with probability } 1-p,
          \end{cases}
\]
and given $X$, let the function $G(f;X,n)$ in equation \eqref{GG:E:GrowthRaten} be the growth rate of capital over the first $n$ bets, assuming starting capital $W_0$:
\begin{align}
  G(f;X,n) =& \frac{1}{n} \, log \left( \frac{W(n; f, X)}{W_0} \right) \nonumber \\
           =& \bar{X}_n log (1 + f) + (1 - \bar{X}_n) log (1 - f), \label{GG:E:GrowthRaten}
\end{align}
where $\bar{X}_n = \frac{1}{n} \sum_{i=1}^{i=n} X_i$ and 
\begin{align}
  W(n; f, X) &= W_0 \; \prod_{i=1}^{i=n} (1 + f)^{X_i} (1-f)^{1 - X_i} \label{GG:E:WealthProcess} \\
             &= W_0 \; (1 + f)^{\sum_{i=1}^{i=n} X_i} (1 - f)^{n - \sum_{i=1}^{i=n} X_i}. \nonumber
\end{align}
Next, consider controlling risk using either of two \emph{loss functions} $\mathscr{L}(W,n)$: (1) the \emph{\gls{worst.loss.function}}\index{loss function!worst loss} $\mathcal{L}(W,n)$ or (2) the \emph{\gls{drawdown.loss.function}}\index{loss function!drawdown} $\mathcal{D}(W,n)$. Both of these are expressed in terms of a generic wealth process $W_n$, $n \ge 0$. These loss functions are defined as follows:
\begin{enumerate}
\item{The ``Worst Loss'' Function.}
\begin{equation}
  \mathcal{L}(W,n) = log(W_0) - \, \underset{0 \le s \le n}{min} log(W_s), \label{GG:E:WorstLossFunctionPM}
\end{equation}
which is the log of the greatest relative loss up to time $n$, and $W_0$ is the initial capital. Note that \eqref{GG:E:WorstLossFunctionPM} is always nonnegative provided that $W_n > 0$ for all $n$.

\item{The ``(Peak-to-Trough) Drawdown'' Loss Function.} 
\begin{equation}
   \mathcal{D}(W,n) = \underset{0 \le s \le n}{max} \left\{ \underset{0 \le u \le s}{max} \, log(W_u) \, - \, log(W_s) \right\}, \label{GG:E:DrawdownLossFunctionPM}
\end{equation}
which is the log of the greatest relative loss from a previous high up to time $n$. Note that \eqref{GG:E:DrawdownLossFunctionPM} is always nonnegative provided that $W_n > 0$ for all $n$.
\end{enumerate}

\vspace{2mm}
\noindent With this notation, the optimal betting fraction $f$ should be a solution to the optimization problem 
\begin{align}
\text{ Maximize}   \quad & E[ \, G(f;X,n) \, ] \quad \text{with respect to } f \nonumber \\
\text{ subject to} \quad & 0 \le f \le 1 \text{  and} \label{GG:E:DiscreteGrationality} \\
                         & P[\, \mathscr{L}(W(f; X,n),n) \ge d \, ] \le u, \nonumber
\end{align}
with $\mathscr{L}(W,n) = \mathcal{L}(W,n)$ or $\mathcal{D}(W,n)$ or the asymptotic version as of \eqref{GG:E:DiscreteGrationality} as $n \rightarrow \infty$.

For prices following a geometric Brownian motion, this problem for  $\mathscr{L}(W,n) = \mathcal{D}(W,n)$ and $u = 0$ was solved in \cite{RePEc:bla:mathfi:v:3:y:1993:i:3:p:241-276}, but for discrete price series, \cite{RePEc:eee:stapro:v:74:y:2005:i:3:p:245-252} showed their solution to be suboptimal. Incidentally, the Grossman-Zhou solution is essentially an algorithm for \emph{portfolio insurance}.

Four glaring differences between this simple set up and the real world are that (a) real-world wagers do not have simple $\pm 1$ gains, (b) most wagers (= investments) occur in \glspl{continuous.auction.market}\index{continuous auction market} that allow trading at any time the venue is open, (c) many systems bet a varying number of times, as is the case if a position is held for a random number of days, and (d) wagers are usually part of a portfolio the \glspl{position}\index{position} of which can be entered and exited asynchronously.

\section{Miller's Critique of Efficient Market Theory}

This section has important background material needed in the development of the \gls{SAFM}\index{SAFM}. It presents a critique of EMT published in \cite{Miller77} that presented an alternative, constructive approach to the evolution of market prices under specific conditions. Although Miller's model has largely been ignored, it has lost none of its relevance. The reason is that its constructive approach describes \emph{why} and \emph{how} prices are formed, and demonstrates straightforward explanations for several well-known EMT ``anomalies.'' It is an early model of the type to which the SAFM aspires. 

Miller based his analysis on the ``\glslink{winner's.curse}{Winner's Curse}''\index{winner's curse} in auctions. The idea of the winner's curse is simple; in the presence of heterogeneous estimates of value, the most optimistic bidder wins the prize, and therefore ``overpays.'' The use of ``overpay'' is not necessarily a statement about ``fair value,'' since overpaying in this instance means that a subsequent auction conducted immediately will clear at a lower price. Astute traders intuitively know about the winner's curse, and in fact, have a rule to avoid it --- when investors clamor after a security, the best long term value, other things being equal, is obtained by selling it to them.

\begin{figure}[h!]
  \centering
  \includegraphics[scale=0.6]{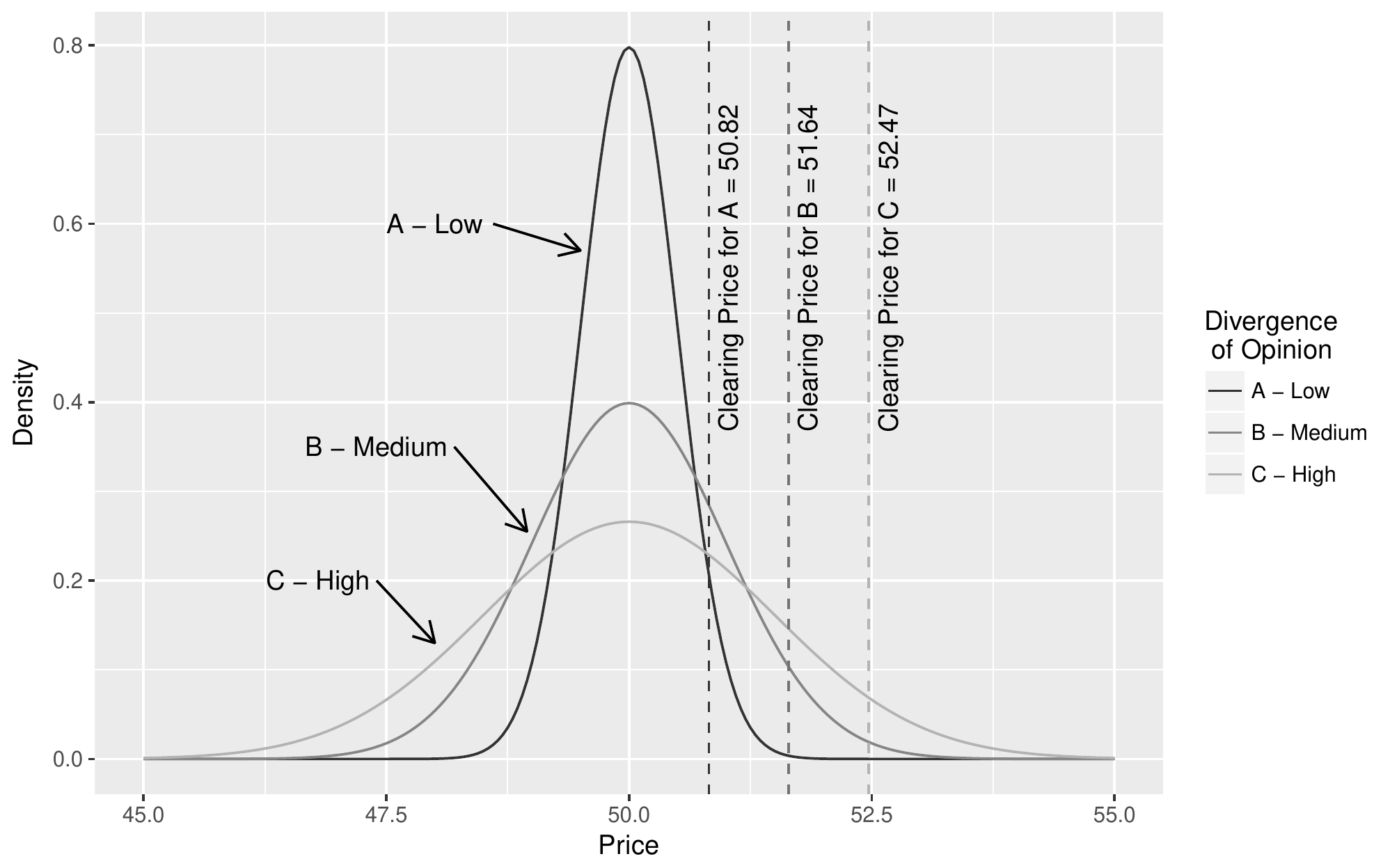}
  \caption{\small{A plot that explains the Winner's Curse effect in asset pricing. Assuming an asset that has small turnover and a large group of potential buyers and for which there is no short selling, Miller argues that the clearing price will be set by the most optimistic buyers, that is, those who have the highest estimates of its value. The figure shows three distributions of divergence of opinion, the curve with the highest peak showing low divergence, the middle curve showing medium divergence and the flattest curve showing the greatest divergence. The three vertical lines, reading from the left, mark the 95th percentiles of the low, medium and high divergence of opinion  distributions.}}
  \label{G:MillerDivergenceOfOpinionDistributions}
\end{figure}

Consider Figure \ref{G:MillerDivergenceOfOpinionDistributions}, which depicts three distributions of price estimates for the security. The ``average'' opinion in each case is \$50, but the dispersions are different. 
Under conditions that Miller specifies, that (1) the number of sellers  of the stock, $N$, is fixed, (2) that the number of potential buyers $M \gg N$ is large compared to demand so that $N/M \ll 1$, and (3) that there is no short selling, Miller argues that only the most optimistic of potential buyers will determine the clearing price. Thus, the clearing price will exceed average price estimate among potential buyers.

Miller opines that this \glslink{winner's.curse}{winner's curse effect}\index{winner's curse} can explain a number of EMT anomalies:
\begin{enumerate}[label=\bfseries {(\thesection.\arabic*)},ref={(\thesection.\arabic*)}]
\item{\bf Risk and return will be negatively correlated.}

Provided that risk in a stock is proxied by greater divergence of opinion, prices of risky stocks should be relatively greater than average ones and therefore have sub par returns. This is the opposite of the EMT prediction that purchasers of risky securities will be compensated for those investments with appropriate risk premiums. This EMT ``anomaly'' is the basis of the recent popularity of ``low volatility'' strategies. See \cite{10.2307/2330270} and \cite{1509106219910101}.

\item{\bf IPO's will have short term price appreciation and poor long term returns.}

IPO's are an insider's game. Underwriters typically price the issue by offering it to institutional clients (``insiders'') before the IPO. That price is typically too low in order to (1) allow insiders a chance to profit, (2) protect the underwriter against losses if the entire issue can't be placed, and (3) encourage speculators to overpay when listed trading begins.\footnote{Listed trading begins after all insiders have already bought; some of them sell in a price fixing auction that is held prior to listed trading in order to set the opening price, or in later trading on the listed market.} This explains the reason that IPO's typically rally after the offering. Miller argues that initial optimism leads to high divergence of opinion, but the eventual removal of uncertainty about the stock's prospects leads to a narrowing of that divergence, hence poor future performance relative to other stocks. Thus the old joke about an IPO ... it stands for ``It's Probably Overpriced.'' See \cite{RePEc:cup:jfinqa:v:35:y:2000:i:04:p:499-528_00}.

\item{\bf High turnover will lead to low returns.}

It has been shown that high turnover (high volume) is associated with low returns in \cite{FIRE:FIRE31}. Provided that high turnover is associated with divergence of opinion, then this result is consistent with Miller's model.

\item{\bf Short selling will attenuate the winner's curse effect in riskier securities.}

Short selling effectively increases the number of shares. Since, short sellers are often pessimistic on a stock's prospects, the supply side of the supply and demand curve is shifted to the right, lowering the clearing price. In view of the widespread shorting of stocks as part of index arbitrage, however, it isn't clear if this anomaly still holds.
\end{enumerate}

\section{The Pursuit of Profits Paradigm\\(POPP)}

There is only one goal common to all trading strategies --- they seek to profit. In the previous section, we reviewed the \cite{Miller77} model, which showed, inter alia, that heterogeneous estimates of prices combined with limited supply of stock led to inefficient clearing. This section concerns a similar model, in which uncoordinated attempts to profit lead to life cycles in markets and strategies, a paradigm called the \emph{\glslink{Pursuit.of.Profits.Paradigm}{Pursuit of Profits Paradigm} (POPP)}\index{Pursuit of Profits Paradigm (POPP)}. The POPP is based on the unremarkable premise that investors are mainly motivated by a desire for wealth, and that the impatient among them ``chase after riches.'' Perhaps surprisingly (except to \gls{Buffett.Warren}\index{Buffett, Warren}), the predictability of the chase leads chasers to losses!\footnote{... which is reminiscent of an old trader's saying: ``The pigs get fat and the hogs get slaughtered.''}

\subsection{The Strategic Point of View}\label{SS:TheStrategicPointOfView}

\vspace{3mm}
\noindent Description of the POPP requires two new terms,
\begin{description}
  \item{\bf \gls{strategic.plan}.}\index{strategic plan}\index{SAMM!strategic plan} A general plan of action for trading that lacks specifics.
  \item{\bf \gls{strategy}.}\index{strategy} A specific plan that can be used to generate trades. One that is specific enough to be programmed is called a \gls{trading.algorithm}\index{trading algorithm}.
\end{description}

In trading jargon, a \gls{strategy}\index{strategy} is a set of rules or an \glslink{trading.algorithm}{algorithm}\index{trading algorithm} that a trader, investor or computer uses to buy and sell. Strategies are often semi-formal schemas for action. For example, a strategy such as ``In a bull market, buy popular, high-beta stocks; in a bear market, selectively short formerly popular stocks but hold them only for a short period,'' requires a determination of the type of market, bear or bull, and of criteria to determine ``popularity'' of stocks and holding periods for trades. While this form can be useful for \glslink{discretionary.trade}{discretionary traders}\index{discretionary trading}, it is not specific enough for \glslink{algorithmic.trade}{algorithmic traders}\index{trading system!algorithmic}\index{algorithmic trading}. We call a strategy that lacks the specificity necessary for algorithmic trading, but which by a selection of specific \emph{parameters} can be traded algorithmically, a \emph{\gls{strategic.plan}}.\index{strategic plan|textbf}\index{SAMM!strategic plan|textbf} 

To convert a \gls{strategic.plan}\index{strategic plan}\index{SAMM!strategic plan} into a \gls{trading.algorithm}\index{trading algorithm}, the best way is to perform analysis on market data. In the strategic plan of the previous paragraph, for example,  three sets of criteria are needed, (1) ones that identify the market type as bull, bear or neither, (2) ones that identify a stock as popular, and (3) ones that determine for a given market type and stock, when to buy and when to sell or short. If as the result of statistical analysis, we specify (1), (2) and (3) by 
\begin{enumerate}[label=\bfseries {(\arabic*)},ref={\thesection (\arabic*)}]
\item Defining a bull market as one having positive 6 month returns, a bear market as one having negative two month returns on the \acrlong{aSnP} and ``neither bull nor bear'' if these criteria point in opposite directions, 
\item Selecting \acrshort{aNASDAQ} stocks with 6-month beta and weekly \gls{turnover.ratio} in the upper decile of their respective distributions, and 
\item In bull markets, buying $\$10,000,000$ in \acrshort{aNASDAQ} popular stocks and selling when the bull ends, and in bear markets, shorting $\$7,000,000$ in \acrshort{aNASDAQ} formerly popular stocks for at least a week and remaining short as long as one month returns are negative and buying back when this criterion ends,
\end{enumerate}
then the \gls{strategic.plan}\index{strategic plan}\index{SAMM!strategic plan} has given rise to a \gls{trading.algorithm}\index{trading algorithm}. Of course, this plan is not entirely specific, i.e. the allocation to stocks is not detailed, beta can be specified in several ways using daily or weekly prices, the time and place of buying and selling is not specified, and so on. Using the strict requirement that a \gls{trading.algorithm} be produced, only a code snippet implementing the algorithm would convert the plan into an algorithm. 

\subsection{State Variables in the POPP}\label{SS:StateVariablesInThePOPP}

The POPP describes the evolution of \glspl{strategic.plan}\index{strategic plan}\index{SAMM!strategic plan} through the changes in \emph{state variables}. Here is a list, along with their abbreviations:
\begin{description}
\item[RET:] Capitalization-weighted returns.
\item[VOL:] Capitalization-weighted volatility (= standard deviation) of returns.
\item[SR:] \gls{Sharpe.ratio} = \{ RET - (Risk-free Rate) \} / VOL.
\item[SP\$:] Aggregate capitalization. 
\item[POP:] Number of implementors. 
\item[LEV:] Capitalization-weighted  \gls{leverage}\index{leverage}. 
\item[SDIV:] \glslink{strategy.diversity}{Strategy diversity}\index{strategy!diversity} as a measure of heterogeneity in the population of strategies.  
\item[SLINK:] \glslink{strategy.linkage}{Strategy linkage}\index{linkage!strategy} with a range from weak (= few links) to strong (= many links), considered as the degree to which the strategies in the plan have ``cascading'' links to similar and dissimilar strategies.
\item[SROB:] Strategy \gls{robustness}\index{robustness} (robust $\rightarrow$ \gls{fragile}\index{fragility}) expressing the probability of a very large loss in a short time.
\end{description}

\begin{figure}[!ht]
  \begin{center}
    \begin{tikzpicture}

      \draw [thick,black] (150pt,10pt)  -- (150pt,150pt) -- (250pt,150pt) -- (250pt,10pt)  -- (150pt,10pt);
      \draw [thick,black] (25pt,110pt)  -- (25pt,250pt)  -- (125pt,250pt) -- (125pt,110pt) -- (25pt,110pt);
      \draw [thick,black] (150pt,210pt) -- (150pt,350pt) -- (250pt,350pt) -- (250pt,210pt) -- (150pt,210pt);
      \draw [thick,black] (275pt,110pt) -- (275pt,250pt) -- (375pt,250pt) -- (375pt,110pt) -- (275pt,110pt);

      \draw [thick,black] (200pt,140pt) node {\textbf{A: Eureka!}};
      \draw [thick,black] (150pt,130pt) -- (250pt,130pt);
      \draw [thick,black] (200pt,118pt) node {\textbf{RET:   ++}};
      \draw [thick,black] (200pt,106pt) node {\textbf{VOL:   +}}; 
      \draw [thick,black] (200pt,94pt)  node {\textbf{SR:    +}};
      \draw [thick,black] (200pt,82pt)  node {\textbf{SP\$:  -}};
      \draw [thick,black] (200pt,70pt)  node {\textbf{POP:   -}};
      \draw [thick,black] (200pt,58pt)  node {\textbf{LEV:   -}};
      \draw [thick,black] (200pt,46pt)  node {\textbf{SDIV:  +}};
      \draw [thick,black] (200pt,34pt)  node {\textbf{SLINK: ?}};
      \draw [thick,black] (200pt,22pt)  node {\textbf{SROB:  +}};

      \draw [thick,black] (80pt,240pt)  node {\textbf{B: Early Copycat}};
      \draw [thick,black] (25pt,230pt)  -- (125pt,230pt);
      \draw [thick,black] (75pt,218pt)  node {\textbf{RET:   +}};
      \draw [thick,black] (75pt,206pt)  node {\textbf{VOL:   =}}; 
      \draw [thick,black] (75pt,194pt)  node {\textbf{SR:    +}};
      \draw [thick,black] (75pt,182pt)  node {\textbf{SP\$:  =}};
      \draw [thick,black] (75pt,170pt)  node {\textbf{POP:   =}};
      \draw [thick,black] (75pt,158pt)  node {\textbf{LEV:   =}};
      \draw [thick,black] (75pt,146pt)  node {\textbf{SDIV:  +}};
      \draw [thick,black] (75pt,134pt)  node {\textbf{SLINK: ?}};
      \draw [thick,black] (75pt,122pt)  node {\textbf{SROB:  +}};

      \draw [thick,black] (200pt,340pt) node {\textbf{C: Late Copycat}};
      \draw [thick,black] (150pt,330pt) -- (250pt,330pt);
      \draw [thick,black] (200pt,318pt) node {\textbf{RET:   -}};
      \draw [thick,black] (200pt,306pt) node {\textbf{VOL:   -}};
      \draw [thick,black] (200pt,294pt) node {\textbf{SR:    -}};
      \draw [thick,black] (200pt,282pt) node {\textbf{SP\$:  ++}};
      \draw [thick,black] (200pt,270pt) node {\textbf{POP:   ++}};
      \draw [thick,black] (200pt,258pt) node {\textbf{LEV:   ++}};
      \draw [thick,black] (200pt,246pt) node {\textbf{SDIV:  -}};
      \draw [thick,black] (200pt,234pt) node {\textbf{SLINK: ?}};
      \draw [thick,black] (200pt,222pt) node {\textbf{SROB:  -}};

      \draw [thick,black] (325pt,240pt) node {\textbf{D: Crash}};
      \draw [thick,black] (275pt,230pt) -- (375pt,230pt);
      \draw [thick,black] (325pt,218pt) node {\textbf{RET:   NA}};
      \draw [thick,black] (325pt,206pt) node {\textbf{VOL:   NA}};
      \draw [thick,black] (325pt,194pt) node {\textbf{SR:    NA}};
      \draw [thick,black] (325pt,182pt) node {\textbf{SP\$:  NA}};
      \draw [thick,black] (325pt,170pt) node {\textbf{POP:   NA}};
      \draw [thick,black] (325pt,158pt) node {\textbf{LEV:   NA}};
      \draw [thick,black] (325pt,146pt) node {\textbf{SDIV:  NA}};
      \draw [thick,black] (325pt,134pt) node {\textbf{SLINK: NA}};
      \draw [thick,black] (325pt,122pt) node {\textbf{SROB:  NA}};

      \draw[->,thick] (150pt,50pt)  to [bend left] (75pt,110pt);  
      \draw[->,thick, dotted] (75pt,250pt) to [bend left] (150pt,310pt); 
      \node[rectangle] at (125pt,320pt) {\large\textbf{\textit{Bubble}}};
      \draw[->,thick,dotted] (75pt,250pt) -- (75pt,295pt); 
      \node[rectangle] at (75pt,305pt) {\large\textbf{\textit{Fizzle}}};
      \draw[->] (250pt,310pt) to [bend left] (325pt,250pt); 
      \draw[->,thick,dotted] (250pt,310pt) -- (295pt,310pt); 
      \node[rectangle] at (320pt,310pt) {\large\textbf{\textit{Fizzle}}};
      \draw[->,thick,dashed] (325pt,110pt) to [bend left] (250pt,50pt); 

      \draw [thick,black] (150pt,210pt) -- (150pt,350pt) -- (250pt,350pt) -- (250pt,210pt) -- (150pt,210pt);

    \end{tikzpicture}
  \end{center}
   \caption{\small{A Strategic Lifecycle Diagram for the Pursuit of Profit Paradigm. See the text for an explanation.}}
   \label{TSAOMM:StrategicEvolution:LifeCycle.pdf}
\end{figure}
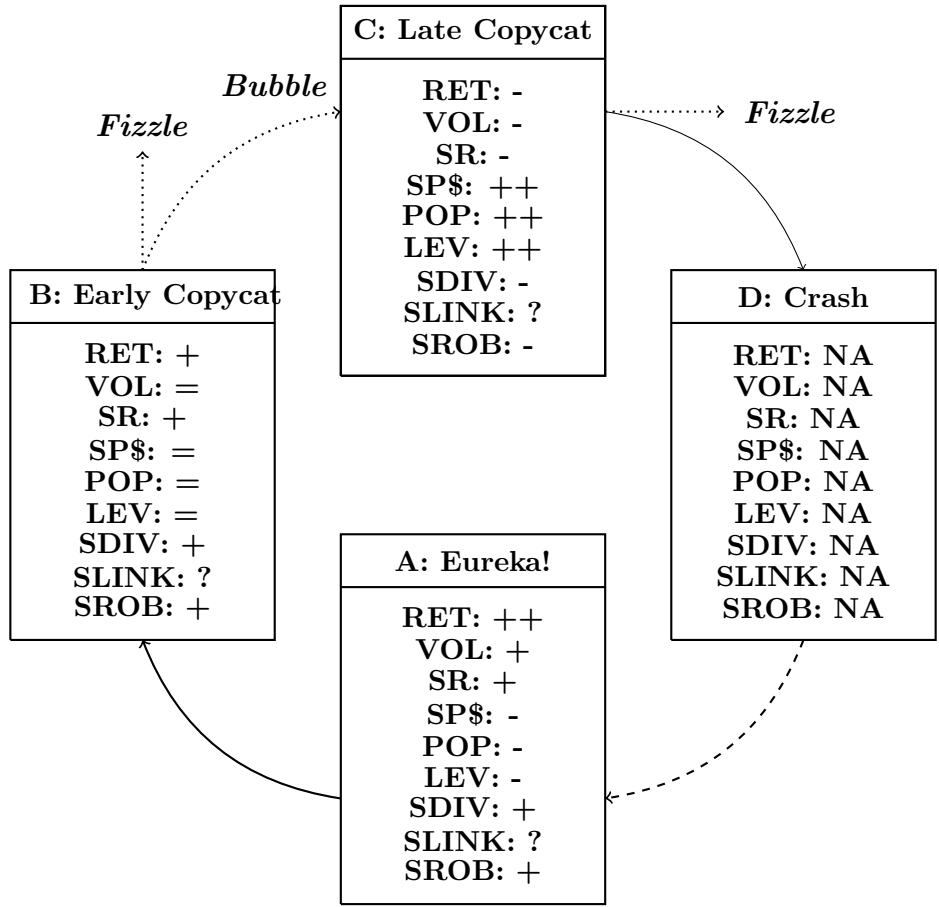

\subsection{The \glslink{Pursuit.of.Profits.Paradigm}{POPP} Lifecycle}\index{Pursuit of Profits Paradigm (POPP)}\label{SS:ThePOPPLifecycle}

Figure \ref{TSAOMM:StrategicEvolution:LifeCycle.pdf} shows the cyclical evolution of the \emph{\glslink{Pursuit.of.Profits.Paradigm}{POPP}}\index{Pursuit of Profits Paradigm (POPP)}, starting with box ``A'' at the bottom and proceeding clockwise. The four phases of the POPP are (1) Phase (A) - Eureka!, (2) Phase (B) - Early Copycat, (3) Phase (C) - Late Copycat, and (4) Phase (D) - Crash. The $+$'s ($++$) and $-$'s ($--$) indicate positive (extreme) and negative (extreme) expected values. Thus the expected Sharpe Ratio (SR) is positive in the \emph{Eureka!} and \emph{Early Copycat} phases, negative in the \emph{Late Copycat} phase and not applicable in the \emph{Crash} phase.

In Phase (A), a new trading idea is ``born'' when one or more innovators devise a \gls{strategic.plan}\index{strategic plan}\index{SAMM!strategic plan} and/or develop \glspl{strategy}\index{strategy} that implement it. Returns (RET) and Sharpe Ratios (SR) are high, but most other state variables besides SROB are low (which is good). The strategic plan is quite profitable in this phase, but perhaps surprisingly, may not be as profitable as Phases (B) or (C). Phase (B), the Early Copycat begins when (a) copycats, attracted by high Sharpe Ratios or the rumor mill, begin trading the plan, (b) serendipitous discoverers enter the arena, and/or (c) existing players expand their allocations. In any case, the effect is an expansion of funds (\emph{SP\$}). The number of users (\emph{POP}) at the start is small, but grows toward the end as discussed further below. Although \gls{robustness}\index{robustness} (\emph{SROB}) remains high, it becomes smaller as the phase proceeds. The expected (in contrast to realized) Sharpe Ratio (\emph{SR}) reaches an apex due to improvements in \gls{strategy}\index{strategy} implementations outweighing increased \emph{SP\$}, but thereafter declines for the duration of the evolution. Phase (B) ends when the \gls{strategic.plan.capacity}\index{capacity!strategic plan}\index{strategic plan capacity} is reached. The arrow from Phase (B) to (C) indicates that allocations begin to exceed the \gls{strategic.plan.capacity}\index{strategic plan capacity}\index{capacity!strategic plan} because there is \emph{overinvestment} or \emph{\gls{crowding}}\index{crowding}. The difference between this and the previous phase lies in the dominance of new copycats that destabilize the system and/or lead to an excessive level of funding. In phase (C) all that is required for a correction or crash is an event that triggers a reversal in the generators of profit, revealing previously occult \glslink{fragile}{fragility}\index{fragility} in the system. The final phase is a market correction or crash, brought on by financial \glslink{fragile}{fragility}\index{fragility} that results when the \gls{strategic.plan.capacity}\index{strategic plan capacity}\index{capacity!strategic plan} is exceeded. The reason that this phase is so much shorter than the preceding ones is that the system has become increasingly susceptible to contagion as a result of high \gls{leverage}\index{leverage} (\emph{LEV}), decreased diversity (\emph{SDIV}), high linkage with other strategies that implement the strategic plan (\emph{SLINK}), and low \gls{robustness}\index{robustness} (\emph{SROB}). Each of these developments makes the likelihood of contagion and the extent of its severity  greater.

A few observations about the transition from Phase (C) to the crash are insightful for trading. First, it is common that there is high volatility in returns near the end of Phase (C). In this model the main reason this happens is that ``enough'' traders try to liquidate or reduce their positions, that the markets can't clear easily. Of course, there can be many reasons for reduction or liquidation --- margin calls, losses in other strategies that mandate overall reduction in the portfolio, excess  \gls{leverage}\index{leverage}, poor performance of the strategies, believing that a correction is coming, and so on. Second, the crash always entails a reduction in liquidity.  While small traders can escape, larger traders cannot, and this suggests that traders ought to measure their risk by calculating the number of trades or days required to fully exit under normal circumstances, with allowances for reduced liquidity during crises. Reduced liquidity has the further effect of making larger moves in short periods more likely. Third, assuming that other markets are performing normally prior to the crash, contagion is likely to occur only if strategies are linked to other strategies in a network. Such links need not be based on fundamentals, e.g. losses in one strategy can trigger reductions in others. Fourth, the longer Phase (C) lasts, the greater the likelihood of a severe correction or crash. The reasons: (1) positions are much larger, making if difficult to unwind them in a short period of time, resulting often in a ``stampede to the exits,'' and (2) links to external markets grow in the chase after riches, increasing the likelihood of contagion in mature \glslink{Pursuit.of.Profits.Paradigm}{POPP}\index{Pursuit of Profits Paradigm (POPP)}'s.

We restate the premise of the \glslink{Pursuit.of.Profits.Paradigm}{POPP}\index{Pursuit of Profits Paradigm (POPP)} --- that strategic evolution is impelled by ``chases after riches.'' If one accepts this paradigm, it says something profound about markets, for won't there always be investors who ``chase after riches?'' Here are a couple of implications. First, in markets characterized by short-termism, there will be lots of bubbles, corrections and crashes because an accelerated chase after riches quickens the pace of \glslink{Pursuit.of.Profits.Paradigm}{POPP}\index{Pursuit of Profits Paradigm (POPP)} cycles. This suggests that a longer-term winning strategy consists of buying ``value.'' Second, the wider market's \glslink{fragile}{fragility}\index{fragility} will be a function of ``strategy linkage,'' as developed in \cite{moffitt2017V1}. 

\vspace{2mm}
\small
\begin{exmp}[POPP in Action: Pairs Trading]\label{TSOAMM:Exmp:PairsTrading}

Pairs trading is really simple --- buy undervalued securities and sell overvalued securities in equal dollar amounts, then exit both sides simultaneously when prices adjust sufficiently. The pairs method we use is presented in \cite{moffitt2017V2}.

The dark line in Figure \ref{TSAOMP:G:PairsCumPnL-19720825-20090413} cumulates the average daily return over pairs and therefore represents the return that a single representative pair would have. There is no adjustment for commissions or slippage, so the curve is NOT TRADABLE. But for purposes of understanding the dynamics of the trade, it's simpler to exclude the distorting effects of commissions and slippage. The gray line is the \acrshort{aDJIA} index. Our intention is to identify market events associated the curve's gains and losses, and to use them to describe the strategic evolution of the \glslink{pairs.trading.strategy}{pairs trade}\index{pairs trading strategy} as an instance of the \emph{\glslink{Pursuit.of.Profits.Paradigm}{Pursuit of Profits Paradigm (POPP)}}\index{Pursuit of Profits Paradigm (POPP)}. To match the dates to market events, refer to Table \ref{TSAOMM:Tbl:DatesInPairsTrade}.
\begin{figure}[ht!]
  \begin{center} .
    \includegraphics[scale=0.52]{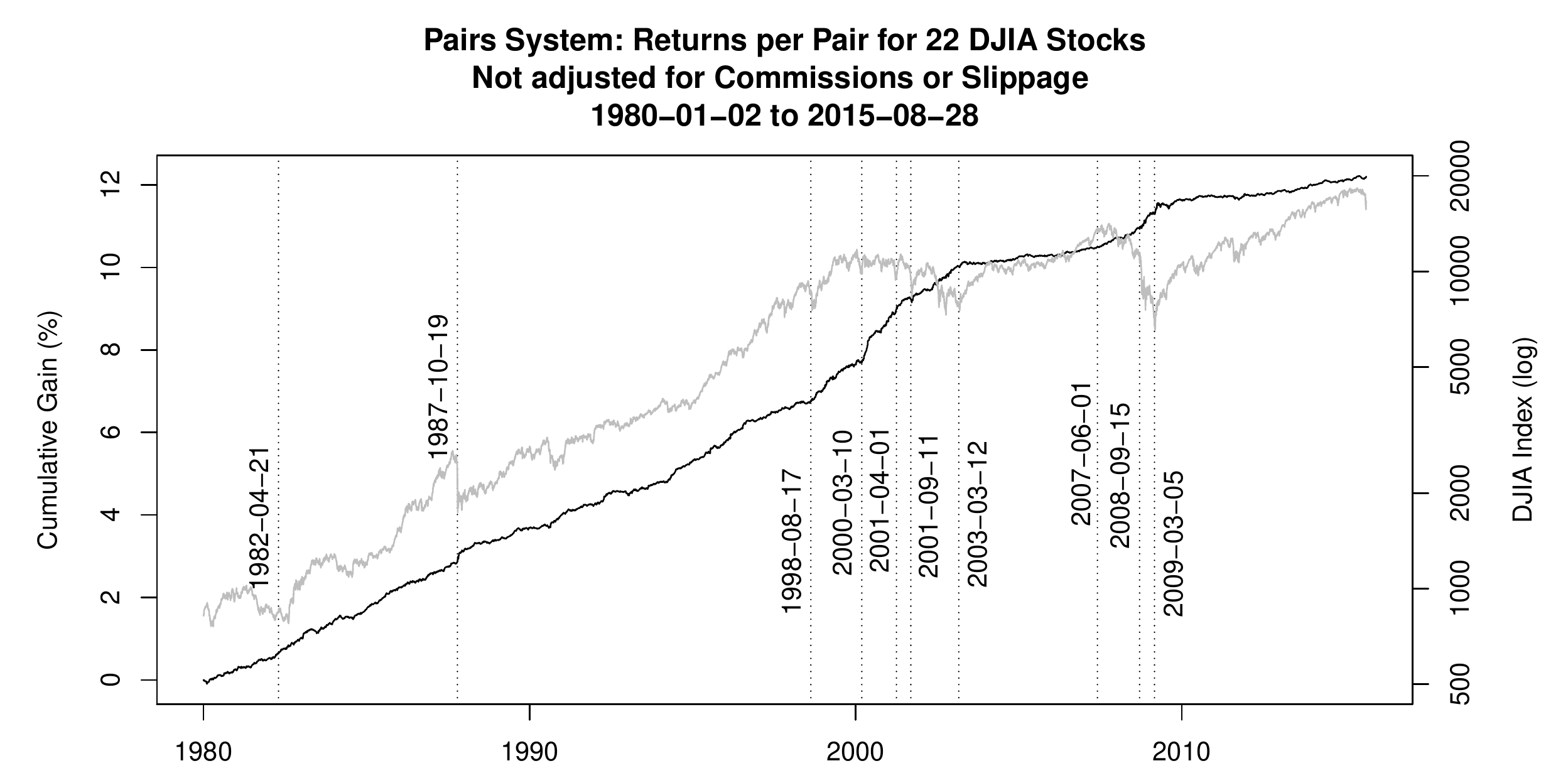}
    \caption{\small{P\&L Plot for a Pairs Trading System: 1969-12-31 to 2015-08-28. See text for details.}}
    \label{TSAOMP:G:PairsCumPnL-19720825-20090413}
  \end{center}
\end{figure}

\begin{table}[!ht]
  \begin{center}
    \caption{\small{Significant Dates for Pairs Trading.}}
    \label{TSAOMM:Tbl:DatesInPairsTrade}
    \small
    \begin{tabular} {ll}
      \rule[-4pt]{0pt}{10pt} \\
      \hline
      \rule[-4pt]{0pt}{10pt} \\
               Date   & \multicolumn{1}{c}{Event Description} \\
      \rule{0pt}{2pt} \\
      \hline
      \rule{0pt}{4pt} \\
      1982-04-21 &  CME S\&P 500 futures begin trading. \\
      1987-10-19 &  Black Monday, the Worldwide Crash of 1987. \\
      1998-08-17 &  Russian Financial Crisis, currency devaluation \& default. \\
      1998-09-23 &  \glslink{LTCM}{LTCM}\index{LTCM} declares bankruptcy (not shown). \\
      2000-03-10 &  The beginning of the Dot-com Bubble crash. \\
      2001-04-01 &  U.S. stock markets begin trading in pennies. \\
      2001-09-11 &  World Trade Center attacks. \\
      2003-03-12 &  U.S. stock market bottoms out. \\
      2007-06-01 &  Beginning of the subprime CDO and CMO collapse.  \\
      2008-08-15 &  Lehman Brothers insolvency. \\
      2009-03-01 &  Bottom of the Global Credit Crisis of 2008. \\
      \rule{0pt}{4pt} \\
      \hline
    \end{tabular}
  \end{center}
\end{table}

At this level of presentation, the most important dates are 1987-10-19 (the Worldwide Crash of 1987), 1998-08-17 (the Russian Financial Crisis), 2000-03-10 (the apex of the \gls{Dot-com.Bubble}\index{crashes!.COM Crash of 2000}\index{bubbles!Dot-Com Bubble of 1996-2000}), 2001-09-11 (the \gls{911} attacks) and 2003-03-12 (the nadir of the post-911 correction).
  
In Figure \ref{TSAOMP:G:PairsCumPnL-19720825-20090413} observe the largely linear appearance from 1980-01-02 to 1998-08-17. Except for the dramatic jump the day of the Crash of 1987, there is no other change like that which occurs during that period. About a month later on 1998-09-23 and directly as a result of the Russian default, the giant hedge fund \emph{Long Term Capital Management (LTCM)} failed. On that date \glslink{LTCM}{LTCM}\index{LTCM} refused a rescue offer by Goldman Sachs, AIG and \gls{Buffett.Warren}\index{Buffett, Warren}, but was rescued instead by a group of creditors in a deal put together by the New York Federal Reserve. It's important to understand that at the time LTCM was the shining star of the hedge fund world, and the fear was that if it wasn't rescued, the ensuing turmoil might lead to a credit crisis (LTCM had massive positions in credit markets). The main question for us is: what increased the profitability of the \glslink{pairs.trading.strategy}{pairs trade}\index{pairs trading strategy} after 1998-08-17? 

Note first that before 1998-08-17, there was a tendency for the \glslink{pairs.trading.strategy}{pairs trade}\index{pairs trading strategy} to perform better during and immediately after a market correction or crash. \cite{RePEc:arx:papers:cond-mat/0211039} and \cite{citeulike:9370415} have documented that declines generally happen faster than market rallies, and that downside correlations among stocks are stronger than upside correlations. Although there is no accepted explanation for these asymmetries, they seem to provide an explanation for the better performance of the \glslink{pairs.trading.strategy}{pairs trade}\index{pairs trading strategy} during downdrafts, provided one accepts that increased correlations favor the \glslink{pairs.trading.strategy}{pairs trade}\index{pairs trading strategy}. Better performance after downdrafts is observable at various points of Figure \ref{TSAOMP:G:PairsCumPnL-19720825-20090413} prior to 1998-08-17. 

A window of understanding is afforded by the experience of \glslink{LTCM}{LTCM}\index{LTCM} around the time it failed. That date (1998-09-23) is not marked on the equity curve, but it is near the bottom of the DJIA drop that occurred after the Russian default on 1998-08-17. According to \cite{chincarini2012crisis}, pgs 58-60, LTCM near the time of its failure had a massive short position in equity index straddles, the net vegas being -$\$25$MM in U.S. equities and about the same for Europe equities. This means that for each $1\%$ increase in implied volatility, LTCM would lose $\$50$MM, assuming that the U.S. and Europe had equal volatility increases. LTCM's entire position in this trade consisted of short straddles on stock indexes and short-term hedges maintained dynamically using futures and options to neutralize portfolio delta and gamma. The long term options were written with banking counterparties and those counterparties had the right to demand more collateral. But LTCM's entire position in this trade depended on the five year volatility of $20$ being too high relative to the historical $13$. In a nutshell, LTCM was short volatility using illiquid, long-dated options with short-term hedges that dynamically maintained delta and gamma neutrality.

As rumors circulated that \glslink{LTCM}{LTCM}\index{LTCM} was in trouble, LTCM's counterparty banks starting demanding more collateral for their long-dated straddles. The situation became so serious that the straddles LTCM had purchased at implied volatility $20$ were being priced at $45$. The cost of that repricing to LTCM: $25 * \$50$MM = \$1BB! Under ordinary circumstances collateral of \$1BB might have been manageable, but at the time LTCM had lots of other positions that were also unraveling. 

To understand how \glslink{LTCM}{LTCM}\index{LTCM}'s problems might affect the wider market, consider the strategic possibilities of a representative counterparty bank. They are long a straddle from LTCM. If LTCM collapses, that in-the-money short straddle goes in the worst case to zero, meaning that the bank doesn't get paid what it's owed. That's why they were demanding more collateral, in order to recoup some of the loss after a default. But now consider what happens if LTCM fails and the LTCM short straddle really does go to zero. If the bank hedged or transferred the position to other traders, then they and those they hedged with are collectively long those straddles and a dynamic delta hedge by holders of those long straddles requires purchasing stocks when prices are dropping and selling when they're rising.  But this \gls{concave.strategy}\index{strategy!concave}\index{concave strategy} favors the \glslink{pairs.trading.strategy}{pairs trade}\index{pairs trading strategy}! 

Whenever a significant credit-related event occurs, the price impact on the wider market depends to a large degree on the web of creditor/debtor relations. In the case of the abrupt P\&L increase in the \glslink{pairs.trading.strategy}{pairs trade}\index{pairs trading strategy} after 1998-08-17 we surmise that there was a \gls{concave.strategy}\index{strategy!concave}\index{concave strategy} at work, one that was precipitated by the Russian default and its aftermath. Further, that effect was greater than the temporary effects after market downdrafts, as an examination of the chart between 1998-08-17 and 2003-03-12 should convince. Based on personal experience, I believe part of the effect was due to many new trading firms' discoveries of the \glslink{pairs.trading.strategy}{pairs trade}\index{pairs trading strategy} around that time.\footnote{I consulted with two groups looking into the trade around years 1998 to 2000. I even supervised an MBA project on pairs then.} As more traders used the \glslink{pairs.trading.strategy}{pairs trade}\index{pairs trading strategy}, it had the impact of reducing spreads in the pairs, thus creating better profitability for extant pairs.

Finally, what caused the accelerating profitability of the \glslink{pairs.trading.strategy}{pairs trade}\index{pairs trading strategy} after the \gls{Dot-com.Bubble}\index{crashes!.COM Crash of 2000}\index{bubbles!Dot-Com Bubble of 1996-2000} burst on 2000-03-10? Since the turn in the market coincides with the top of the Dot-Com Bubble, it seems clear that the Dot-com crash is somehow involved. Here is a guess based on work I was doing at the time. Many options traders I knew were long puts at incredibly high prices, and the majority of them delta hedged those positions. So as we discussed earlier, they were employing a \gls{concave.strategy}\index{strategy!concave}\index{concave strategy}. And that, I believe, was enough to have an effect on the pairs strategy. Ironically the put buying strategy made no money according to people I knew, because the drop in NASDAQ stocks was so gradual that eroding options premium ate up all the directional gains. Was this due to concave hedging strategies as well? 

Can one detect footprints of the POPP in Figure \ref{TSAOMP:G:PairsCumPnL-19720825-20090413}? Here is what I think. From 1982 to about 1995, the strategy wasn't well known and high commissions precluded all but a few professionals from using it, Phase (A). There is a marked increase in returns from 1995 to 1997 and this happens during the ascendancy of hedge funds, the \gls{Dot-com.Bubble}\index{crashes!.COM Crash of 2000}\index{bubbles!Dot-Com Bubble of 1996-2000} and ending with the collapse of \glslink{LTCM}{LTCM}\index{LTCM}. I suspect that hedge funds had more to with this quickening than other factors; this period is Phase (B). From March 2000 to March 2003 is Phase (C). Note the rapid ascent of the P\&L in the very last bubble phase with low volatility. We cannot know the outcome had \gls{911} not occurred, but I suspect the strategy would have failed to continue its gains even earlier. Note that after 9/11, the gains and losses of the pairs strategy are inverse to the \acrshort{aDJIA}, much as they had been before 1982. 
\begin{figure}[ht!]
  \begin{center} .
    \includegraphics[scale=0.52]{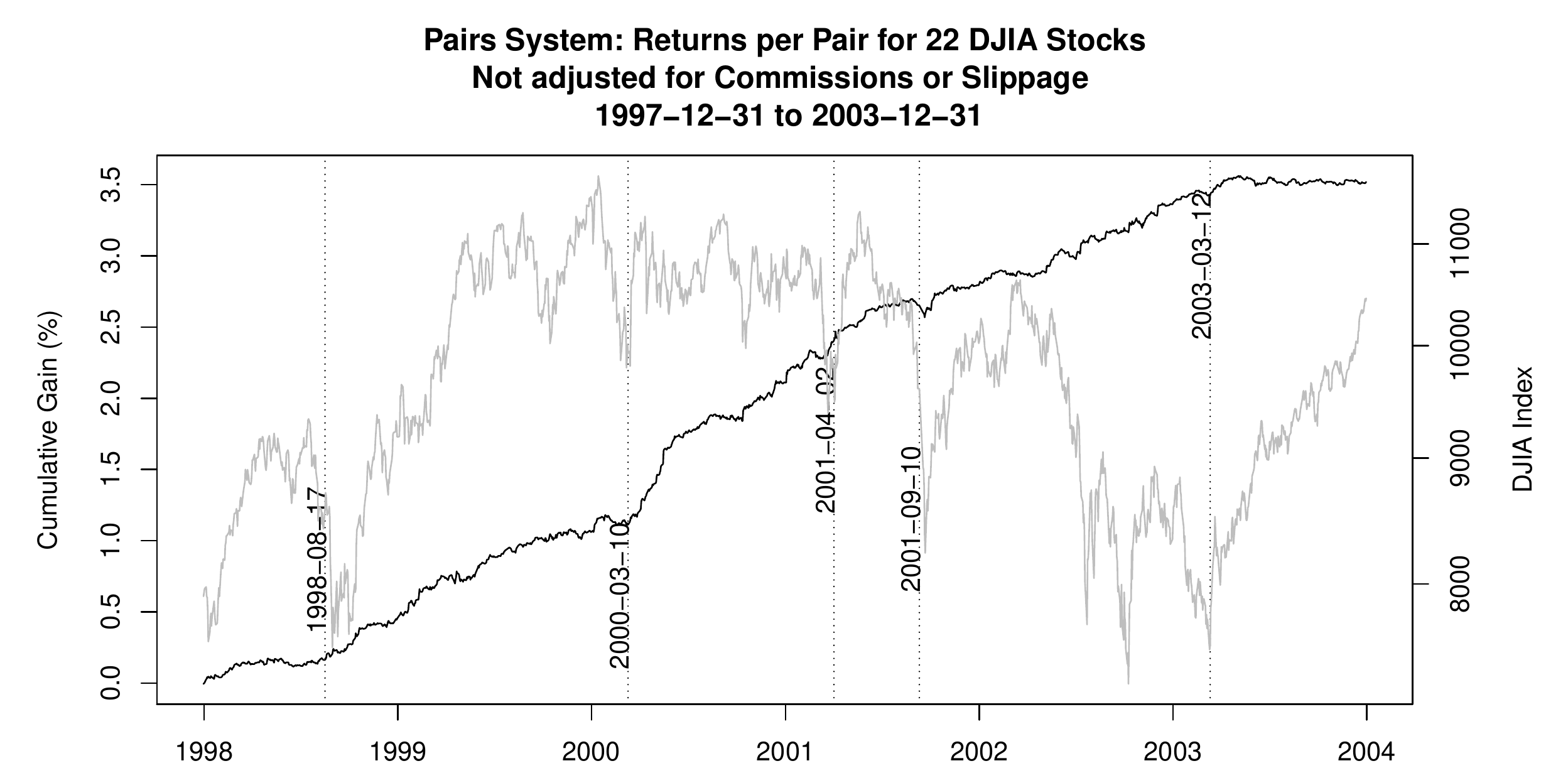}
    \caption{\small{P\&L Plot for a Pairs Trading System: 1997-12-31~2003-12-31. See text for details.}}
    \label{TSAOMP:G:PairsCumPnL:19971231:20010102}
  \end{center}
\end{figure}

Figure \ref{TSAOMP:G:PairsCumPnL:19971231:20010102} shows Phase (C) in more detail. Note the steady rise after the Russian default (1998-08-17) and the collapse of \glslink{LTCM}{LTCM}\index{LTCM} (1998-09-23) to the beginning of the \gls{Dot-com.Bubble}\index{crashes!.COM Crash of 2000}\index{bubbles!Dot-Com Bubble of 1996-2000} collapse that started on 2000-03-10. Although Figure \ref{TSAOMP:G:PairsCumPnL-19720825-20090413} shows a modest increase in the equity curve after March 2003, that gain is illusory. As is shown in \cite{moffitt2017V2}, the costs of commissions and slippage cause a steady loss up to 2008, a gain during the crisis of 2008, and a steady loss thereafter.

Incidentally, what caused the acceleration in Figure \ref{TSAOMP:G:PairsCumPnL-19720825-20090413} in $2007$-$2008$? Well, obviously the Credit Crisis of 2008, but also the collapse of many hedge funds. During that period, the extinction of the simple pairs strategy was involved, since the acceleration started in the early part of $2007$ when many hedge funds were already under pressure. Does the pairs strategy forecast crashes?

\noindent $\blacksquare$
\end{exmp}
\normalsize 

Consider the impact of \emph{\glslink{Pursuit.of.Profits.Paradigm}{POPP}}\index{Pursuit of Profits Paradigm} phases on Kelly betting. In the discrete setting, the Kelly fraction increases linearly with the success probability $p$. Thus, one should make the largest bets in Phases (A) and (B). And if can't make accurate predictions of, or efficiently escape from phase (D), bets in Phase (C) should be smaller due to lower $p$. We note that this prescription has important exceptions that are discussed in \cite{moffitt2017V1}.

In summary, the POPP is a useful model for traders, at least conceptually. Although not discussed here, there are different paths that the evolution can take. For example, the lifecycle can fizzle at any stage. A strategy can die with a whimper instead of a crash. Some thoughtful traders have addressed this, albeit in terms quite different from the \gls{SAFM}\index{SAFM}. An example is the ``reflexive'' model of \gls{Soros.George}\index{Soros, George}.  But the biggest problem with the POPP is the issue of estimating the phase of a \gls{strategic.plan}\index{strategic plan}\index{SAMM!strategic plan}, a problem we don't address here. 
 
\section{The Importance of Predictability}

Two players, \(1\) and \(2\), each wager \$1 and give it to a referee who will pay the winner.  Then they privately write either the letter \(H\) or the letter \(T\) on pieces on paper, seal them in envelopes, and then hand them to the referee.  The referee opens the envelopes and awards player \(1\) with the \$2 if the letters are the same, and player \(2\) the \$2 if the letters are different.  In tabular form, the players' net gains are \\
\mbox{ } \\
\begin{tabular}{rl}
  \((+1,-1)\) & if \((H,H)\) occurs, \\
  \((-1,+1)\) & if \((H,T)\) occurs, \\
  \((-1,+1)\) & if \((T,H)\) occurs, and \\
  \((+1,-1)\) & if \((T,T)\) occurs.
\end{tabular}
\mbox{ } \\ \\
where the notation \((i,j)\) means that player \(1\) wins \(i\) and player \(2\) wins \(j\).
\mbox{ } \\

\noindent \textbf{Discussion:} While this game is superficially one of coin tossing, it differs from coin tossing in important ways that facilitate a discussion of real games as opposed to artificial mathematical games. An obvious difference is the intrusion of human decision making into what is otherwise a random, mechanical process. Thus the human element is an integral part of this game, which changes everything.

The letters \(H\) and \(T\) will denote the choices made by \(1\) and \(2\). Thus each player has a choice between two ``strategies,'' \(H\) and \(T\).  Table \ref{GG:Tbl:MatchingPennies} shows the payoffs classified by the players' strategic choices.
\begin{table}[htp]
  \begin{center}
    \caption{\small{Normal Form of the Matching Pennies Game.}} 
    \label{GG:Tbl:MatchingPennies}
    \small
    \begin{tabular} {cccc}
      \rule[-4pt]{0pt}{10pt} \\
      \hline
      \rule[-4pt]{0pt}{10pt} \\
                              &            & \multicolumn{2}{c}{\textbf{Player \(2'\)s Strategies}} \\
      \rule{0pt}{2pt} \\
                              &            & \textbf{H}  & \textbf{T}  \\
      \rule{0pt}{2pt} \\
      \hline
      \rule{0pt}{4pt} \\
      \textbf{Player \(1'\)s} & \textbf{H} & \((+1,-1)\) & \((-1,+1)\)  \\
      \textbf{Strategies}     & \textbf{T} & \((-1,+1)\) & \((+1,-1)\)  \\
      \rule{0pt}{4pt} \\
      \hline
    \end{tabular}
  \end{center}
\end{table}

The display of the players' payoffs in Table \ref{GG:Tbl:MatchingPennies} is called the \emph{\gls{normal.form}}\index{games \& game theory!normal form} or \emph{\gls{strategic.form}}\index{games \& game theory!strategic form} of the game.  Note that in this game, the net amount won in all cases is zero -- games having this property are called \emph{\gls{zero.sum}} games\index{games \& game theory!zero-sum}.  If the referee charged a fee for his services, the game between \(1\) and \(2\) would be \emph{\gls{negative.sum}}\index{games \& game theory!negative-sum}. Because of the strategic symmetry of this simple game, neither player has any a priori advantage. Thus the game should be ``fair,'' meaning that in some sense the expected return to either player ought to be zero.

And indeed, that this game is fair and that the expectation to either player \emph{who plays perfectly} is zero, follows from a theorem in \cite{citeulike:494140}. Roughly speaking, that theorem states that there always exists a strategy for each player that guarantees the best expected return (\cite{citeulike:105659, myerson1997game}).\footnote{ This expected return is  called the \emph{\gls{value}}\index{games \& game theory!value} of the game. In general, such a strategy is not unique.} In the case of the guessing game, a perfect strategy consists of using the result of the flip of a fair coin to write \(H\) or \(T\), thus ensuring that the expected return is \(0\) to a player using this strategy. Note the subtlety of this result; if one player flips a coin and the other always chooses \(H\), an obviously exploitable strategy, the expected return to either player is still zero! And that is pretty much where game theory leaves it --- neither player has an advantage.

\begin{enumerate}
\item Since nothing was said about how many times the game will be played, the game specification is incomplete. Suppose that the players sign a contract to play the game \(100\) times, which can be managed by having each player give the referee \$100 before play begins. After each round of play, the referee awards \$2 to the winner; if one player quits the game before \(100\) games have been completed, the other player is awarded the remaining money. Consider the following strategy for \(1\) -- always write \(H\). After a few games, \(2\) will notice that \(1\) always chooses \(H\), and he will begin to play \(T\).  If \(1\) persists in always choosing \(H\), \(2\) wins all the remaining games. Obviously, this is a foolish strategy for \(1\), but the gist of this example is that \(1\) might have an exploitable nonrandom pattern of choice. For example, suppose that without realizing it, player \(1\) writes \(H\) about \(60\%\) of the time. By randomly choosing \(T\) a fraction \(x\) of the time, \(2\) expects to win 
\[
  ((0.6 x + 0.4(1 - x)) - ( 0.4 x + 0.6(1 - x)) * \$200 = (0.4 x - 0.2) * \$200
\]
over the \(100\) games, which is positive provided \( x> 0.5\). Thus \(2\) has a positive expectation in this case. If, for example, \(x = 0.6\), his expected gain is \(\$8\).  Think that this is impossible? Think again ---  \gls{shannon} invented an electronic device\footnote{ Shannon, Claude. ``A Mind-Reading (?) Machine.'' Bell Laboratories Memorandum, March 18, 1953.} that beats most people by detecting simple patterns in their choices!

Another variant is this: player \(1\) as he writes his number, tells player \(2\), ``I'm going to write \(H\) with probability \(0.75\).'' It is then possible for \(2\) to gain an edge by writing \(H\). Of course, it is assumed here that \(1\) is honest. When one of the players makes some statement or takes some action that implies a choice of \(H\) with probability \(p\) such that \(p \ne 0\), \(p \ne 0.5\), and \(p \ne 1\), it is called a \emph{\gls{partial.disclosure}}\index{partial disclosure}\index{games \& game theory!partial disclosure} action. When the player specifies or implies \(p = 0\) or \(p = 1\), it is called a \gls{full.disclosure}\index{full disclosure}\index{games \& game theory!full disclosure}\footnote{ The terms \emph{\gls{full.disclosure}} and \emph{\gls{partial.disclosure}} are not standard in game theory. They are sufficient, however, for the limited purposes of this presentation.} action. Partial or full disclosure actions in this game are exploitable.

\item Suppose player \(2\) hires a spy to watch player \(1\) write a choice -- 
and the spy lets player \(2\) know what was written before \(2\) writes his 
choice. What is \(2'\)s return in this case? Obviously, it's \(+1\) per bet, 
since \(2\) will always write the other letter, thus winning \(\$1\) each time 
the game is played. 

\item Cheating by influencing the referee is also a possibility. Either 
player can bribe the referee by offering him or her a payment of less than \(\$1\). 
This only works if the honest player does not detect the cheating. Situations 
that might arise in such cases can be imagined, i.e., the referee can produce a 
forged version of the cheater's choice if necessary.
\end{enumerate}

These methods of gaining advantage are quite general in \gls{zero.sum} games\index{games \& game theory!zero-sum}. The first case occurs because of \emph{skill}\index{games \& game theory!skill} in playing the game. Game theory generally studies best play by both parties within the framework of the rules, yet is strangely silent on outcomes due to differences in skill and cheating. In most athletic games such as tennis, golf and slalom skiing as well as in board games like chess, poker and backgammon, skill in the talented can be developed by hard work. The development of expertise is the whole point of these games.  

In the case of partial or \gls{full.disclosure}\index{full disclosure}\index{games \& game theory!full disclosure} actions, one player has information that the other does not have, obtained in the example above by pattern recognition, voluntary disclosure by an opponent or by immoral conduct, spying or bribing the referee. This case illustrates a phenomenon that occurs in real \gls{zero.sum} games\index{games \& game theory!zero-sum}, \emph{\gls{competition.for.the.second.move}}. Competition for the second move in \gls{zero.sum} games\index{games \& game theory!zero-sum} occurs because the player that moves second with full knowledge of the first player's choice, essentially a \gls{full.disclosure}\index{full disclosure}\index{games \& game theory!full disclosure} action, generally gains an advantage.  

You're probably wondering what this example has to do with financial markets. Actually, each of these paradigms has analogs in financial markets.

For example, when participants in financial markets reveal information about their strategies, they allow others the potential to profit from that information, essentially by having the advantage of the second move. Consider, for example, the fact that some mutual fund managers sell losing stocks prior to quarterly reports to avoid investor backlash. This information opens the possibility of profiting by buying those stocks just before the end of a quarter. Why? Because excessive selling by the funds can ``oversell'' those stocks and drive them below their ``fair'' values. Of course, it is assumed here that excessive supply moves a market below its ``fair'' price. 

An interesting example occurred many years ago in the most popular financial show of the era, ``Wall Street Week,'' hosted by Louis Rukheyser. The program featured a witty, attention-getting monologue by Rukheyser in which he poked fun at market follies of the previous week. Later, he and market guru guests would chat in a homey room with modestly appointed furniture, lamps and fixtures. Guests would be asked for their market wisdom, especially for any hot tips. In a series of articles written between \(1985\) and \(1993\), well-respected market analysts \cite{Moffitt:Dorfman:NYMag:19850114} and \cite{Moffitt:BaltimoreSun:19930425} analyzed these recommendations. Their analyses revealed that recommended stocks would generally rise starting about three weeks before the program, continue for a few days after its airing and decline significantly until \(3\) to \(6\) weeks after the show, and these moves were excessive relative to market averages. When experienced investors were asked to explain this pattern, they suggested that savvy investors would inquire as to the show's guests in upcoming weeks, learn what stocks they liked and begin buying them. They would then sell into anticipated demand after the show. Another explanation was that guests were generally invited because the sectors they specialized in were ``hot,'' and were due to reverse that trend. And, of course, there is the rational possibility that gurus would share their picks with other gurus, either returning a favor or expecting the favor to be returned later.

Spying in the markets happens all the time. One form is the hiring of a trader, software developer or other significant player in order to ``steal'' proprietary trading ideas. While theft of trading ideas is governed in the U.S. by intellectual property law, such agreements have a short life-span (usually a year) and are difficult to enforce because improvements or reprogramming can get around the agreement. 

I don't have any direct knowledge of ``bribing the referee,'' but instances are widely reported. For instance, \cite{johnson201013} reported that several investment banks ``worked with'' rating firms to gain lenient treatment of mortgage derivatives. Yes, this was indeed a blatant conflict of interest by the rating companies, but to date no one has suffered criminal penalties. Though it does not easily fit into this category, there are well-publicized instances of ``rigging the game.'' A recent, and particularly egregious case of this sort is the ``Eurodollar scandal of 2012'' (\cite{wiki:LiborScandal}). The scandal involved price rigging by a number of market-maker institutions.\footnote{LIBOR rigging was akin to alleged point-spread fixing by umpires and referees in sports betting. When bookies have tremendous exposure to a game's outcome, it helps to have a ``friendly'' referee. Fixing is especially easy in basketball, since a few strategic calls can flip the point spread without changing the winner.} In a quote attributed to Andrew Lo at MIT by \cite{OToole:CNNMoney:2012-07-10},
\begin{quote}
  ``This dwarfs by orders of magnitude any financial scam in the history of 
  the markets.''
\end{quote}

The foregoing discussion of market realities should not obscure the message of this section --- that in zero-sum games, partial or full disclosure actions confer the advantage of the second move to astute players. In these cases, then, the astute have potentially winning strategies in otherwise fair games.

\section{Getting an Edge: Human Behavior and Price Distorters}

Up to this point, it's been assumed that our \gls{grational}\index{grationality} investor has an \gls{edge}\index{edge}\footnote{That is, the expected returns on his portfolio of bets exceeds risk-free returns.} which is exploited efficiently using grational principles and the POPP. ``Fine,'' you say, ``but how do I get an edge?'' In fact, if that question were to go unanswered, the \glslink{SAFM}{SAFM's}\index{SAFM} value would be minimal. Therefore, beginning with this section and for the rest of this article, an answer is provided.

One popular answer is: learn how people make mistakes using investor psychology, or as it's known in the scholarly literature, ``behavioral finance'', and derive an edge therefrom. Indeed, for our \gls{grational}\index{grationality}  investor, knowledge of investor psychology is essential. But investor psychology alone does not an edge make --- that takes strategic thinking and gamesmanship.

\subsection{Mostly Human Behavior}

This section presents briefly some of the important findings about human decision making discussed in \cite{moffitt2017V1}.

Popular theories model the mind as a  \emph{dual process}\index{dual process theory of the mind}, \emph{\gls{associative.machine}}\index{associative machine}. The associative machine organizes memories, beliefs and constructs of reality in a massive network with links formed according to nearness in time, attribute classifications, emotional reactions, likes and dislikes, etc. When presented with a sensory or mentally-generated stimulus, the associative machine attempts to find a \emph{coherent} explanation based on its extant web of associations. Learning cannot be effective if the mind completely revises its beliefs with every dissonant piece of information; it follows therefore that people revise their beliefs by discounting new information. The big question is: how does this discounting affect Bayesian updating of beliefs? Many studies provide an answer, that humans revise their beliefs slower that Bayesian rules do, a bias known as the \emph{\gls{belief.revision.bias}}\index{belief revision bias}. 

The \emph{dual process} theory models thinking as a product of collaboration of two minds, the \emph{reflexive}, primitive \emph{\gls{System1}}\index{System 1} and the \emph{reflective}, conscious \emph{\gls{System2}}.\index{System 2} System 1 is a massively parallel system that takes sensory and brain-generated information to make effortless, rapid judgments below the conscious threshold. System 1, however, is difficult to program consciously, especially since it makes some decisions using emotions and \gls{affect}\index{affect} (deciding purely by ``liked'' or ``disliked''). System 2 on the other hand, is the conscious, effortful, programmable, plodder portion of the mind. It is involved in such things as mental calculations and abstract reasoning, operations that drain mental energy. While this simplistic introduction suffices for this article, serious quantitative traders should read the definitive introduction to this material: \cite{citeulike:9935672}, ``Thinking, Fast and Slow.''

In the late 1960's, Daniel Kahneman and Amos Tversky became interested in the ``rational man'' model that had become popular in economics in the 1960's, and began conducting experiments of its basic quantitative decision making model, \emph{\gls{utility.theory}}\index{utility theory}. By the mid 1980's, they had discovered that people do not decide using utility theory, and in \cite{citeulike:99680} proposed \emph{\gls{prospect.theory}}\index{prospect theory}\footnote{And proposed later in \cite{citeulike:790903} an improved model, \emph{\gls{cumulative.prospect.theory}}} to approximate human decision making. But they are best known for discovering \emph{heuristics} and \emph{biases} that lead to violations of utility theory. The best known are (1) \emph{\gls{representativeness}}\index{representativeness}, an heuristic that chooses the best matching pattern, (2) \emph{\gls{availability}}\index{availability}, an heuristic that chooses using the most easily recalled instances, and (3) \emph{\glslink{anchor}{anchoring}}\index{anchoring}, an heuristic that biases choices toward a reference point. For examples, see \cite{citeulike:9935672} or \cite{moffitt2017V1}.

More important for finance, however, are the biases discovered by Kahneman, Shefrin, Statman and Thaler: (1) the \emph{\gls{status.quo.bias}}\index{status quo bias}, (2) the \emph{\gls{endowment.effect}}\index{endowment effect}, (3) \emph{\gls{loss.aversion}}\index{loss aversion}, (4) \emph{\gls{mental.accounting}}\index{mental accounting}, (5) the \emph{\gls{disposition.effect}}\index{disposition effect} and (6) \gls{framing}\index{framing} effects. Briefly, the status quo bias expresses the human preference for the status quo (no change) to superior alternatives. The endowment effect is the human tendency to value the same object more if it is owned than if it is not. Loss aversion is the near universal preference to avoid losses, even if such action gives away considerable \gls{edge}. Mental accounting refers to a compartmentalization of financial activity into \emph{mental accounts} in which gains or losses in any one account are not aggregated with gains or losses in the others. The disposition effect is the tendency for investors to sell too early and hold too long, violating the long-standing trader's injunction to ``cut your losers and let your winners run.'' Framing refers to the manner in which a problem is posed; a problem posed in terms of number of lives saved and the same problem posed in terms of lives lost, will often evoke two different decisions.

A larger catalog of heuristics and biases important for trading is contained in Chapters 6 \& 7 of \cite{moffitt2017V1}; here we observe (tautologically) that human behavior lies behind all market anomalies and patterns. The view of this article is that while knowledge of behavior is necessary for system developers, it is not sufficient --- that requires the analysis of trading strategies and their interactions.

We conclude this section with a brief discussion of two important pieces of research that are seldom mentioned in the behavioral literature. The first is the study of heuristics and biases in monkeys. Somewhat surprisingly, monkeys exhibit loss aversion, the \gls{status.quo.bias}\index{status quo bias}, the endowment effect and framing effects! Implication --- in humans these are probably \gls{System1}\index{System 1} biases that are \emph{very} hard to reprogram. The second is research by \cite{Hilbert:PsychBulletin:2012:138:2:201203} which shows that a Shannon information-theoretic model can explain the \emph{\gls{belief.revision.bias}}\index{belief revision bias}. The implication is that imperfect human storage and retrieval of memories gives rise to the \emph{belief revision bias}, and therefore that the belief revision bias is an integral part of human learning!

These and other human heuristics, biases and behaviors lead to partly predictable strategic behavior, and therefore to \emph{\glslink{potentially.profitable.gambling.system}{potentially profitably gambling systems}}\index{PPGS}! 

\subsection{Price Distorters}

A price distorter\index{price distorter} is any exogenous news, constraints on trading, widely held beliefs, widely used strategies, or other regular market phenomena that have potential \glspl{price.impact}. It is difficult to appreciate this concept without some examples, but the core idea is this. All exogenous or endogenous happenings important to a market have the potential to ``distort'' prices, in the sense that prices would have been higher or lower in their absence. Many market participants unnecessarily limit the scope of this idea to concrete occurrences like dividend announcements, stock splits and so on. But it has much wider scope. For example, some strategies are price distorting, e.g. portfolio insurance or risk aversion. Structural factors such as the prohibition on shorting by mutual funds can be distorting. Conventions established in the effort to organize markets (rules) such as futures and options expirations are potentially price distorting. The introduction of a new financial instrument, e.g. the S\&P 500 index future in August, 1982 can be price distorting. And so on.

We identify three classes of price distorters that occur commonly: (1) event, (2) strategic and (3) rule-constrained. A well-known event type was discovered by \cite{Bernard:Thomas:xxx:jacct:v:27:y:1989:i:1:p:1-36}. That study found that after surprising earnings announcements, there was price drift for at least 60 trading days in the direction indicated by the surprise. A clear-cut example of a strategic price distorter is portfolio insurance (\cite{moffitt2017V1}), the strategy of writing dynamically replicated puts on portfolios. Portfolio insurance is widely believed to have been the main cause of the \acrlong{aWSMC87}. See Example \ref{TSAOMP:Exmp:PriceDistortionFactor:EndOfYearTaxSelling} for a rule-constrained distorter. Distorters are often hybrids, e.g. the Tax Day Trade described in \cite{moffitt:TaxDayTrade} that has event and rule-constrained origins.

Many other examples of price distorters are contained in \cite{moffitt2017V1}. The following example involves a factor whose origins are in U.S. tax law that produces end-of-year tax selling, and which can be exploited by a known strategy.

\vspace{2mm}
\small
\begin{exmp}[Price Distorter: End of Year Tax Selling]\label{TSAOMP:Exmp:PriceDistortionFactor:EndOfYearTaxSelling}

U.S. tax law allows deductions of no more than $\$3,000$ per tax year against income, so that if, say, a net of $\$9,000$ were lost from stock sales, only $\$3,000$ could be used as a deduction against current income. In subsequent tax years the additional $\$6,000$ could be used to offset capital gains in stocks or as offsets against income. For this reason, investors who want to realize stock gains often prefer to sell them together with losing stocks in a joint transaction that loses no more than $\$3,000$. But the law has a further provision that prevents the strategy of tax selling followed by immediate repurchase --- the wash sale rule. It states that stocks sold as part of an offset cannot be repurchased for $30$ days after the sale (with exceptions that are unimportant to this discussion; if you need to do tax selling, you need to read the tax law carefully or consult a financial professional).

In practice, brokers and tax professionals advise using these rules as follows. After a tax sale involving the sale of stocks with capital gains of $\$x$ and the sales of stocks with capital losses of $\$x + \$3,000$  or less, the investor bullish on one of the stocks should find a ``close'' substitute (the IRS has rules for this) and buy it. Then after $30$ days, sell that stock and repurchase the original. This strategy is so common that is has a name: a \emph{tax swap}. The strategy makes sense for an investor who has a large capital gain in a stock that he or she is bullish on, because the repurchase establishes a new, higher basis that reduces the future tax burden. And contrary to the optimal \glslink{efficient.market.hypothesis}{EMT}\index{efficient market theory (EMT)} strategy for using these rules (\cite{RePEc:ecm:emetrp:v:51:y:1983:i:3:p:611-36}) investors typically defer decisions on tax sales and tax swaps until December. Such last-minute-ism is characteristic of investors, especially when taxes are involved. 

This tax-selling strategy has been implicated in the \gls{January.effect}\index{January effect}, first published by \cite{RePEc:eee:jfinec:v:9:y:1981:i:1:p:3-18} who discovered that small stocks had large returns when bought in the old year and held only for a short period in the new year. The reason, some speculated, is that end of year tax selling differentially affected small stocks. See also Example \ref{Exmp:TheTaxDayTrade}, the \emph{Tax Day Trade}. 

\noindent $\blacksquare$

\end{exmp}
\normalsize

\section{The Strategic Analysis of Markets Method\\(SAMM)}\label{S:TheStrategicAnalysisOfMarketsMethod}

The SAMM\index{SAMM} has six steps:
\begin{enumerate} [label=\bfseries {(\thesection.\arabic*)},ref={(\thesection.\arabic*)}]
\item Select potential \glslink{price.distorter}{price distorters}\index{price distorter} for trading systems. \label{TSAOMM:Enum:SAMZM:GeneratePriceDistortionFactors}

\item Assemble datasets to investigate the \glspl{price.impact}\index{price impact} associated with the price distorters.\index{price distorter} \label{TSAOMM:Enum:SAMM:InvestigatePriceImpacts}

\item If Step (2) shows significant price impacts\index{price impact}, formulate \glslink{strategic.plan}{strategic plans}\index{strategic plan}\index{SAMM!strategic plan} therefrom. \label{TSAOMM:Enum:SAMM:FormulateStrategicPlans}

\item Convert each strategic plan\index{strategic plan}\index{SAMM!strategic plan} to a \gls{trading.algorithm}\index{trading algorithm}.  \label{TSAOMM:Enum:SAMM:ConvertStrategicPlanToTradingSlgorithm}

\item Perform backtesting on the trading algorithms\index{trading algorithm} and iterate Steps \ref{TSAOMM:Enum:SAMM:FormulateStrategicPlans}-\ref{TSAOMM:Enum:BacktestTradingAlgorithm} as necessary. \label{TSAOMM:Enum:BacktestTradingAlgorithm}

\item Upon the completion of iteration, locate the strategic plan in the \glslink{Pursuit.of.Profits.Paradigm}{POPP}\index{Pursuit of Profits Paradigm (POPP)} cycle and follow the evolution forward. \label{TSAOMM:Enum:POPPLocation}

\end{enumerate}
These steps are unremarkable except for the first and last--- all successful systems developers employ some version of them. The value of the \glslink{Strategic.Analysis.of.Markets.Method}{SAMM}\index{Strategic Analysis of Markets Method (SAMM)}\index{SAMM} lies in its systematic approach to generating hypothetical trading ideas without any data analysis, using what we call \glspl{price.distorter}\index{price distorter}, and in its requirement of following the strategic life cycle using \glslink{Pursuit.of.Profits.Paradigm}{POPP}\index{Pursuit of Profits Paradigm (POPP)}. But the efficient employ of distorting factors comes at a price --- it requires a good working knowledge of market history and \emph{\gls{market.ecology}}\index{market ecology}.

\vspace{2mm}
\small
\begin{exmp}[The Tax Day Trade]\label{Exmp:TheTaxDayTrade}

Example \ref{TSAOMP:Exmp:PriceDistortionFactor:EndOfYearTaxSelling} was an example of a rule-constrained distorter associated with tax law. Since tax law affects most U.S. citizens, this potentially large group could easily move stock markets if tax and stocks were affected. The table below shows major changes in tax law since 1954 (compliments of Wikipedia):
\begin{enumerate}[label={(\roman*)},ref={\thethm~(\roman*)}]

  \item{\textbf{Capital Gains Tax Rate:}}
    \begin{description}
      \vspace{-1mm}
      \item{\textbf{1954-1967:}} Flat $25\%$.
      \vspace{-1mm}
      \item{\textbf{1968-1977:}} Variable, but significantly increased
      \vspace{-1mm}
      \item{\textbf{1978-1980:}} Variable, but maximum $28\%$.
      \vspace{-1mm}
      \item{\textbf{1981-1986:}} Flat $20\%$.
      \vspace{-1mm}
      \item{\textbf{1987-1986:}} Variable, with maximum at $28\%$.
    \end{description}

  \item{\textbf{Dividend Taxation:}}
    \begin{description}
      \vspace{-1mm}
      \item{\textbf{1954-1984:}} At individual tax rate.
      \vspace{-1mm}
      \item{\textbf{1985-2002:}} At individual tax rate with maximum of $50\%$.
      \vspace{-1mm}
      \item{\textbf{2003-2015:}} Flat 15\%.
    \end{description}

  \item{\textbf{Other Tax Changes since 1974:}}
    \begin{description}
      \vspace{-1mm}
      \item{\textbf{1974:}} Employee Retirement Income Security Act of 1974 (ERISA) \\
                            * Established \emph{Individual Retirement Accounts} (IRAs) with restrictions.\\
                            * Deadline for deposit, tax day.
      \vspace{-1mm}
      \item{\textbf{1981:}} Economic Recovery Tax Act of 1981 (Kemp-Roth Act). \\
                            * Allowed all working taxpayers to establish \emph{Individual Retirement Accounts} (IRAs) \\
                            * Deadline for deposit, tax day. \\
                            * Expanded provisions for employee stock ownership plans (ESOPs)
      \vspace{-1mm}
      \item{\textbf{1986:}} Tax Reform Act of 1986. \\
                            * Lowered individual tax rates for upper brackets, raised for lower. 
    \end{description}
\end{enumerate}
Capital gains rates will have an impact, making stocks held for the mandated minimum period (one year at this writing) more attractive. Thus lowering of rates would be expected to boost stock prices. Dividend taxation would affect the attractiveness of dividend-paying stocks, favoring such stocks when dividend tax rates are below the ordinary income rate. Neither of these would seem to have immediate impacts except when the law changes. However, the ERISA law of 1974 and Kemp-Ross Act of 1981 have potential to impact markets, since establishment of individual retirement accounts could result in increased demand for stocks.

The law that would be expected to have the greatest impact is Kemp-Ross, since it allowed average tax payers to create account on or before the deadline for individual income tax payments. Furthermore, one might expect procrastination in establishing such accounts, just as investors procrastinated in end of year tax selling. Thus one would expect large numbers of accounts to be established at the deadline (usually April 15). This suggests testing stock returns around the deadline each year.

This trade is an hybrid of \emph{event} and \emph{rule-constrained} types. For event types, one locates all events at time zero (in this case, tax day), forms an aligned dataset of events and their nearby prices or returns and performs a multivariate analysis to discover tradable patterns. Performing such an analysis, we find that a major positive return occurs the day after tax day, as shown in Figure \ref{TSAOMM:G:IRSTaxDayStrategy}. 
\begin{figure}[h!]
  \centering
  \includegraphics[scale=0.5]{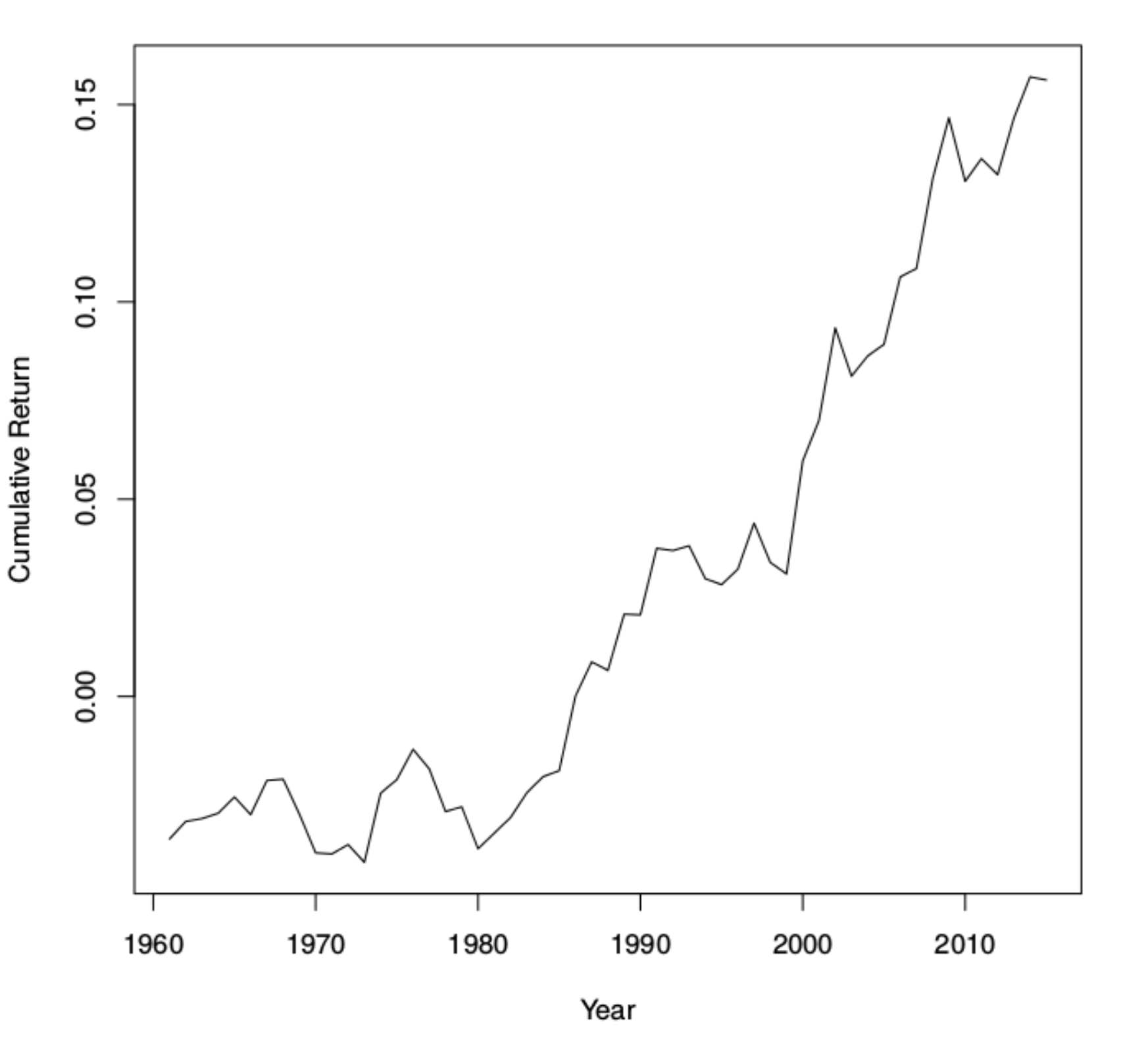}
  \caption{\small{Cumulative returns from buying the S\&P 500 index on the close of U.S. tax day and selling on the close one day later. Commissions and slippage not included.}}
  \label{TSAOMM:G:IRSTaxDayStrategy}
\end{figure}

The figure shows cumulative returns for the day after tax day for the $56$ years $1960$-$2015$. This strategy can be realized approximately (omitting commissions and slippage) by buying the close of S\&P 500 futures\footnote{Of course, S\&P 500 futures did not exist until April $21$, $1982$. Details, details \ldots, luckily irrelevant in this case. We also note that commissions were far higher than today during much of this period, making the trade unprofitable for non-professionals until lower commissions were generally available.} on tax day each year, and selling on the close one day later. Note the strikingly different behavior of cumulative returns after $1984$ --- the curve ascends steadily. Despite being a one day trade, it has averaged about $1/2 \%$ per year since $1980$.

Using the hypothesis of concerted establishment of IRA's on the deadline day due to investor procrastination, many brokers would be too busy with the paperwork on tax day, and defer buying until the next day!

\noindent $\blacksquare$
\end{exmp}
\normalsize 

We give the results of another SAMM analysis in an example which is fully developed in Chapter 14 of \cite{moffitt2017V2}. The price distorter of Step 1 is an event type: expirations of Eurodollar futures. 

\vspace{2mm}
\small
\begin{exmp}[Two Systems for Eurodollar Futures Developed Using the SAMM]\label{Exmp:TwoSystemsForEurodollarFutures}

Eurodollar-System 1 holds long or short Eurodollar positions and should be profitable with efficient execution. Its Cumulative P\&L curve before commissions is shown in Figure \ref{TSYC:G:TSYC-SEGMENTED:123-1990-CUMPNL}.
\begin{figure}[!ht]
  \begin{center}
    \includegraphics[scale=0.35]{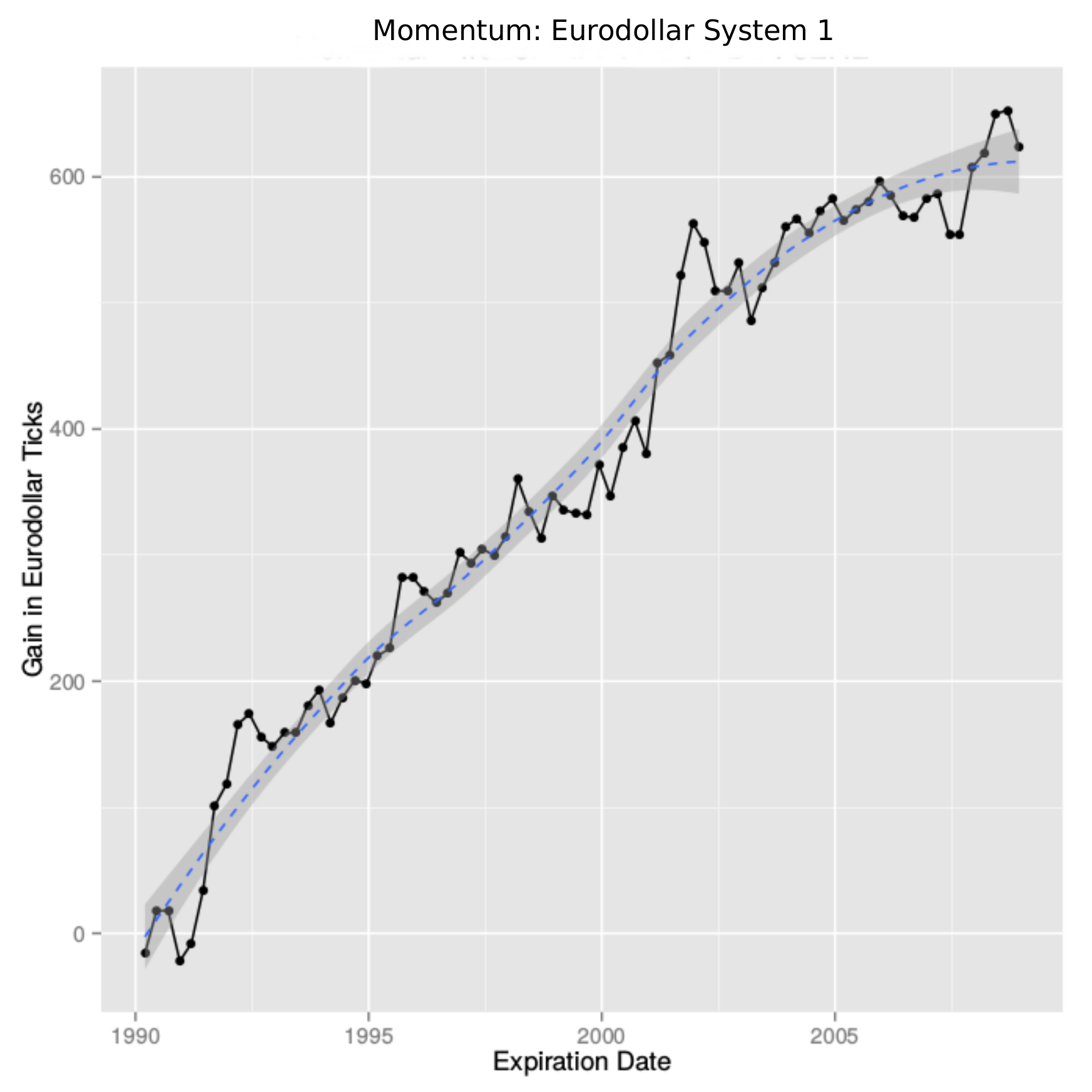}
  \end{center}
  \caption{\small{Cumulative gains for the Eurodollar-System 1 trade. Source: \cite{moffitt2017V2}}} 
  \label{TSYC:G:TSYC-SEGMENTED:123-1990-CUMPNL}
\end{figure}
and a summary of the system's characteristics is shown in the Table below. Separate lines are shown for (1) long trades, (2) short trades, and (3) all trades. Column meanings are shown in Table \ref{TSYC:Tbl:ColumnDescriptionForTradingSystemSummaryYC}.

\newpage
\scriptsize
\begin{verbatim}
            np npi   maxdd pnlpp    ir pnltot sdpnl winpct runs runspvu 
      Long  41  41  -65.00 10.96  0.45 449.52 24.38  65.85   21    0.70 
      Short 32  32 -112.47  5.44  0.19 173.98 28.07  53.12   14    0.19 
      All   76  76  -77.49  8.20  0.32 623.50 25.55  57.89   38    0.74 
\end{verbatim}
\small

The total gain for the $19$ year period of analysis is $623$ Eurodollar ticks on a one lot, for an average gain per year of $32.8$ Eurodollar ticks. This is not an insubstantial gain! In fact, the target position from days $32$ to $42$ (two weeks) is held only 4 times per year, for a total of $8$ weeks in $52$ ($8/52 \sim 15\%$) and averages a gain of about $\$820$ per year on a one lot. The initial margin on a one year Eurodollar has varied over the years but probably has averaged about $\$1,500$ per contract. If, say, funds of $\$5,812$ per contract are required (see reasoning below), then the return per year would be about $14\%$ per year before fees and \gls{slippage}\index{slippage}. Thus this preliminary estimate of system profitability is encouraging, but that does not make the system viable. 
\begin{table}[h!]
  \begin{center}
    \caption{\small{Column descriptions for trading system \gls{profit.and.loss} summary.}}
    \label{TSYC:Tbl:ColumnDescriptionForTradingSystemSummaryYC}
    \small
    \begin{tabular} {ll}
      \rule[-4pt]{0pt}{10pt} \\
      \hline
      \rule[-4pt]{0pt}{10pt} \\
         \multicolumn{1}{c}{\textbf{Column}} & \multicolumn{1}{c}{\textbf{}} \\
         \multicolumn{1}{c}{\textbf{Name}}   & \multicolumn{1}{c}{\textbf{Description}} \\
      \rule{0pt}{2pt} \\
      \hline
      \rule{0pt}{4pt} \\
           np      &  The total number of periods. \\
           npi     &  The number of periods ``in the market,'' that had a position. \\
           maxdd   &  Maximum peak-to-trough \gls{drawdown}\index{drawdown} \\
           pnlpp   &  P\&L per period (\%). \\
           ir      &  Information ratio = (mean ret)/(sd ret). \\
           pnltot  &  Total of trade returns for the entire period (not compounded). \\
           sdpnl   &  Standard deviation of P\&L per period. \\
           winpct  &  Winning percentage. \\
           runs    &  Number of runs. \\
           runspvu &  P-value for too few runs. \\
      \rule{0pt}{4pt} \\
      \hline
    \end{tabular}
  \end{center}
\end{table}

A second system, Eurodollar-System 2, produces the following summary table and cumulative P\&L Figure \ref{TSYC:G:TSYC-CumulativeReturnsFromQuarterlyCrossoverOf3-6MonthContracts}. This system has quite low commissions and slippage, and has produced annual returns exceeding 10\%.

\scriptsize
\begin{verbatim}
             np  npi   maxdd pnlpp    ir  pnltot sdpnl winpct runs runspvu 
Sys2 Long  2410 2410 -138.49  1.05  0.12 2533.54  8.50  56.72 1097     
Sys2 Short 1894 1894 -232.51  0.22  0.03  418.37  7.83  52.48  932     
Sys2 All   4730 4304 -205.50  0.69  0.08 2951.91  8.22  54.86 2026       0 
\end{verbatim}
\small

\vspace{3mm}
\begin{figure}[!ht]
  \begin{center}
    \includegraphics[scale=0.6]{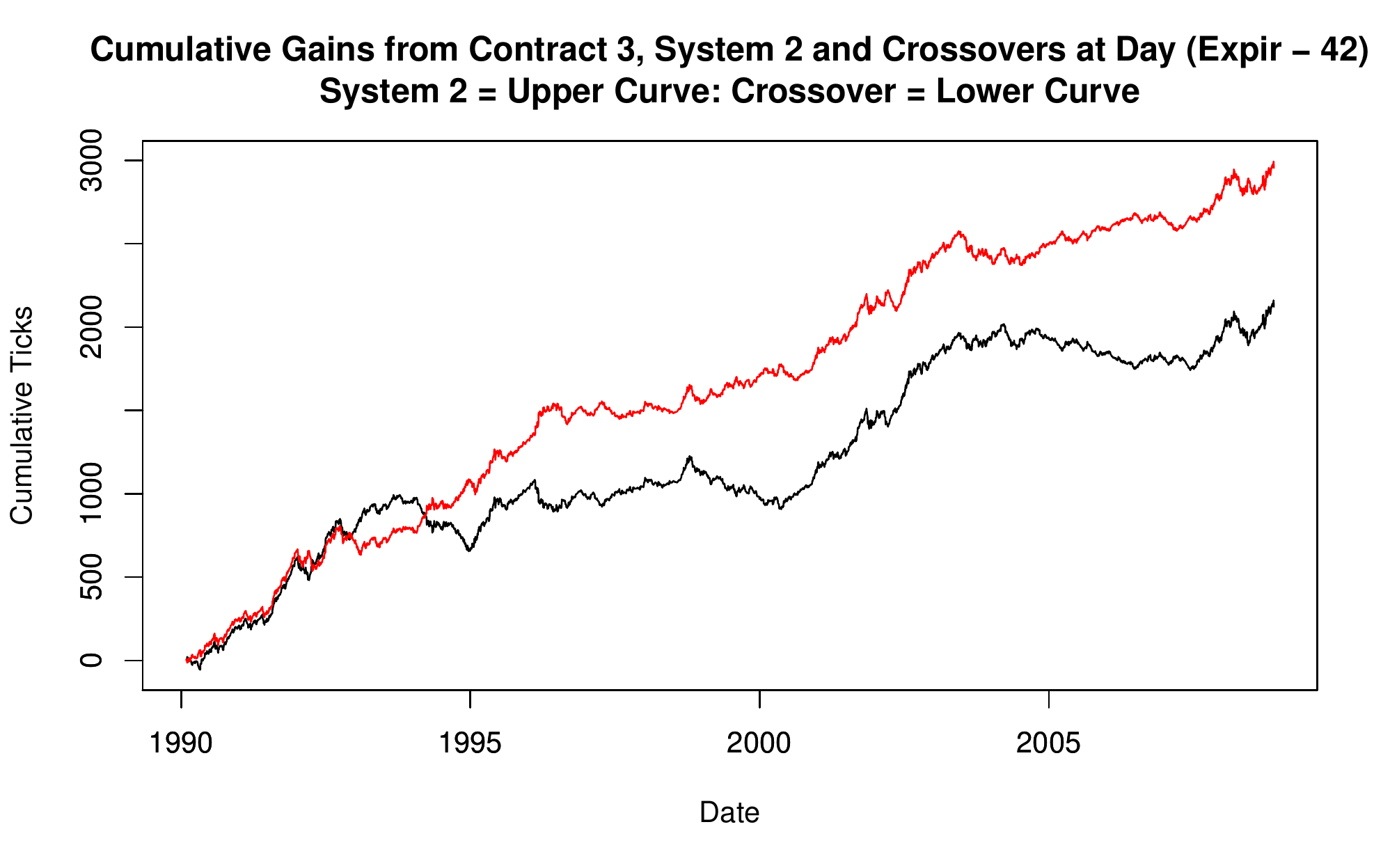}
  \end{center}
  \caption{\small{Cumulative Returns from the Roll Strategy (lower line) vs. \emph{Eurodollar-System 2} (upper line).}} 
  \label{TSYC:G:TSYC-CumulativeReturnsFromQuarterlyCrossoverOf3-6MonthContracts}
\end{figure}

\noindent $\blacksquare$
\end{exmp}
\normalsize 

\begin{appendices}
\section{Calculation of the Kelly Fraction}\label{A:CalculationOfTheKellyFraction}

Using the setup of independent outcomes $X_i = \{0,1\}$ paying $1 - f$ for $X_i = 0$ and $1 + d \cdot f$ for $X_i = 1$, the growth rate of wealth after $n$ flips is 
\begin{equation}
  G_n = \frac{1}{n} \left( \left( \sum_{i=1}^{i=n} X_i \right) log(1 + d \cdot f) - \left( n - \sum_{i=1}^{i=n} X_i \right) log(1-f) \right), \label{FractionalGrowth}
\end{equation}
where the notation $d \cdot f$ has been used to avoid confusion with the differential $df$. By the law of large numbers, \eqref{FractionalGrowth} converges strongly to 
\begin{equation}
  G_{\infty} = p \; log(1 + d \cdot f) \, - \, (1-p) log(1-f). \label{KellyStrongConvergence}
\end{equation}
Differentiating \eqref{KellyStrongConvergence} with respect to $f$ gives
\begin{equation}
  \frac{dG_{\infty}}{df} = \frac{d \cdot p}{1 + d \cdot f} - \frac{1-p}{1 - f}. \label{FirstOrderKelly}
\end{equation}
Setting \eqref{FirstOrderKelly} to $0$ and solving yields
\[
  f = p -q/d,
\]
and this can be shown to maximize \eqref{FractionalGrowth}.

\end{appendices}

\newpage 
\printglossaries
\newpage
\bibliographystyle{apalike}
{\footnotesize 
\bibliography{WMAI}
\printindex

\end{document}